\documentclass[ reprint,
showpacs,aps,pre,superscriptaddress,
floatfix,
]{revtex4-1}
\usepackage{amsmath,amsfonts,amsthm}
\usepackage{dsfont}
\usepackage{braket}
\usepackage{graphicx}
\usepackage{epstopdf}
\usepackage{dcolumn}
\usepackage{bm}
\usepackage{float}
\usepackage[caption=false]{subfig}
\usepackage{color}
\usepackage{multirow}
\usepackage[bookmarks=true,colorlinks,linkcolor=OrangeRed,urlcolor=NavyBlue,citecolor=RoyalBlue]{hyperref}
\usepackage[dvipsnames]{xcolor}
\definecolor{mygreen}{rgb}{0.1,0.5,0.1}

\begin{document}
 
\title{Detecting quantum phase transitions in the quasi-stationary regime of Ising chains}

\author{Ceren B.~Da\u{g}}
\email{ceren.dag@cfa.harvard.edu}
\affiliation{ITAMP, Harvard-Smithsonian Center for Astrophysics, Cambridge, Massachusetts, 02138, USA}
\affiliation{Department of Physics, Harvard University, 17 Oxford Street Cambridge, MA 02138, USA}
\affiliation{Department of Physics, University of Michigan, Ann Arbor, Michigan 48109, USA}

\author{Philipp Uhrich}
\affiliation{INO-CNR BEC Center and Department of Physics, University of Trento, Via Sommarive 14, I-38123 Trento, Italy}

\author{Yidan Wang}
\affiliation{Department of Physics, Harvard University, 17 Oxford Street Cambridge, MA 02138, USA}

\author{Ian P.~McCulloch}
\affiliation{School of Mathematics and Physics, University of Queensland, St. Lucia, Queensland 4072, Australia}

\author{Jad C.~Halimeh}
\affiliation{INO-CNR BEC Center and Department of Physics, University of Trento, Via Sommarive 14, I-38123 Trento, Italy}

\date{\today}
\begin{abstract}
Recently, single-site observables have been shown to be useful for the detection of dynamical criticality due to an emergence of a universal critically-prethermal temporal regime in the magnetization [\href{https://arxiv.org/abs/2105.05986}{arXiv:2105.05986}]. Here, we explore the potential of single-site observables as probes of quantum phase transitions in integrable and nonintegrable transverse-field Ising chains (TFIC). We analytically prove the requirement of zero modes for a quasi-stationary temporal regime to emerge at a bulk probe site, and show how this regime gives rise to a non-analytic behavior in the dynamical order profiles. Our $t$-DMRG calculations verify the results of the quench mean-field theory for near-integrable TFIC both with finite-size and finite-time scaling analyses. We find that both finite-size and finite-time analyses suggest a dynamical critical point for a strongly nonintegrable and locally connected TFIC. We finally demonstrate the presence of a quasi-stationary regime in the power-law interacting TFIC, and extract local dynamical order profiles for TFIC in the long-range Ising universality class with algebraic light cones.
\end{abstract}

\maketitle
\tableofcontents
%\onecolumngrid

\section{\label{intro}Introduction}
Phase transitions of matter are among the most prevalent physical processes in nature, during which certain properties of matter can change discontinuously \cite{landau1937theory,cardy1996scaling}. The notions of universality and scaling allow a unified description of microscopically different systems by means of universality classes, whereby the systems in the same class exhibit identical critical behavior independently of their microscopic details \cite{Kadanoff_book,Zinn-Justin:572813,sachdev2001quantum}. Far from equilibrium, universality and scaling are well-understood in classical systems \cite{Hohenberg_review}, unlike in quantum many-body systems, despite of the exploration of various analogous concepts, e.g.,~dynamical quantum phase transitions and criticality \cite{RevModPhys.83.863,Mori_2018,Heyl_2018,PhysRevLett.96.136801,PhysRevLett.106.227203,PhysRevLett.98.180601,Calabrese_2012,PhysRevLett.110.136404,PhysRevLett.110.135704,PhysRevB.88.201110,PhysRevB.91.220302,PhysRevB.96.134427,PhysRevE.96.022110,PhysRevLett.123.115701,PhysRevA.100.031601,PhysRevB.101.245148,wu2020nonequilibrium,wu2020dynamical,2020arXiv200412287D,halimeh2020local,PhysRevX.11.031062,halimeh2019dynamical,2020arXiv200506481T,PhysRevLett.121.016801,PhysRevLett.123.140602,PhysRevB.97.235134,PhysRevB.101.104415,https://doi.org/10.1002/andp.201900270,PhysRevLett.126.200602,bandyopadhyay2021universal}, and the experimental observation of out-of-equilibrium critical behavior in various quantum synthetic matter (QSM) setups \cite{PhysRevLett.115.245301,Jurcevic2017,Zhang_2017,PhysRevLett.113.045303,Flaeschner2018,PhysRevA.100.013622,Muniz2020,PhysRevLett.124.250601,Xueaba4935,doi:10.1063/5.0004152}.

With the advent of QSM experiments enjoying a high level of precision and control \cite{Bloch2008}, including single-site addressing techniques \cite{2009Natur.462...74B}, the prospect of probing equilibrium and dynamical quantum phase transitions following quantum quenches has become realistic and appealing. Although it may sound counterintuitive, extracting equilibrium criticality through quench dynamics can be an experimentally more viable and simpler scheme than the challenging procedure of cooling a quantum many-body system to its ground state. Instead, the system is initialized in an easily accessible product state, and subsequently quenched through a control parameter. This method has enabled the extraction of quantum critical points and universal scaling laws in various models \cite{PhysRevB.88.201110,PhysRevLett.115.245301,PhysRevB.91.220302,PhysRevB.96.134427,PhysRevE.96.022110,PhysRevA.97.023603,PhysRevLett.121.016801,PhysRevA.100.013622,PhysRevLett.123.115701,PhysRevB.101.245148,2020arXiv200412287D,PhysRevX.11.031062}.

In this vein of probing quantum phase transitions and criticality through quench dynamics, some of us proposed in Ref.~\cite{letter} to use single-site observables close to an edge in open-boundary and short-range many-body models. When boundaries are introduced to a suddenly quenched $1$D transverse-field Ising chain (TFIC) an onset of a quasi-stationary (q.s.) temporal regime appears \cite{PhysRevLett.106.035701}, thus interrupting an otherwise exponential decay to zero in time.
Specifically, Ref.~\cite{letter} analytically and numerically revealed a universal and critical prethermal regime in the vicinity of the quantum critical point (QCP) with single-site observables, regardless of probe location, initial states and weak perturbations to the TFIC. This was shown by means of exact analytical and numerical calculations where the latter is based on a Pfaffian formalism as well as quench mean-field theory (qMFT). Ref.~\cite{letter} also derived a scaling function of the dynamical order parameter in the vicinity of the QCP and showed that the long-lived prethermal regime makes the scaling function nonlinear. The presence of a nonlinear dynamical scaling function was found to be independent of the probe site and initial state, and to be robust to weak integrability breaking.

In this long paper, our theme is the detection of QCP in integrable and nonintegrable TFIC through single-site observables near the edge of the chain $r \ll N/2$. We first determine the QCP in the integrable TFIC by locating a nonanalyticity in the q.s.~value of single-site observables. We find consistently for all sites sufficiently near the edge that the location of this nonanalyticity coincides with the equilibrium QCP. We also derive an analytical expression for the edge magnetization in the integrable TFIC quenched from a polarized state, which is used to characterize the critically prethermal regime in Ref.~\cite{letter}.

We detail the aforementioned Pfaffian formalism behind the exact results on single-site observables in open-boundary TFICs, and how MFT is implemented in quench dynamics, e.g.,~qMFT method. An analytical form of the latter was first employed in Ref.~\cite{PhysRevLett.123.115701} for periodic TFIC. Since we lack space translational symmetry in open-boundary chains, we modify the qMFT method with the help of the cluster theorem to make it suitable for numerics. By benchmarking the qMFT method against exact diagonalization (ED) and time-dependent density matrix renormalization group ($t$-DMRG) results, we show that the MFT is reliable within times accessible to $t$-DMRG. Then we use our qMFT method in TFIC with weak interactions, and detect a non-analyticity in the dynamical order profile slightly shifted from the location of the QCP. By following Ref.~\cite{PhysRevLett.123.115701}, we call this non-analytic point a dynamical critical point (DCP).

Furthermore, we perform extensive $t$-DMRG calculations ranging from near-integrable to strongly nonintegrable and locally connected TFIC, as well as long-range power-law interacting TFIC. We provide finite-size and finite-time analyses on the $t$-DMRG data for the near-integrable model up to system sizes $N=120$, showing that they support the MFT result on the location of the DCP. Remarkably, evidence of a DCP in the thermodynamic limit, namely a crossing between the order profiles at different system sizes up to $N=48$, remains in the $t$-DMRG data for a strongly nonintegrable and locally connected TFIC, albeit shifted significantly from its corresponding QCP. To our knowledge, this is the first evidence of a quench measurement of magnetization that suggests a DCP for the strongly nonintegrable TFIC with ferromagnetic next nearest neighbor (n.n.n.) couplings $\Delta/J=1$, where $J$ is the fixed energy scaling set in the calculation. Previously, a dynamical order profile was generated in Ref.~\cite{PhysRevLett.121.016801} with out-of-time-order correlators at $\Delta/J=0.5$ with no clear evidence of a crossing point between the order profiles at different system sizes. Ref.~\cite{PhysRevLett.123.115701} predicted a DCP, by using two-point correlators, up to $\Delta/J=0.3$ for chains at $N=25$, and Ref.~\cite{PhysRevX.11.031062} has recently predicted a QCP with the same correlator, but for antiferromagnetic n.n.n~couplings. 
Additionally, for this strongly nonintegrable TFIC, we observe that the finite-time analysis of our $t$-DMRG data underestimates the location of the DCP, thereby shifting it further away from the QCP. This reveals a breakdown of the validity regime of the finite-time analysis, and we elaborate on a possible reason of this observation. 

We show the presence of a q.s. temporal regime in TFIC with power-law decaying interactions, whose equilibrium criticality properties belong either to short- or long-range Ising universality class with, respectively, $\alpha \geq 3$ and $\alpha < 3$, where $\alpha$ is the power-law exponent of the interaction strength.
We reveal that single-site observables still reflect a difference in the quench dynamics between the sites near the edges and in the middle when the model is long-range interacting $\alpha < 3$, up until $\alpha=2$ where there is no longer an onset of a q.s. regime. Upon closer study of the model with $\alpha=2.5$, which is long-range with algebraic light cones \cite{PhysRevLett.114.157201}, we reveal that the local order profiles at different $r$ are consistent with each other and suggest a crossover. 

Because the q.s.~regime is the basis of the criticality detection in our paper, we analytically show why the zero modes are vital in the origin of this temporal regime. Besides the requirement of zero modes, the origin of the q.s. regime can also be explained with translational symmetry breaking and making an asymmetric measurement with respect to the symmetry center of the chain, e.g.,~the middle of the chain. This differentiates the two chain edges as either near or far from the perspective of quasi-particles in the integrable models, or of the wavefronts in operator spreading in the nonintegrable models. Hence, the locality of the underlying model, e.g.,~the linear or sublinear light cones in the operator spread, rather than the integrability seems to be a reason for the presence of the q.s. regime. In fact, we explicitly show instances of q.s. behaviour appearing in strongly nonintegrable many-body models and its direct application in probing QCP. Consistently, the q.s. regime does not occur when the model becomes long-range with logarithmic light cones \cite{PhysRevLett.111.207202}. Furthermore, we find the presence of a q.s. regime even when the hard edges are smoothed out.
This implies that our theoretical predictions could in principle be tested even in imperfect conditions where hard boundaries might not be present, e.g.,~in trapped atomic and ionic gases.

Although we focus on the dynamics of single-site observables, since we aim to dynamically probe ferromagnetic order in our paper, we emphasize that our method could be easily adjusted to probe other quantum orders, e.g.,~anti-ferromagnetic order, by increasing the unit cell of the underlying lattice. Therefore, our work and Ref.~\cite{letter} are the first examples of \textit{spatially minimal measurements} that could probe quantum phase transitions and nonequilibrium criticality. 

This paper consists of the following sections. In Sec.~\ref{qsSec}, we describe the locally-connected TFIC with open boundaries and introduce its q.s.~temporal regime. Then we focus on the detection of QCP in the nearest-neighbor (n.n.) integrable limit of TFIC where we also explain the formalism behind our calculations, express the cluster theorem and prove the requirement of zero modes to observe a q.s.~regime in Sec.~\ref{integrableTFICSec}. Later in Sec.~\ref{NNNmodel}, we detail our qMFT method and its results, and systematically discuss the $t$-DMRG results on our locally-connected nonintegrable models. In Sec.~\ref{longrange}, we focus on the quench dynamics of single-site observables, and local order profiles in the power-law interacting TFIC. We conclude in Sec.~\ref{Conc} with an outlook.

\section{\label{qsSec}The quasi-stationary regime in the short-range TFIC}

The TFIC with n.n.n.~interaction strength $\Delta$ reads
\begin{eqnarray}
H = -J \sum_{1}^{N-1} \sigma_r^z \sigma_{r+1}^z - \Delta  \sum_{1}^{N-2} \sigma_r^z \sigma_{r+2}^z + h \sum_{1}^N \sigma_r^x, \label{Hamiltonian}
\end{eqnarray}
where $\sigma_r^{x,z}$ are the Pauli spin matrices on site $r$, $h$ is the transverse-field strength, $N$ is the length of the open-boundary chain, and we fix the n.n. coupling strength $J=1$ as the energy scale. The TFIC is a paradigmatic model of quantum phase transitions, hosting a critical point at $h_c=1$ when $\Delta=0$ which separates a spontaneously broken symmetry (ferromagnetic) phase from a symmetry preserving (paramagnetic) phase. When $\Delta > 0$, the low-energy properties remain the same, but the QCP shifts to favor order $h_c > 1$.

\begin{figure}
\subfloat[]{\label{fig1a}\includegraphics[width=0.24\textwidth]{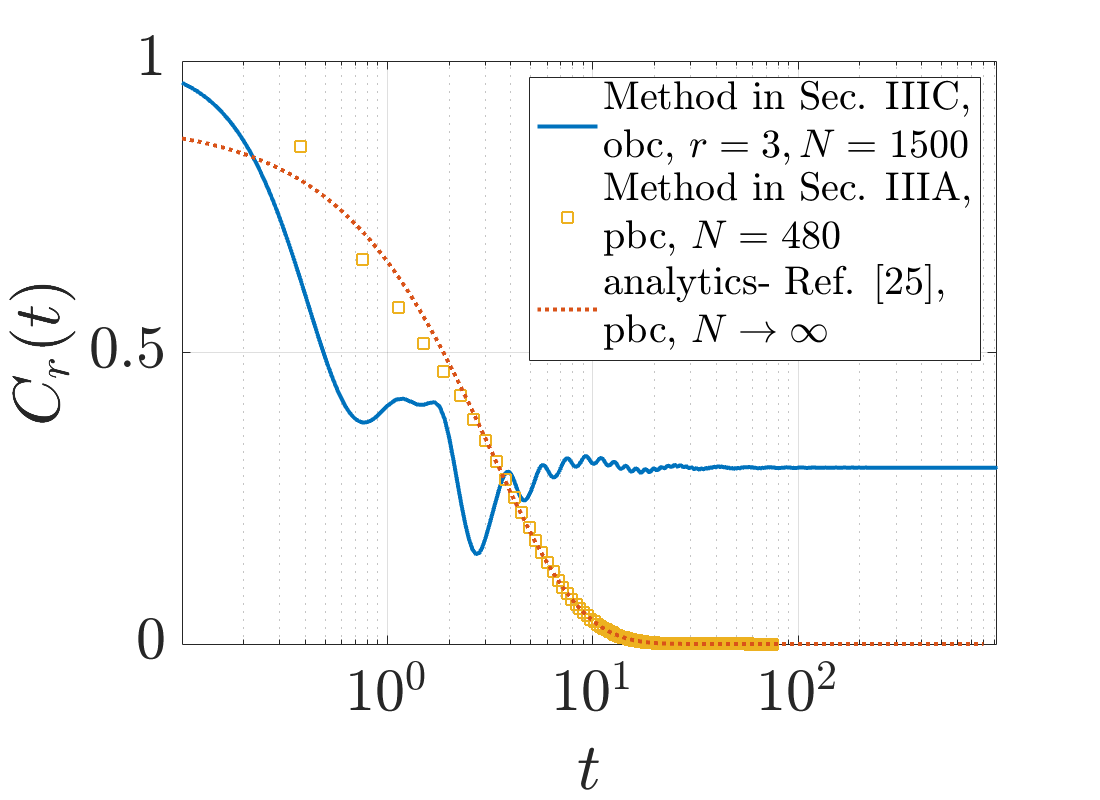}}\hfill
\subfloat[]{\label{fig1b}\includegraphics[width=0.24\textwidth]{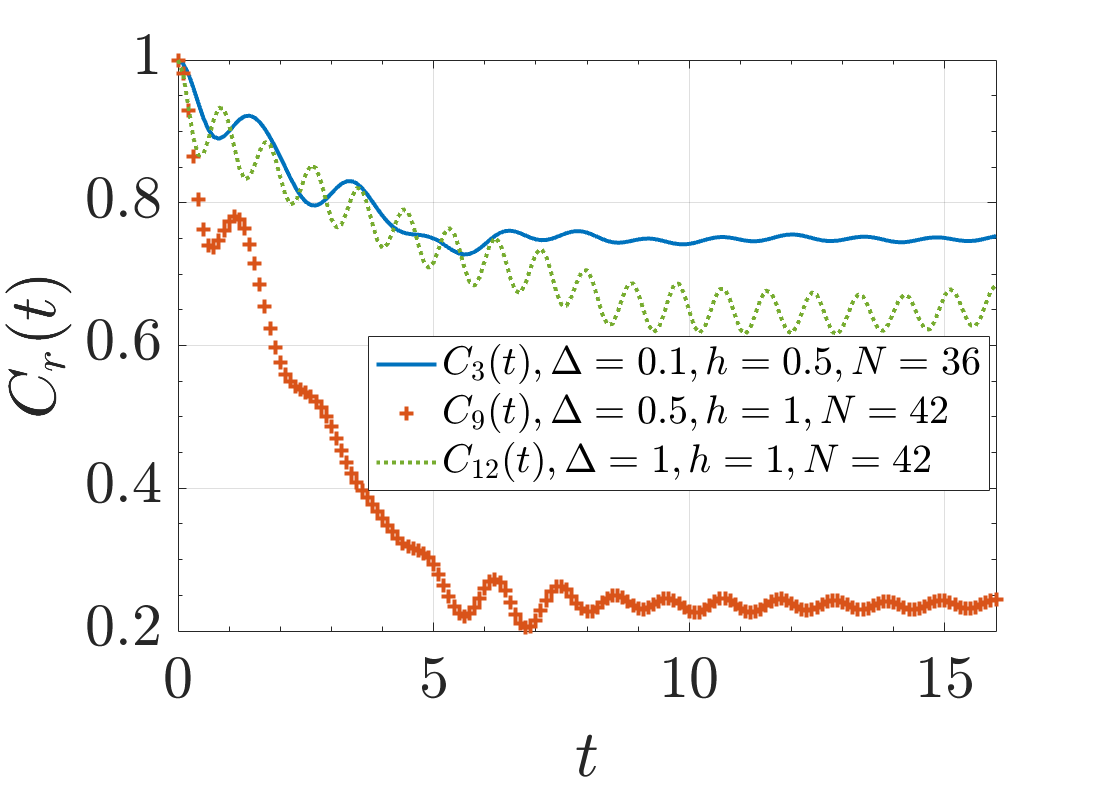}}\hfill
\caption{(a) The local magnetization $C_{r}(t)$ in the integrable TFIC with open-boundary (solid-blue) and periodic-boundary (red-dotted and yellow-squares) conditions given in Eq.~\eqref{Hamiltonian} after a quench in the transverse-field strength from $h_i=0$ to $h=0.8$. See Secs.~\ref{clusterThmSec} and~\ref{zeroModeSec} for the numerical methods used in yellow-squares and solid-blue. The red-dotted line is the analytical equation for the periodic chain derived in Ref.~\cite{2020arXiv200412287D}. (b) The local magnetization $C_{r}(t)$ in open-boundary nonintegrable TFIC ($\Delta>0$) after a quench in the transverse-field strength from $h_i=0$ and calculated with $t-$DMRG.}
\label{Fig1}
\end{figure}

Let us consider as initial state the ground state $\Ket{\psi_0}$ of $H$ at initial value $h_i$ of the transverse-field strength, and then we quench the latter to a value $h$. While in the case of periodic-boundary conditions the single-site nonequilibrium response $C_r(t)=\Bra{\psi_0}\sigma^z_r(t)\Ket{\psi_0}$ decays exponentially to zero \cite{Calabrese_2012,2020arXiv200412287D} (see Fig.~\ref{fig1a}), open-boundary conditions give rise to a nonzero-valued equilibrium regime after a power-law decay when $h_i < h \leq h_c=1$. This is referred to as the q.s. regime \cite{PhysRevLett.106.035701}. Figure~\ref{fig1a} shows the q.s. regime of $C_{3}(t)$ in the integrable TFIC quenched from $h_i=0$ to $h=0.8$ and at a system size $N=1500$. While one could argue that the origin of the q.s. regime is because of the interference between quasi-particles due to the asymmetric location of the probe site with respect to the symmetry center of the chain that gives rise to reflection from the closest edge, we show in the next section the necessity of zero modes to observe a q.s. regime.

We break integrability in Eq.~\eqref{Hamiltonian} by taking $\Delta > 0$ in Fig.~\ref{fig1b} for different physical parameters (see legend). In all cases, our $t-$DMRG calculations in open-boundary chains (see Sec.~\ref{t-DMRG} for the method) show the emergence of a q.s. regime in relatively early times as opposed to periodic chains where the local magnetization is expected to exponentially decay to zero \cite{2020arXiv200412287D}. Whether this early time q.s. plateau has a finite lifetime and the system eventually thermalizes is an interesting and important question. Because we focus on whether one could utilize the value of the q.s. regime to probe QCP, an answer to this question is beyond the scope of our paper. We will characterize the q.s. regime that emerges in open-boundary nonintegrable TFIC in Secs.~\ref{NNNmodel} and~\ref{longrange}, respectively for short-range and long-range interacting models.

\section{\label{integrableTFICSec}The integrable TFIC}

In this section, we present our analytical and numerical techniques on the integrable TFIC with $\Delta=0$, which includes (A) the cluster theorem technique, (B) a proof on the origin of the q.s. regime; (C) the derivation of a series expression for the edge magnetization; and finally (D) the numerically extracted single-site nonequilibrium phase diagrams based on the cluster theorem technique.

\subsection{\label{clusterThmSec}Quench dynamics of single-site magnetization via the cluster theorem}

\begin{figure}
\centering
\includegraphics[width=0.45\textwidth]{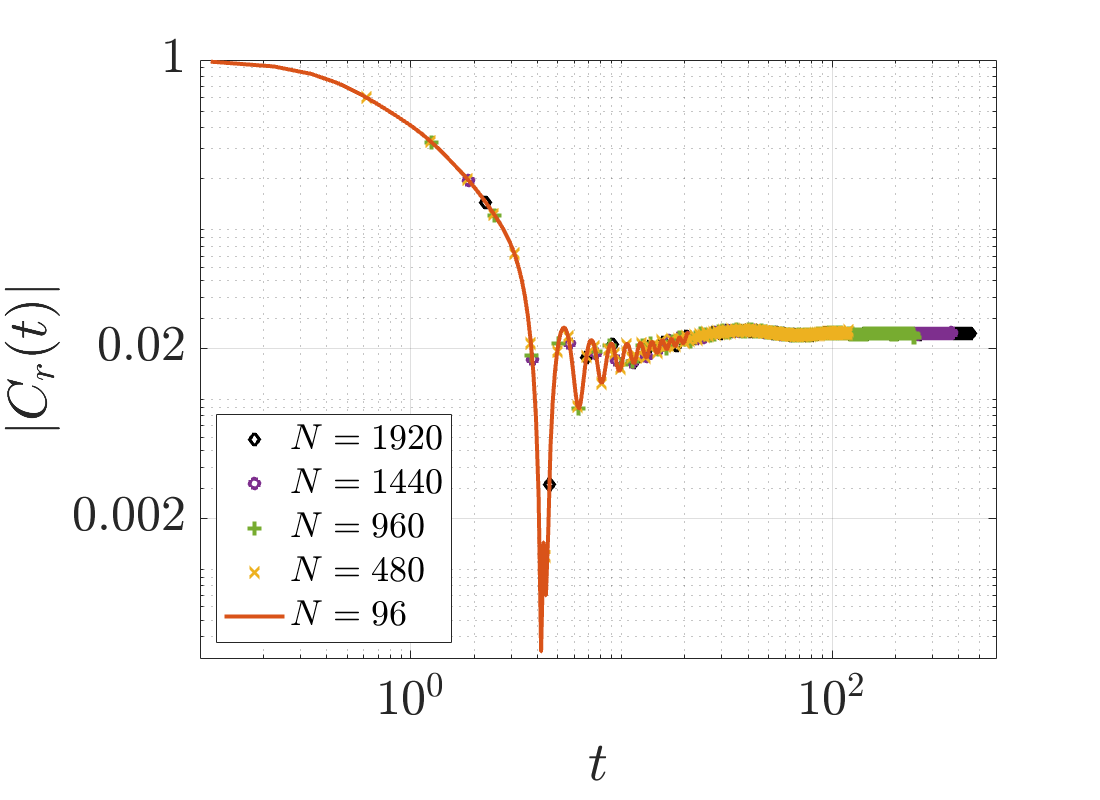}\hfill 
\caption{Single-site magnetization $|C_r(t)|$ at $r=6$ for different system sizes (see legend) at $h=0.95$. The magnetization is plotted until corresponding $t_l$, the breakdown time of the cluster theorem, and in this region the data for different system sizes collapse on each other. This suggests that we effectively simulate the thermodynamic limit with a finite-size system at the expense of a breakdown time.}\label{figN3}
\end{figure}

We start by mapping the integrable TFIC to a noninteracting fermionic model in $1$D via the Jordan--Wigner transformation \cite{sachdev2001quantum},
\begin{eqnarray}
\sigma^z_r &=& - \prod_{s<r} \left(1-2c_s^{\dagger}c_s \right)\left(c_r+c_r^{\dagger}\right),\label{mapping} \\
\sigma^x_r &=& 1-2c_r^{\dagger}c_r,\notag \\
\sigma^y_r &=& -i \prod_{s<r} \left(1-2c_s^{\dagger}c_s \right)\left(c_r-c_r^{\dagger}\right),\notag
\end{eqnarray}
to obtain the noninteracting Hamiltonian
\begin{eqnarray}
H &=&  \sum_r \big[-J\big( c^{\dagger}_r c_{r+1} + c_r^{\dagger} c_{r+1}^{\dagger} + \text{H.c.}\big)+2 h c_r^{\dagger} c_r \big)\big]. \label{nonInteractingH}
\end{eqnarray}
Expressing the local magnetization of a bulk spin $\Braket{\sigma_r^z(t)}$ in this noninteracting picture brings a string of operators, and hence is not tractable. Instead, one could utilize the Wick theorem and Pfaffian formalism for two-point correlators to encode the string of operators into an anti-symmetric matrix \cite{sachdev2001quantum,Calabrese_2012}. This direction of solution requires rewriting a single-site observable in terms of two-point correlators \cite{2020arXiv200412287D} and we therefore invoke the cluster theorem \cite{Calabrese_2012},
\begin{eqnarray}
\Braket{\sigma_{r}^z(t)\sigma_{N-r+1}^z(t)} \approx \Braket{\sigma_{r}^z(t)} \Braket{\sigma_{N-r+1}^z(t)}, \label{clusterTheorem}
\end{eqnarray}
where $r \ll N/2$ is close to the left boundary. The cluster theorem breaks down outside of the lightcone, and the corresponding breakdown time $t_l$ can be understood in the context of operator spread between two well separated sites $r$ and $N-r+1$ that eventually get correlated with each other. This time can be estimated based on the maximum quasiparticle velocities $v_q$, as $t_l = \Delta x/(2v_q)$ where $\Delta x = N-2r+1$ is the distance between two spins that are equidistant from the symmetry center of an open-boundary chain in Eq.~\eqref{clusterTheorem}, which is the middle of the chain. For such symmetrically placed sites, the non-equilibrium response is the same, and one can thus write
\begin{eqnarray}\label{e:bulk_from_sop}
\Braket{\sigma_{r}^z(t)} = \sqrt{\Braket{\sigma_{r}^z(t)\sigma_{N-r+1}^z(t)}} \equiv C_r(t).
\end{eqnarray}
Therefore, we can extract the dynamical evolution of a spin at site $r$ from the equal-time two-point correlators of sites $r$ and $N-r+1$. The latter can be written in terms of auxiliary operators $\phi^{\pm}_r=c_r^{\dagger}\pm c_r$ as
\begin{eqnarray}
\Braket{\sigma_{r}^z(t)\sigma_{N-r+1}^z(t)} &=& 
\bigg\langle \phi_r^-(t) \bigg( \prod_{s=r+1}^{N-r}  \phi_s^+(t) \phi_s^-(t) \bigg) \notag \\
&\times  & \phi_{N-r+1}^+(t) \bigg\rangle. \label{SOP}
\end{eqnarray}
This is, in fact, the expectation value of the so-called string order parameter (SOP) \cite{PhysRevB.93.085416}. It can be calculated by invoking Wick's theorem, which allows one to re-express the above expectation value as a sum over products of elementary contractions, which in turn is the Pfaffian of an appropriately constructed antisymmetric matrix $T(t)$ i.e. $(C_r(t))^2 = \mathrm{Pf}(T(t))$ \cite{sachdev2001quantum,Calabrese_2012}. Although this is in general a complex number, it follows from Eq.~\eqref{clusterTheorem} that for $t<t_l$ and the ordered phase we actually have $(C_r(t))^2\in\mathbb{R+}$, so that we may compute $C_r(t) = \lvert \mathrm{Pf}(T(t)) \rvert^{1/2} = \lvert \sqrt{\mathrm{det}(T(t))} \rvert^{1/2}$. This is advantageous since it is numerically more efficient to calculate determinants as compared to Pfaffians. However note that in the disordered phase and close to QCP in the ordered phase, $C_r(t) \in \mathbb{R}$. In these cases, we compute $|C_r(t)|$ with the cluster theorem method. To construct the matrix $T(t)$ we only need to calculate all possible elementary two-point contractions $\Braket{\phi_a^p(t) \phi_b^q(t)}$ where $p,q=\pm$ and $r\leq a \leq b \leq N-r+1$. Now we briefly review how to implement a sudden quench in this picture by following Ref.~\cite{PhysRevB.101.245148}.

Quenching from an initial Hamiltonian $H_i$ with transverse field $h_i$, we first solve
\begin{eqnarray}
H_i = \sum_k E_k^i \alpha_k^{\dagger}\alpha_k, 
\end{eqnarray}
where $E^i_k$ are the single particle energies, $\alpha_k$ and $\alpha_k^{\dagger}$ are the new annihilation and creation operators, respectively, after diagonalization where the ground state $\ket{\psi_0}_{\alpha}$ satisfies $\alpha_k \ket{\psi_0}_{\alpha}=0$. The solution, in a general form,  follows as \cite{1961AnPhy..16..407L}
\begin{eqnarray}
\left(\begin{array}{c}
\alpha \\
\alpha^{\dagger}
\end{array} \right) = \left( \begin{array}{c c}
G_i & F_i \\
F_i & G_i
\end{array} \right) \left(\begin{array}{c}
c_i \\
c^{\dagger}_i
\end{array} \right),
\end{eqnarray}
where $c_i=(c_1, c_2, \cdots, c_N)^{T}$ and similarly for the creation operator $c_i^{\dagger}$. The expressions for the $G_i$ and $F_i$ block matrices will follow shortly. By solving the eigensystem of
\begin{eqnarray}
\left[(A_i - B_i)(A_i + B_i) \right] \Ket{\Phi_k^i} = (E_k^i)^2 \Ket{\Phi_k^i},
\end{eqnarray}
we obtain the eigenenergies $E_k^i$ and eigenvectors $\Ket{\Phi_k^i}$. Note that $A_i$ and $B_i$ are the nearest neighbor hopping and pairing terms of the Hamiltonian, respectively. Then the Hamiltonian could be written as,
\begin{eqnarray}
H_i = \left( \begin{array}{c c}
A_i & B_i \\
B_i^{\dagger} & -A_i
\end{array} \right),
\end{eqnarray}
in the $(c \hspace{1mm} c^{\dagger})^T$ basis. Next we use the eigensystem $(E_k^i,\Ket{\Phi_k^i})$ to obtain
\begin{eqnarray}
\Ket{\Psi_k^i} = \frac{1}{E_k^i}\left[\Bra{\Phi_k^i} (A_i - B_i) \right]^T.
\end{eqnarray}

Let us emphasize that the expression above would not work for zero edge modes where $E_{k}^i=0$, and handling the case with zero mode is crucial for us. This is because we are primarily interested in the quenches from and to the ordered phase in an open-boundary TFIC. In this case, we set $\Ket{\Psi_k^i}=-\Ket{\Phi_k^i}$ \cite{1961AnPhy..16..407L} and make sure that the resulting state is linearly independent from the other modes. Finally we calculate the matrices $G_i$ and $F_i$ in terms of $\Ket{\Phi_k^i}$ and $\Ket{\Psi_k^i}$. Defining matrices
\begin{eqnarray}
\Phi_i &=& \left[\Ket{\Phi_1^i} \hspace{1mm} \Ket{\Phi_2^i} \cdots \Ket{\Phi_N^i} \right], \notag \\
\Psi_i &=& \left[\Ket{\Psi_1^i} \hspace{1mm} \Ket{\Psi_2^i} \cdots \Ket{\Psi_N^i} \right], \notag
\end{eqnarray}
the block matrices follow
\begin{eqnarray}
G_i = \frac{1}{2}\left(\Phi_i^T + \Psi_i^T\right), \hspace{1mm} F_i = \frac{1}{2}\left(\Phi_i^T - \Psi_i^T\right).
\end{eqnarray}
Similarly for the final Hamiltonian $H_f$ with transverse field $h$, one could write
\begin{eqnarray}
\left(\begin{array}{c}
\beta \\
\beta^{\dagger}
\end{array} \right) = \left( \begin{array}{c c}
G_f & F_f \\
F_f & G_f
\end{array} \right) \left(\begin{array}{c}
c_f \\
c^{\dagger}_f
\end{array} \right),
\end{eqnarray}
together with corresponding $\Phi_f$ and $\Psi_f$. We calculate the transfer matrices with the help of these block matrices,
\begin{eqnarray}
T_1=G_f G_i^T + F_f F_i^T, \notag \\
T_2=G_f F_i^T + F_f G_i^T. \notag
\end{eqnarray}
These transfer matrices are used to finally calculate the matrix elements ${}_{\alpha}\Bra{\psi_0} \left[\phi^p_a \phi^q_b\right]_{\beta}\Ket{\psi_0}_{\alpha}$ of the antisymmetric matrix $T(t)$, where Greek subscripts denote in which basis we have the states and the operators.
Because we would like to make use of $\alpha_k \Ket{\psi_0}_{\alpha}=0$, we use the above transfer matrices to write $\left[\phi^p_a \phi^q_b\right]_{\beta}$ in the $\alpha$ basis as
\begin{eqnarray}
\left[\phi^{\pm}_b\right]_{\beta} \Ket{\psi_0}_{\alpha}  &=& \left[c_b^{\dagger}(t)\pm c_b(t)  \right]_{\beta}\Ket{\psi_0}_{\alpha}, \notag \\
&=& \left[(G_f^T \pm F_f^T)\left( e^{i \mathcal{E} t} T_1 \pm e^{-i \mathcal{E} t} T_2 \right) \alpha^{\dagger}\right]_b\Ket{\psi_0}_{\alpha}, \notag
\end{eqnarray}
where $\mathcal{E}$ is a diagonal matrix with eigenenergies of the final Hamiltonian as the entries, $\mathcal{E}=\text{diag}[E_1^f \hspace{1mm} E_2^f \cdots E_N^f]$. Based on this formulation, we construct matrices $\mathcal{M}_{q}(t)$ explicitly,
\begin{eqnarray}
\mathcal{M}_+(t) &=& \Phi_f \left( e^{-i \mathcal{E} t} T_1 + e^{i \mathcal{E} t} T_2 \right),\notag \\
\mathcal{M}_-(t) &=& \left(T_1^T e^{i \mathcal{E} t} - T_2^T e^{-i \mathcal{E} t} \right)\Psi_f^T,
\end{eqnarray}
to utilize in the following contractions,
\begin{eqnarray}
\Braket{\phi^+_a(t) \phi^+_b (t)} &=& [\mathcal{M}_+(t)\mathcal{M}_+(t)^{\dagger}]_{ab}, \notag\\
\Braket{\phi^-_a(t) \phi^-_b(t)} &=& -[\mathcal{M}_-^{\dagger}(t)\mathcal{M}_-(t)]_{ab}, \notag\\
\Braket{\phi^+_a(t) \phi^-_b(t)} &=& [\mathcal{M}_+(t)\mathcal{M}_-(t)]_{ab}, \notag \\
\Braket{\phi^-_a(t)\phi^+_b(t)} &=& -[\mathcal{M}_-^{\dagger}(t)\mathcal{M}_+^{\dagger}(t)]_{ab}. \label{contractions}
\end{eqnarray}

Now we can construct the antisymmetric matrix $T(t)$ at time $t$ with the matrix elements $T_{ks}(t)=\Braket{\phi_a^p(t)\phi_b^q(t)}$, where $1 \leq k < s \leq 2\Delta x$, $p=+(-)$ for $k$ even(odd) and $q=+(-)$ for $s$ even(odd). The relation between parameters $a,b$ and $k,s$ reads $a=r + \lfloor k/2 \rfloor$ and $b=r+ \lfloor s/2 \rfloor$, because $r  \leq a \leq b \leq N-r+1$.
Having constructed $T(t)$, one can then extract $C_r(t) = |\sqrt{\det{T(t)}}|^{1/2}$, as discussed below Eq.~\eqref{SOP}. Let us note that this quench formalism for noninteracting fermions was recently employed in Refs.~\cite{PhysRevB.101.245148,2020arXiv200412287D}.  We use this method to obtain the numerical results presented for the critically prethermal regime in Ref.~\cite{letter}.

The breakdown time of the method, $t_l$, is also the time when finite-size effects kick in. Therefore, the method is naturally immune to these finite-size effects, and the local magnetization for different sizes collapses on each other, Fig.~\ref{figN3}. This is similar in spirit to the separation timescale of single-site observables which is the time when the quasi-particles reach the end of the chain \cite{2020arXiv200412287D} and the finite-size effects in this time interval are exponentially suppressed in system size \cite{Wang_2021}.

In the next section, we analytically show that (i) the system needs to support zero modes to stabilize a q.s. regime and (ii) how bulk probe sites are susceptible to these zero modes which then gives rise to the detection of QCP at an arbitrary site.

\subsection{\label{zeroModeSec}Requirement of zero modes to observe a quasi-stationary regime}

\begin{figure}
\subfloat[]{\label{fig22a}\includegraphics[width=0.24\textwidth]{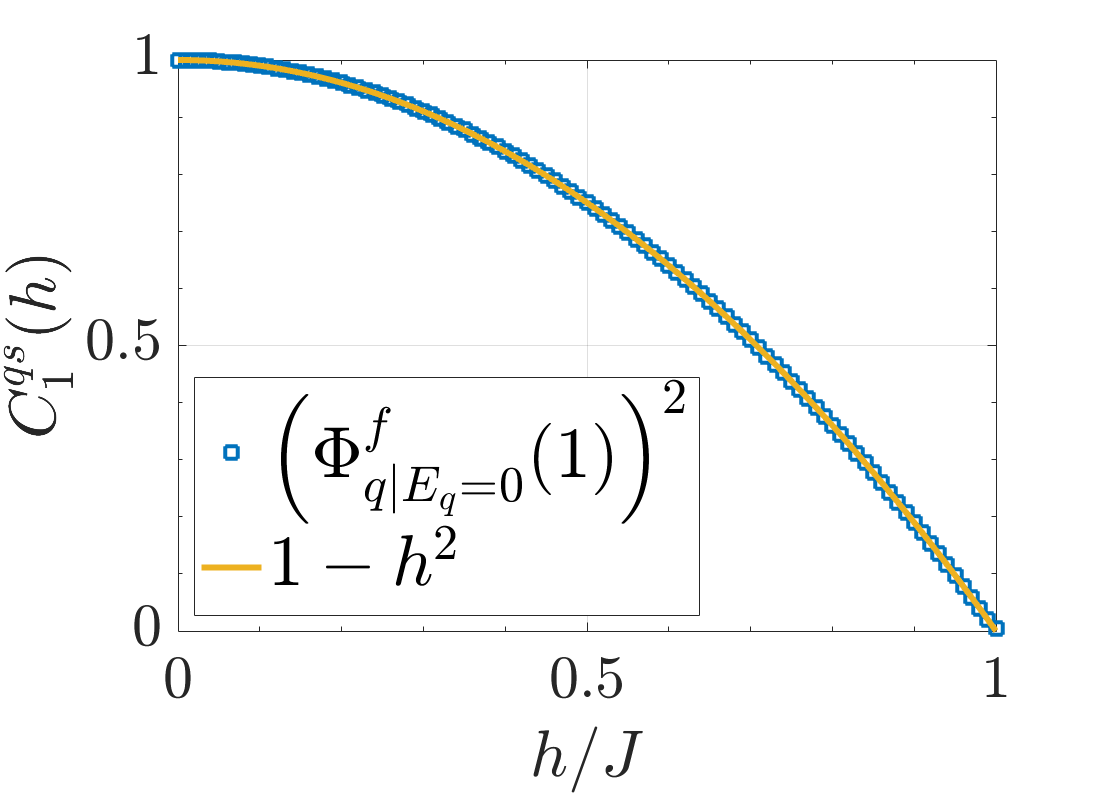}}\hfill
\subfloat[]{\label{fig22b}\includegraphics[width=0.24\textwidth]{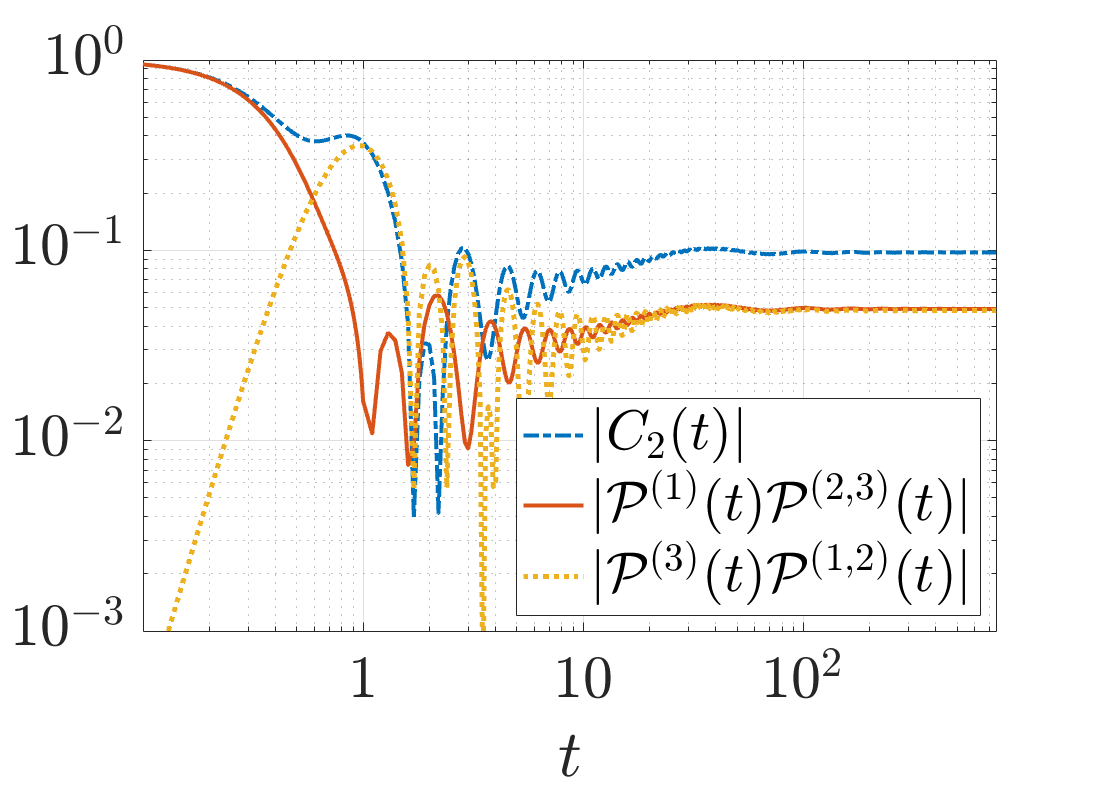}}\hfill
\caption{(a) Eq.~\eqref{edgeQS} plotted with respect to $h$ (blue-squares) which coincides with the function $1-h^2$ (solid-yellow). (b) Absolute values of the first (red-solid) and the third term contributions (yellow-dotted) in Eq.~\eqref{secondSMag1} for $N=1500$ spins at $h=0.95$. Their summation (without the absolute values) gives the entire time evolution for the second spin whose absolute value is plotted with blue dashed-dotted.}
\label{Fig22}
\end{figure}

 In order to show the requirement of zero modes to stabilize a q.s. regime, we have to adopt an approach different than the cluster theorem for studying the quench dynamics. This alternative approach, proposed by Ref.~\cite{PhysRev.85.808}, defines the single-site magnetization as 
%the off-diagonal element between the doubly degenerate ground states of the many-body Hamiltonian
\begin{eqnarray}
C_r(t)&=& (-1)^{r-1} \mathbin{_{\alpha}{\Bra{\psi_0}}} \left(\prod_{k}^{r-1}\phi_k^+(t)\phi_k^-(t)\right) \phi_r^+(t) \Ket{\psi_1}_{\alpha}, \notag
\end{eqnarray}
where $\Ket{\psi_1}_{\alpha}=\alpha^{\dagger}_N\Ket{\psi_0}_{\alpha}$ is the single-particle excited state. The auxiliary operators that are already defined in Sec.~\ref{clusterThmSec} can be written in the basis of the initial Hamiltonian in terms of $\alpha_k$,
\begin{eqnarray}
\phi_{l}^+ &=& \sum_k^N \Phi_k^i(l)(\alpha_k^{\dagger}+\alpha_k) \equiv \phi_{2l-1}, \label{phi+Def} \\
\phi_{l}^- &=& \sum_k^N \Psi_k^i(l)(\alpha_k^{\dagger}-\alpha_k) \equiv \phi_{2l}.\label{phi-Def}
\end{eqnarray}
The parenthesis in $\Phi_k^i(l)$ and $\Psi_k^i(l)$ denote the $l^{\text{th}}$ element of the corresponding vector. The new notation defined in Eqs.~\eqref{phi+Def} and~\eqref{phi-Def} make the following equations easier to follow, 
\begin{eqnarray}
C_{ 1}(t) &=&  \sum_{n,k} \bigg[P_{1,2n-1}(t) \Phi_k^i(n) \mathbin{_{\alpha}{\Bra{\psi_0}}} (\alpha_k^{\dagger}+\alpha_k) \alpha_N^{\dagger} \Ket{\psi_0}_{\alpha} \notag \\
&+& P_{1,2n}(t) \Psi_k^i(n) \mathbin{_{\alpha}{\Bra{\psi_0}}} (\alpha_k^{\dagger}-\alpha_k) \alpha_N^{\dagger} \Ket{\psi_0}_{\alpha}\bigg], \notag \\
&=& \sum_{n}  \left[P_{1,2n-1}(t) \Phi_N^i(n) - P_{1,2n}(t) \Psi_N^i(n)\right], \label{edgeMag1}
\end{eqnarray}
where $P_{1,2n-1}$ and $P_{1,2n}$ are the single-particle propagators of the quench Hamiltonian \cite{Igl_i_2013},
\begin{eqnarray}
P_{2l-1,2k-1}(t) &=& \sum_{q} \cos(E_q t) \Phi_q^f(l)\Phi^f_q(k), \notag \\
P_{2l-1,2k}(t) &=& -\sum_{q} \sin(E_q t) \Phi^f_q(l)\Psi^f_q(k), \notag \\
P_{2l,2k-1}(t) &=& \sum_{q} \sin(E_q t) \Phi^f_q(k)\Psi^f_q(l), \notag \\
P_{2l,2k}(t) &=& \sum_{q} \cos(E_q t) \Psi^f_q(l)\Psi^f_q(k). \notag 
\end{eqnarray}
To be concrete, let us choose the initial state as the ground state of the Hamiltonian with $h_i=0$. This results in
\begin{eqnarray}
\Phi_k^i(n) &=& \begin{cases}
        1, \hspace{5mm} &\text{if }  n=k+1\\ 
        0, \hspace{5mm} &\text{if } n\neq k+1 , \\
        \end{cases} \label{phiInitCond}\\
\Psi_k^i(n) &=& \begin{cases}
        -1, \hspace{5mm} &\text{if }  n=k\\ 
        0, \hspace{5mm} &\text{if } n\neq k .\\
        \end{cases} \label{psiInitCond}  
\end{eqnarray}
When Eqs.~\eqref{phiInitCond} and~\eqref{psiInitCond} are substituted into Eq.~\eqref{edgeMag1}, one obtains
\begin{eqnarray}
C_{ 1}(t) &=&P_{1,1}(t) + P_{1,2N}(t), \label{edgeMag2}\\
&=& \sum_{q} \bigg ( \cos(E_q t) \Phi_q^f(1)\Phi^f_q(1) \notag \\
&-& \sin(E_q t) \Phi^f_q(1)\Psi^f_q(N) \bigg ).\label{edgeMag3}
\end{eqnarray}
In the infinite time limit, $t\to \infty$, $E_q=0$ is the only nonzero contribution. Defining the q.s. value $C_{ 1}(t\rightarrow \infty) \equiv C_{ 1}^{qs}(h)$, we obtain
\begin{eqnarray}
C_{ 1}^{qs}(h) =\sum_{E_q=0} \Phi_q^f(1)\Phi_q^f(1) = 1-h^2. \label{edgeQS}
\end{eqnarray}
This means that the q.s. regime in the edge magnetization is solely due to the presence of edge modes in the Kitaev chain, Eq.~\eqref{nonInteractingH}. We plot Eq.~\eqref{edgeQS} for a system size $N=500$ in Fig.~\ref{fig22a}, where we show that it coincides perfectly with $1-h^2$ (solid-yellow). 

Next, we aim to understand the role of zero modes in the q.s. regime of a bulk spin, e.g.,~the second site $r=2$. 
\begin{eqnarray}
C_{ 2}(t)&=& - \mathbin{_{\alpha}{\Bra{\psi_0}}} \phi_1^+(t)\phi_1^-(t) \phi_2^+(t) \Ket{\psi_1}_{\alpha} \\
&=&\left|
\begin{array}{c c c c}
0 & -\mathcal{P}^{(1,2)}(t) & -\mathcal{P}^{(1,3)}(t) & -\mathcal{P}^{(1)}(t)  \\
\mathcal{P}^{(1,2)}(t) & 0 &  -\mathcal{P}^{(2,3)}(t) & -\mathcal{P}^{(2)}(t) \\
\mathcal{P}^{(1,3)}(t)  & \mathcal{P}^{(2,3)}(t) & 0 & -\mathcal{P}^{(3)}(t)  \\
\mathcal{P}^{(1)}(t) & \mathcal{P}^{(2)}(t) & \mathcal{P}^{(3)}(t) & 0 
\end{array} \right|^{1/2} , \notag
\end{eqnarray}
where we have used Wick's theorem, similar to Sec.~\ref{clusterThmSec}. The definitions of this anti-symmetric matrix's elements read
\begin{eqnarray}
\mathcal{P}^{(2r-1)}(t) &\equiv& \mathbin{_{\alpha}{\Bra{\psi_0}}} \phi_r^{+}(t) \alpha_N^{\dagger} \Ket{\psi_0}_{\alpha}, \label{singleOdd}\\
\mathcal{P}^{(2r)}(t) &\equiv& \mathbin{_{\alpha}{\Bra{\psi_0}}} \phi_r^{-}(t) \alpha_N^{\dagger} \Ket{\psi_0}_{\alpha},\label{singleEven}\\
\mathcal{P}^{(2r-1,2r'-1)}(t) &\equiv & \mathbin{_{\alpha}{\Bra{\psi_0}}} \phi_r^{+}(t) \phi_{r'}^{+}(t) \Ket{\psi_0}_{\alpha},\label{twoPpropagator}\\
\mathcal{P}^{(2r-1,2r')}(t) &\equiv & \mathbin{_{\alpha}{\Bra{\psi_0}}} \phi_r^{+}(t) \phi_{r'}^{-}(t) \Ket{\psi_0}_{\alpha},\notag \\
\mathcal{P}^{(2r,2r'-1)}(t) &\equiv & \mathbin{_{\alpha}{\Bra{\psi_0}}} \phi_r^{-}(t) \phi_{r'}^{+}(t) \Ket{\psi_0}_{\alpha},\notag \\
\mathcal{P}^{(2r,2r')}(t) &\equiv & \mathbin{_{\alpha}{\Bra{\psi_0}}} \phi_r^{-}(t) \phi_{r'}^{-}(t) \Ket{\psi_0}_{\alpha}.\notag
\end{eqnarray}
The two-point correlators, Eq.~\eqref{twoPpropagator} can be calculated with Eq.~\eqref{contractions}. We know the form of Eq.~\eqref{singleOdd} from the expression for edge magnetization, Eq.~\eqref{edgeMag2}. Hence, Eqs.~\eqref{singleOdd} and~\eqref{singleEven} follow similarly,
\begin{eqnarray}
\mathcal{P}^{(2r-1)}(t) &=& P_{2r-1,1}(t) + P_{2r-1,2N}(t), \notag \\
\mathcal{P}^{(2r)}(t) &=& P_{2r,1}(t) + P_{2r,2N}(t).
\end{eqnarray}
Then, it is straightforward to show 
\begin{eqnarray}
C_{ 2}(t) &=& \mathcal{P}^{(1)}(t)\mathcal{P}^{(2,3)}(t) - \mathcal{P}^{(2)}(t)\mathcal{P}^{(1,3)}(t) \notag \\
&+&\mathcal{P}^{(3)}(t)\mathcal{P}^{(1,2)}(t).  \label{secondSMag1}
\end{eqnarray}
In the infinite time limit, $C_2(t\rightarrow \infty)$ brings energy conditions of ${ E_{\alpha}\pm E_{\beta}\pm E_{\gamma}=0 }$ where all energies are nonnegative $E_{\alpha} \geq 0$. Then, so long as (i) the single particle eigenstates are not all product states and (ii) the single particle spectrum is nondegenerate, all of which is satisfied when $h \neq 0$, the energy condition holds only when $E_{\alpha}=E_{\beta}= E_{\gamma}=0$. 
%\begin{eqnarray}
%C_{2}^{qs}(h) &=&  \bigg(\sum_{E_q=0}  \Phi_q^f(1)\Phi^f_q(1) \bigg)\mathcal{P}^{(2,3)}(t \rightarrow \infty) \notag \\
%&-&\bigg(\sum_{E_q=0}  %\Psi^f_q(1)\Psi^f_q(N) %\bigg)\mathcal{P}^{(1,3)}(t \rightarrow \infty) \notag \\
%&+& \bigg(\sum_{E_q=0}  \Phi_q^f(2)\Phi^f_q(1) \bigg)\mathcal{P}^{(1,2)}(t \rightarrow \infty). \label{secondSpinMag}
%\end{eqnarray}
Therefore, to observe nonzero magnetization in the long time limit, even at bulk probe sites, the spectrum has to support zero modes. An important difference in the infinite time limit at $r=2$ from the edge magnetization in Eq.~\eqref{edgeQS} is that there is a nonzero contribution coming from the second site in the fermionic chain in addition to the edge. % which is scaled with two-point correlator $\mathcal{P}^{(2,3)}(t \rightarrow \infty)$. 
While the contribution of the middle term in Eq.~\eqref{secondSMag1} is always zero because $\Psi_{q|E_q=0}^f(1)=0$ in $\mathcal{P}^{(2)}(t)$, the contribution of the first term decreases as the transverse field increases as opposed to increasing contribution from the third term. Close to QCP, these two contributions become equal, which is depicted in Fig.~\ref{fig22b}. These observations suggest that the zero mode spreads across the chain as we approach the QCP, as expected.  Importantly, one can observe the prethermal and q.s. regimes in both terms, shedding light on (i) why an arbitrary bulk spin is still a good probe site to detect QCP; and (ii) why a universal collapse is possible in the prethermal regime of different probe sites \cite{letter}.

Let us note that one could straightforwardly generalize this alternative numerical method to calculate the local magnetization at all sites. Fig.~\ref{fig1a} shows the application of this method to calculate the local magnetization at site $r=3$ for $N=1500$ spins. To mathematically show the origin of zero modes for probe sites $r>2$, one has to add one more condition to the condition list above, which is the requirement of nondegenerate energy gaps \cite{Gogolin_2016}.

\begin{figure}
\centering
\subfloat[]{\label{figN4a}\includegraphics[width=0.24\textwidth]{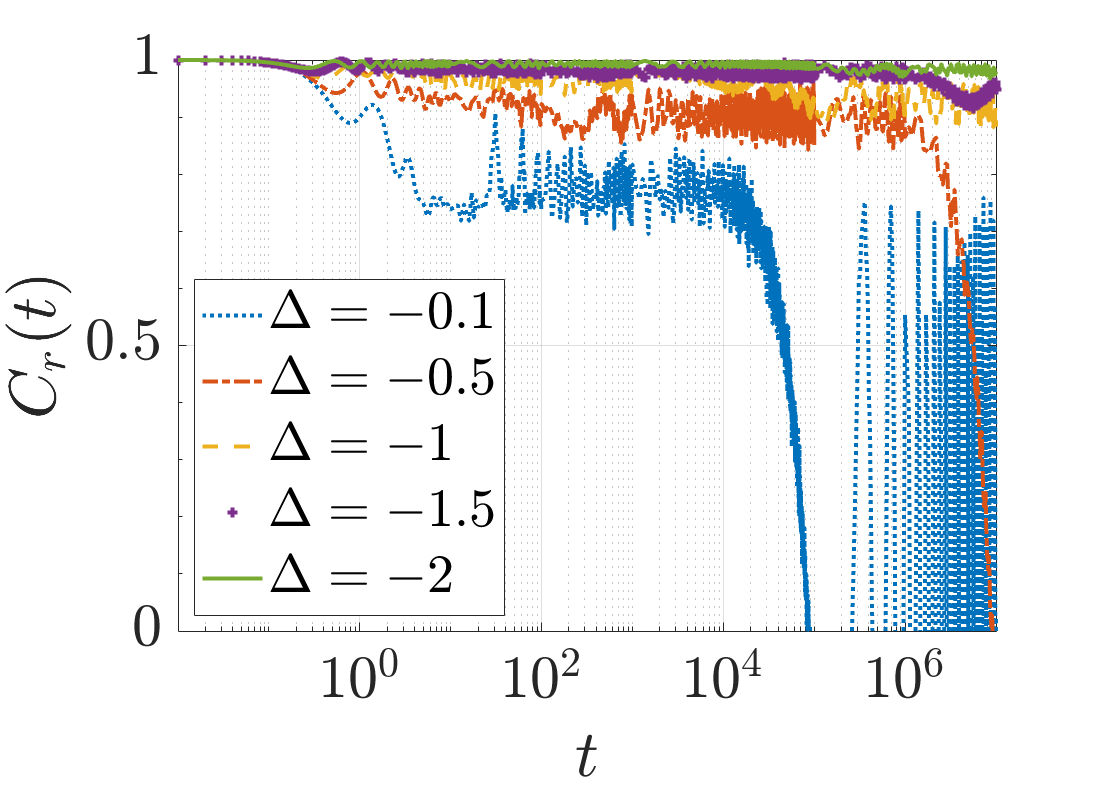}}\hfill 
\subfloat[]{\label{figN4b}\includegraphics[width=0.24\textwidth]{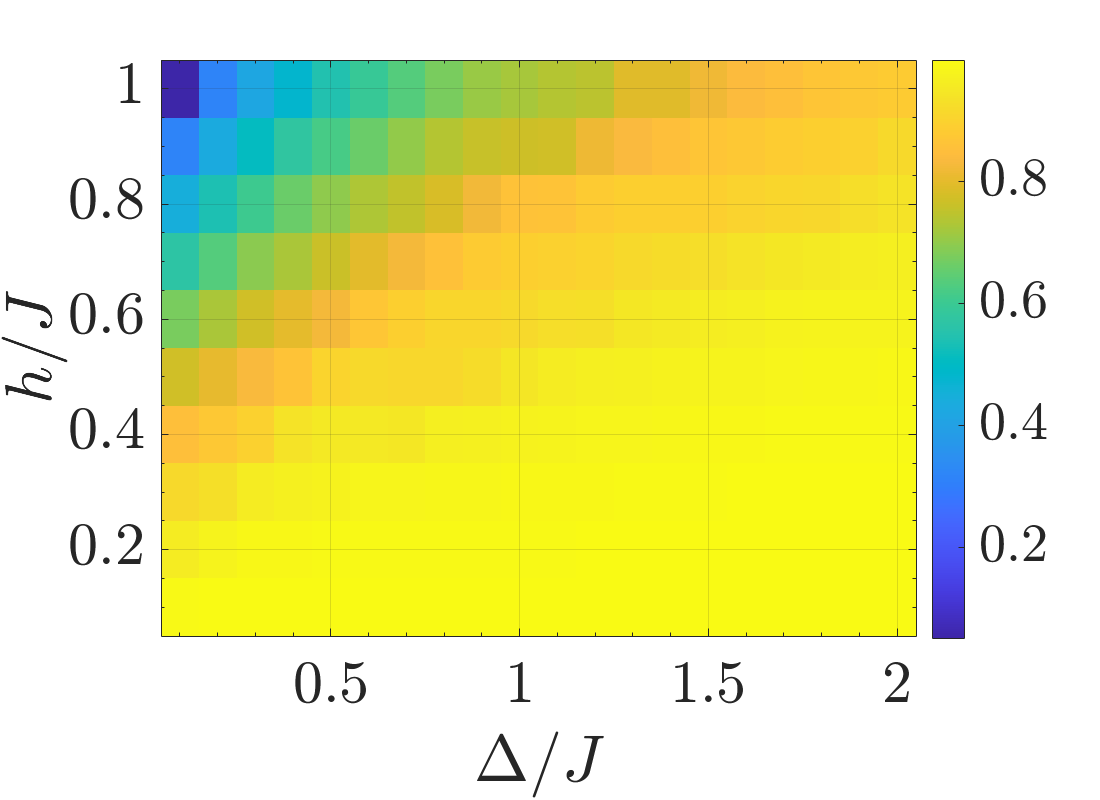}}\hfill 
%\subfloat[]{\label{figS3c}\includegraphics[width=0.33\textwidth]{Fig17c.png}}\hfill 
\caption{Exact diagonalization results for the coherence time of $C_{3}(t)$ at a system size of $N=14$. Subfigure (a) depicts certain nonintegrable models at $h=0.5$ (see legend), whereas (b) gives a two-dimensional color plot of the long-time value of the q.s. regime with respect to external field $h$ and the interaction strength $\Delta$.}
\label{FigN4}
\end{figure}

Finally, we discuss the coherence times of the local magnetization at a bulk site $C_{3}(t)$ (Fig.~\ref{figN4a}), with respect to time for different interaction strengths $\Delta$. Then we map the q.s. values at a field strength $h$ and the interaction strength $\Delta$ in Fig.~\ref{figN4b}. The behavior is monotonous everywhere between $0 < h < 1$ and $0 < \Delta < 2$. An important evidence of strong zero modes \cite{Kemp_2017} is the presence of resonances due to the competition between n.n.~and n.n.n.~terms in Eq.~\eqref{Hamiltonian}, which would result in a non-monotonous trend of the steady state value with respect to $\Delta$. The absence of such behavior seen in Fig.~\ref{FigN4}, despite our analytical proof earlier on the necessity of zero modes to stabilize a q.s. regime, is likely a result of the quench protocol. This is because, in a typical setup where the effects of strong zero modes are demonstrated, e.g.,~infinitely long or finite but long coherence times, the entire many-body spectrum is relevant due to infinite-temperature initial state \cite{Kemp_2017,PhysRevB.97.235134,PhysRevB.101.104415}. Although a quench from a polarized state to an arbitrary $h$ in the ordered phase still generates excitations to higher energy levels, Fig.~\ref{FigN4} clearly demonstrates that only a part of the many-body spectrum is actually relevant in the observation of the q.s. regime. 

\subsection{\label{seriesSec}Derivation of the series expression for the edge magnetization}

In order to write the edge magnetization, Eq.~\eqref{edgeMag3}, in terms of the physical parameters, e.g.,~transverse field $h$, one has to analytically express $\Phi_q^f$ and $\Psi_q^f$. In this section, we will follow an alternative route to derive an analytical expression for $C_1(t)$ in the thermodynamic limit. First we rewrite the Hamiltonian of  Eq.~\eqref{nonInteractingH} at $\Delta=0$ in terms of auxilary operators $\phi_r^{\pm}$,
\begin{eqnarray}
H=\sum_{r} (-J \phi_r^{-} \phi_{r+1}^{+} + h \phi_{r}^{+}\phi_r^{-}). \label{MajoranaH}
\end{eqnarray}
Then we use the Baker–Campbell–Hausdorff formula
\begin{eqnarray}
e^{sA}Be^{-sA}=B+s[A,B]+\frac{s^2}{2!} [A,[A,B]] + \dots ,\label{BCH}
\end{eqnarray}
to calculate $\phi_{1}^{+}(t)=e^{iHt} \phi_{1}^{+} e^{-iHt}$. The following commutators are obtained,
\begin{eqnarray}
[H,\phi_{1}^{+}] &=& 2 h \phi_{1}^{-}, \notag\\
\left[H,\phi_{2}^{+} \right] &=& 2(J \phi_{1}^{-} + h \phi_{2}^{-}),\notag\\
\left[H,\phi_{3}^{+} \right] &=& 2(J \phi_{2}^{-} + h \phi_{3}^{-}),\notag\\
\dots \notag\\
\left[H,\phi_{1}^{-} \right] &=& -2(h \phi_{1}^{+} + J \phi_{2}^{+}),\notag\\
\left[H,\phi_{2}^{-} \right] &=& -2(h \phi_{2}^{+} + J \phi_{3}^{+}),\notag\\
\dots\notag
\end{eqnarray}
and used to calculate the terms of Eq.~\eqref{BCH},
\begin{eqnarray}
[H,[H,\phi_{1}^{+}]]&=&-4(h^2 \phi_{1}^{+} + J h \phi_{2}^{+}), \notag\\
\left[H,\left[H,\left[H,\gamma_1 \right]\right]\right]&=& - 8 [(h^3+J^2 h) \phi_{1}^{-}  + J h^2 \phi_{2}^{-} ] ,\notag \\
\left[H,\left[H,\left[H,\left[H,\gamma_1 \right]\right]\right]\right]&=& 16 [(h^4 + J^2 h^2) \phi_{1}^{+} + (h J^3 \notag \\
&+& 2h^3 J) \phi_{2}^{+} + J^2 h^2 \phi_{3}^{+}], \notag \\
\dots \label{BCHterms}
\end{eqnarray}
Since we calculate $C_{1}(t)=\mathbin{_{\alpha}{\Bra{\psi_0}}} \phi_{1}^{+}(t) \Ket{\psi_1}_{\alpha}$ quenched from a polarized state, the initial state dictates $\Braket{\phi_{1}^{+}}=1$, $\Braket{\phi_{N}^{-}}=-i$ and the remaining expectation values are zero. Hence, Eq.~\eqref{BCHterms} read,
\begin{eqnarray}
\Braket{[H,[H,\gamma_1]]}_{\alpha} &=&-2^2 h^2,\notag \\
\Braket{\left[H,\left[H,\left[H,\gamma_1 \right]\right]\right]}_{\alpha}&=& 0 \notag\\
\Braket{\left[H,\left[H,\left[H,\left[H,\gamma_1 \right]\right]\right]\right]}_{\alpha}&=& 2^4 (h^4 + J^2 h^2) \notag\\
\Braket{\left[H,\left[H,\left[H,\left[H,\left[H,\gamma_1 \right]\right]\right]\right]\right]}_{\alpha}&=& 0 \notag\\
\Braket{\left[H,\left[H,\left[H,\left[H,\left[H,\left[H,\gamma_1 \right]\right]\right]\right]\right]\right]}_{\alpha} &=& -2^6 (h^6 \notag \\
&+& 3 h^4 J^2 + h^2 J^4) \notag\\ 
\dots \notag
\end{eqnarray}
resulting in the following series solution
\begin{eqnarray}
C_{1}(t) =\sum_{m=0}\frac{(-1)^m}{(2m)!} (2t)^{2m} N_m(h^2), \label{seriesSoln}
\end{eqnarray}
where $N_m(x)$ are called the Narayana polynomials
\begin{eqnarray}
N_m(h) = \sum_{n=1}^m N_{mn}h^n, \hspace{3mm} N_{mn} = \frac{1}{m} \binom{ m}{n-1}\binom{m}{n}.\notag
\end{eqnarray}
Let us note that one obtains exactly the same series solution in the calculation of two-time correlators of the edge spin at infinite-temperature \cite{PhysRevB.97.235134}. 

\begin{figure}
\centering
\subfloat[]{\label{figS14a}\includegraphics[width=0.24\textwidth]{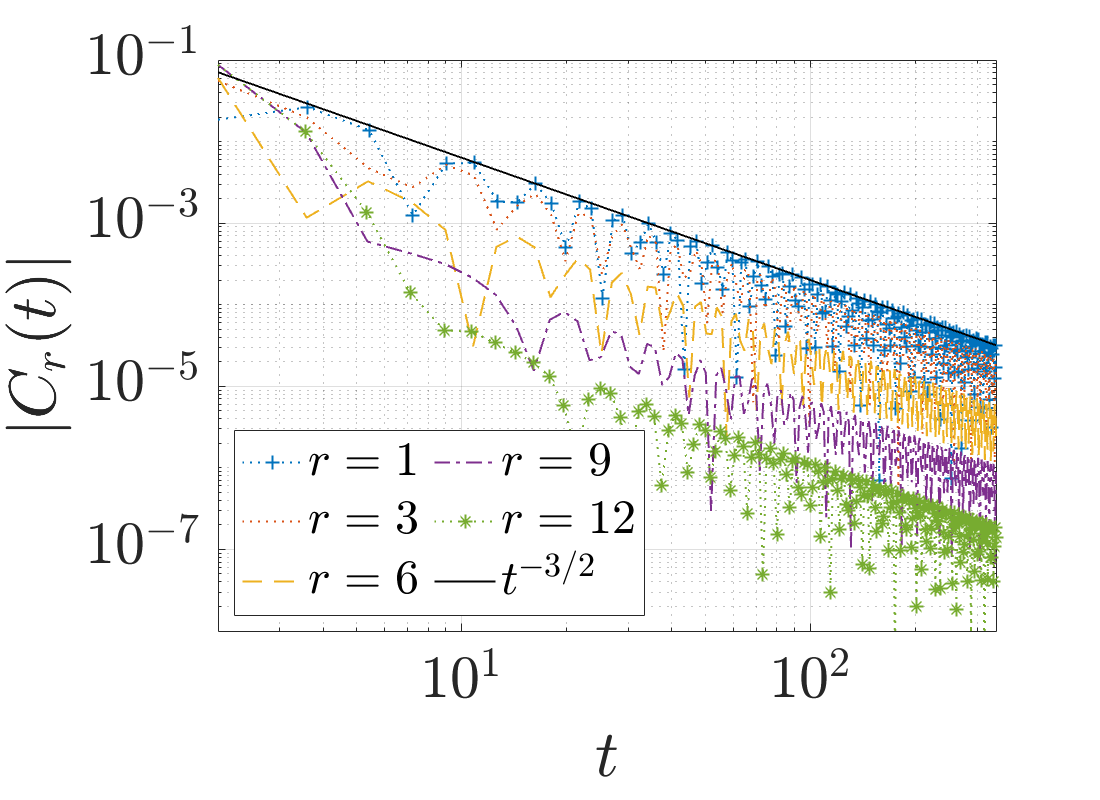}}\hfill 
\subfloat[]{\label{figS14b}\includegraphics[width=0.24\textwidth]{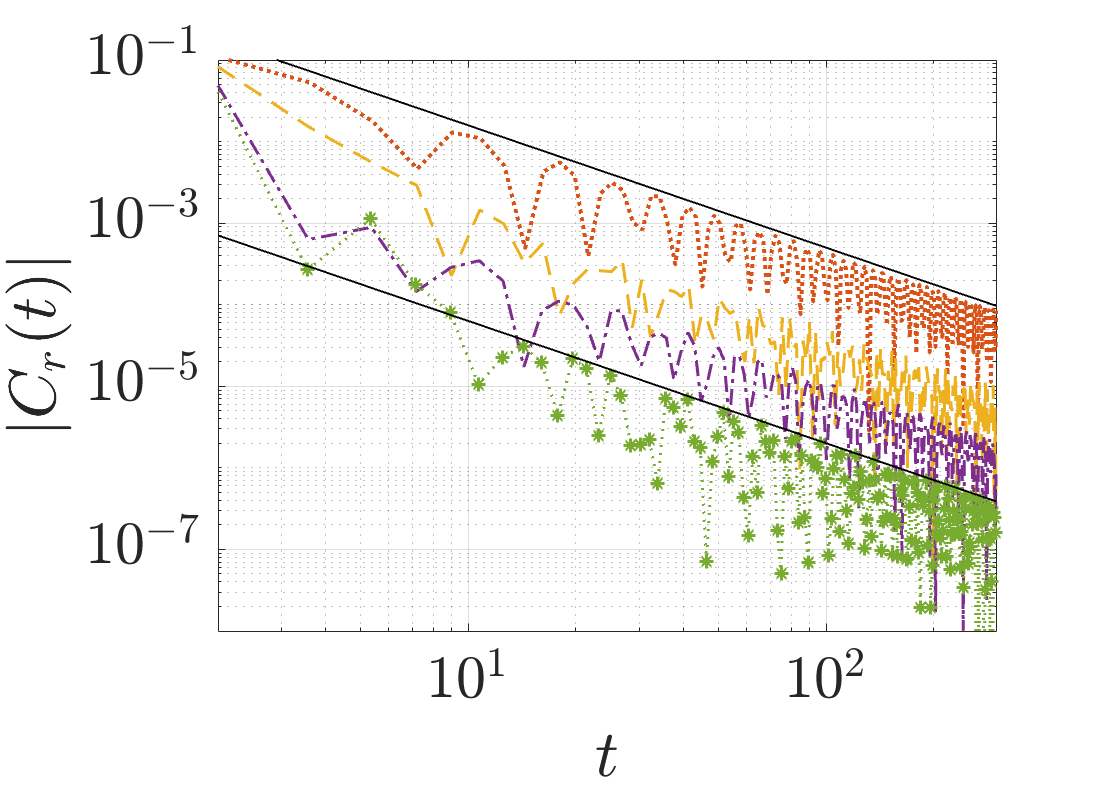}}\hfill 
\caption{The local magnetizations of the integrable TFIC (a) at QCP $h=1$ and (b) in the disordered phase $h=1.2$ for different probe sites (see legend). All decays as a power-law in long time with $t^{-3/2}$.}\label{figS14}
\end{figure}

At the critical point $h=1$, the series solution Eq.~\eqref{seriesSoln} can be written in a closed form,
\begin{eqnarray}
C_{1}(t,h=1) = \frac{J_1(4t)}{2t},
\end{eqnarray}
where $J_1(4t)$ is the Bessel function of the first kind. The long-time asymptotics of $J_1(4t)/2t$ is $\propto t^{-3/2}$ which is exactly what we observe in the numerical calculations, Fig.~\ref{figS14a}, not only for the edge magnetization but also for local magnetization at different probe sites.

Since the Narayana polynomials satisfy the property $N_m(x^{-1})=x^{-m-1}N_m(x)$, $C_{1}(t,h)$ satisfies  %\footnote{This relation is pointed out in Ref.~ \cite{PhysRevB.97.235134} for the two-time correlators of the edge spin at infinite-temperature.} 
\begin{equation}
C_{1}(t,h)=1-h^2+h^2 C_{1}(ht,h^{-1}), \label{eqduality}
\end{equation}
indicating that the edge magnetization behaves in one phase  (transverse-field strength $h$) as it would in the complementary phase (transverse-field strength $h^{-1}$) but with a scaled time $ht$. 

In Ref.~\cite{letter}, some of us proposed a general form for the critical response function  $\delta C_{r}(t,h_n) \equiv C_{r}(t,h_n)-C_{r}(t,h_n=0)$ where $h_n\equiv(h_c-h)/h_c$, 
\begin{eqnarray}
\delta C_{r}(t,h_n)=C^{qs}_r(|h_n|)f_{\Delta,h_i}(h_n t)
\end{eqnarray}
which holds near $h_n= 0$ when $t \gg 1$, and the function $f_{0,0}(h_n t)$ has been derived based on the edge magnetization in Eq.~\eqref{seriesSoln},

\begin{equation}
f_{0,0}(h_n t)= \frac{1}{2}+\frac{(h_nt)}{2}{}_1F_2\left[\left\{\frac{1}{2}\right\};\left\{\frac{3}{2},2\right\};-(h_n t)^2\right] \label{eqdeltaCr1}
\end{equation}
where $_1F_2\left(\left\{a_1\right\};\left\{b_1,b_2\right\};z\right)$ is a generalized Hypergeometric function. %Note that the right hand side of Eq.\ \eqref{eqdeltaCr1} is a function of the rescaled time $h_n t$.  
Even though Ref.~\cite{letter} focuses on dynamics when quenching to the ferromagnetic phase, i.e.,~$h<1 \ (h_n>0)$, the function $f_{0,0}(h_n t)$ given by Eq.\ \eqref{eqdeltaCr1} is a good approximation of $\delta C_{r}(t,h_n)$  in both phases near QCP, $|h_n| \rightarrow 0$. This suggests that there is a corresponding prethermal temporal regime in the disordered phase too where dynamics critically slow down. %Hence, with an additional short-time oscillation term $- \frac{\tilde{h}}{2} J_0(4t) $, we can obtain an ansatz of the critical response that works at any $t$ near the critical point in both phases: $\delta C_{1}(t,h) \approx  - \frac{\tilde{h}}{2} J_0(4t) +\delta \mathcal{C}_{1}(t,h)$. 
\begin{figure}
\centering
\includegraphics[width=0.45\textwidth]{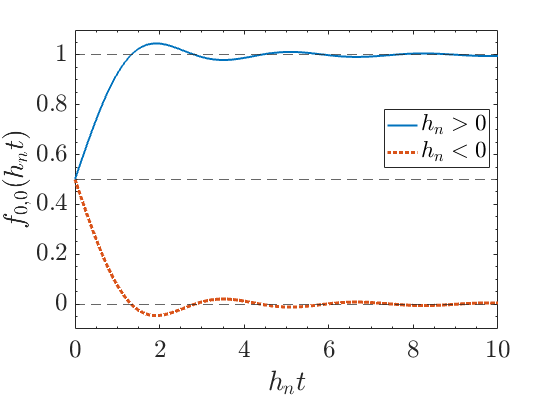}\hfill 
\caption{The rescaled critical response $\tilde{h}^{-1}\delta C_{1}(t,h)$ is plotted with respect to the rescaled time $\tilde{h}t$. The blue and red solid lines show the critical response for the ordered and disordered phases, respectively. Dashed lines are added to guide the eye. }\label{figAnsatz}
\end{figure}

\begin{figure*}
\includegraphics[width=0.99\textwidth]{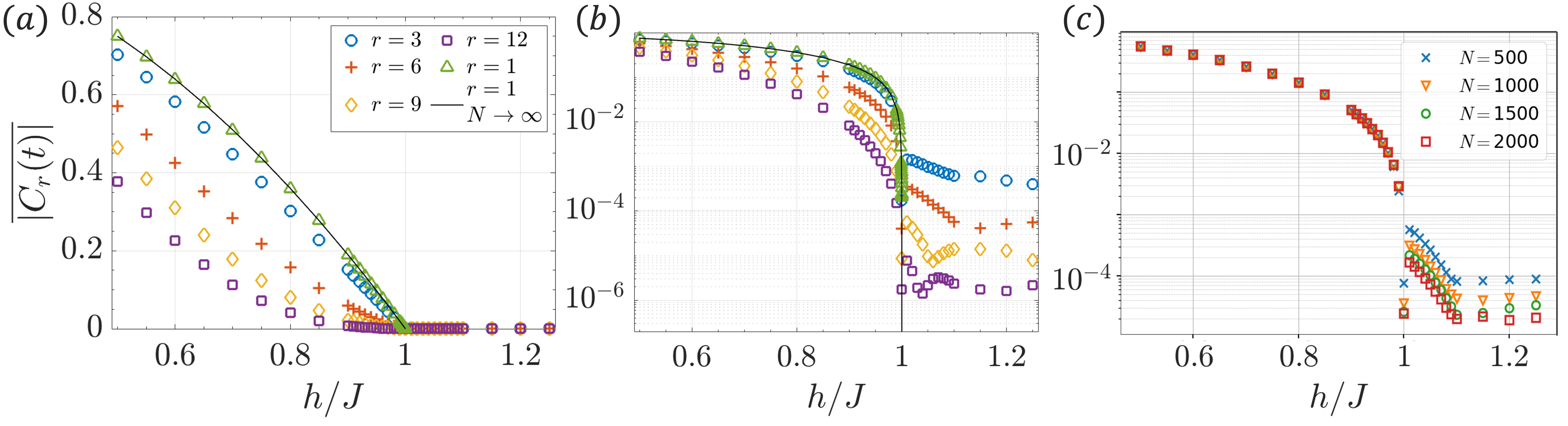}\hfill 
\caption{Nonequilibrium phase diagram for $\Delta=0$ (a) for different sites $r=1,3,6,9,12$ at $N=1440$ (see legend). Solid-black line is the analytic result for edge magnetization in the thermodynamic limit. (b) The same plot with (a) in semi-logarithmic scale, exhibiting a kink behavior at the QCP which is nonanalytic in the thermodynamic limit. The small but finite values at the QCP and in the disordered phase are due to inevitable finite-time cutoff. These values are expected to vanish when $t_l \rightarrow \infty$, see Fig.~\ref{figS14}. (c) For different system sizes, $N=500$~--~$2000$, at site $r=6$. $\overline{C_r(t)}$ at $h_c$ scales with the system size as $1/N$ for all sites $r=3,6,9,12$ (see Appendix~\ref{AppA}).}\label{figN3a}
\end{figure*}

In Fig.\ \ref{figAnsatz}, we plot the function $f_{0,0}(h_n t)$ as a function of $h_n t$. Since the second term on the right hand side of Eq.\ \eqref{eqdeltaCr1} is an odd function in $h_n t$, $f_{0,0}(h_n t)$ is symmetric about the line $\frac{1}{2}$ in Fig.\ \ref{figAnsatz} both for $h_n>0$ and  $h_n < 0$. We observe $\delta C_{1}(t=0,h)=h_n/2$ in both phases when $t\gg 1$ and $h_n t\ll 1$ hold, hence the prethermal dynamics starts at value $h_n/2$ in both phases. $\delta C_{1}(\infty,h < 1)=1-h^2$, while $\delta C_{1}(\infty,h>1)=0$, indicating that the edge magnetization reaches a q.s.~value $1-h^2$ in the ordered phase and  $0$ in the disordered phase as $t\rightarrow \infty$. Therefore, the critical response  shows a non-analyticity at QCP in the long time limit. We will harness this non-analyticity in the next sections in the detection of QCP with single-site magnetizations. 
%The duration of the prethermal regime is inversely proportional to $|\tilde{h}|$, as in ordered phase. 

\subsection{\label{NumericalResults}Detection of QCP}

To probe the q.s.~regime, we study the the time-averaged single-site magnetization,
\begin{eqnarray}
 \overline{|C_r(t,h)|}=\frac{1}{(t_l-t^*)}\int_{t^*}^{t_l} dt\, |C_r(t,h)| \equiv \bar{C}_r(h)
\end{eqnarray}
where $t^*$ is the ultraviolet (short-time, short-distance) cutoff and $t_l$ is the (long-time, long-distance) infrared cutoff \cite{2020arXiv200412287D}. For the numerical results presented in this section, $t_l$ is the evolution time at which the cluster theorem, Sec.~\ref{clusterThmSec}, breaks down, and hence as explained in the previous section $t_l= \Delta x/(2v_q)$. This means that the infrared cutoff $t_l$ is parametric in the transverse field $t_l(h)$, because in the integrable TFIC $v_q=2h$ for $h\leq h_c$ and $v_q=2h_c$ for $h>h_c$ \cite{Calabrese_2012}. 

For quenches sufficiently far away from the vicinity of the QCP, $\overline{|C_r(t)|}$ matches the q.s. value, as then there is no prethermal regime \cite{letter}. Each single-site observable in the TFIC equilibrates around a different value in the q.s. regime. This can be seen in Fig.~\ref{figN3a}a that depicts $\overline{|C_r(t)|}$ for $r=1-12$, all of which have a different functional form of $h$. The analytic expression for the value of the q.s. regime at $r=1$, $C^{qs}_{1}(h)=1-h^2$ (Sec.~\ref{seriesSec}) matches the corresponding numerical result in Fig.~\ref{figN3a}a. Fig.~\ref{figN3a}b shows the same plot in logarithmic scale that exhibits a kink in $\bar{C}_r(h_c)$ regardless of the value of $r$. 

Whether there is a nonanalytic behavior at QCP for all $r$ could be answered through a finite-size analysis. In Fig.~\ref{figN3a}c, we observe that the dynamical order of $\bar{C}_{6}(h)$ is persistent for $h < h_c$ as system size increases from $N=500$ to $2000$, i.e., its value stays the same with increasing system size (see Appendix \ref{AppA} for other $r$). 
Although the local order profiles for each $r$ away from the QCP are different, Fig.~\ref{figN3a}a, they all approach to QCP linearly in $h_n$ in the thermodynamic limit \cite{letter}. For $h \geq h_c$ the dynamic order vanishes for all $r$ in Fig.~\ref{figN3a}c, i.e.,~it has a decreasing trend with increasing system size. In fact, we can prove that the local magnetization should vanish in the thermodynamic limit, based on the simple observation of $t^{-3/2}$ decay of local magnetization for all $r$ both at the QCP, Fig.~\ref{figS14a} and in the disordered phase, Fig.~\ref{figS14b}.
\begin{eqnarray}
\overline{C(t,h \geq h_c)} = \frac{8}{\sqrt{t^*}}\left(\frac{N}{h_c}\right)^{-1} + \mathcal{O}(N^{-3/2}).\label{scalingEqatQCP}
\end{eqnarray}
The derivation of Eq.~\eqref{scalingEqatQCP} and the accompanying numerics of the scaling $1/N$ at the QCP can be found in Appendix~\ref{AppA}. Therefore, the nonanalyticity at $h_c$---the hallmark of a phase transition---is captured by all single-site observables $r\ll N/2$. Consequently, we have demonstrated a DPT for different sites $r \ll N/2$ that reflects the underlying QPT. 

\section{\label{NNNmodel}The nonintegrable TFIC with next nearest neighbor coupling}

In this section, we first lay out the qMFT formalism that has also been applied to obtain the results of weak perturbations in Ref.~\cite{letter}. Subsequently, we present the $t$-DMRG calculations of both weakly and strongly interacting TFIC. 

\subsection{Quench Mean-field theory (qMFT) analysis}

\begin{figure}
\centering
\subfloat{\includegraphics[width=0.5\textwidth]{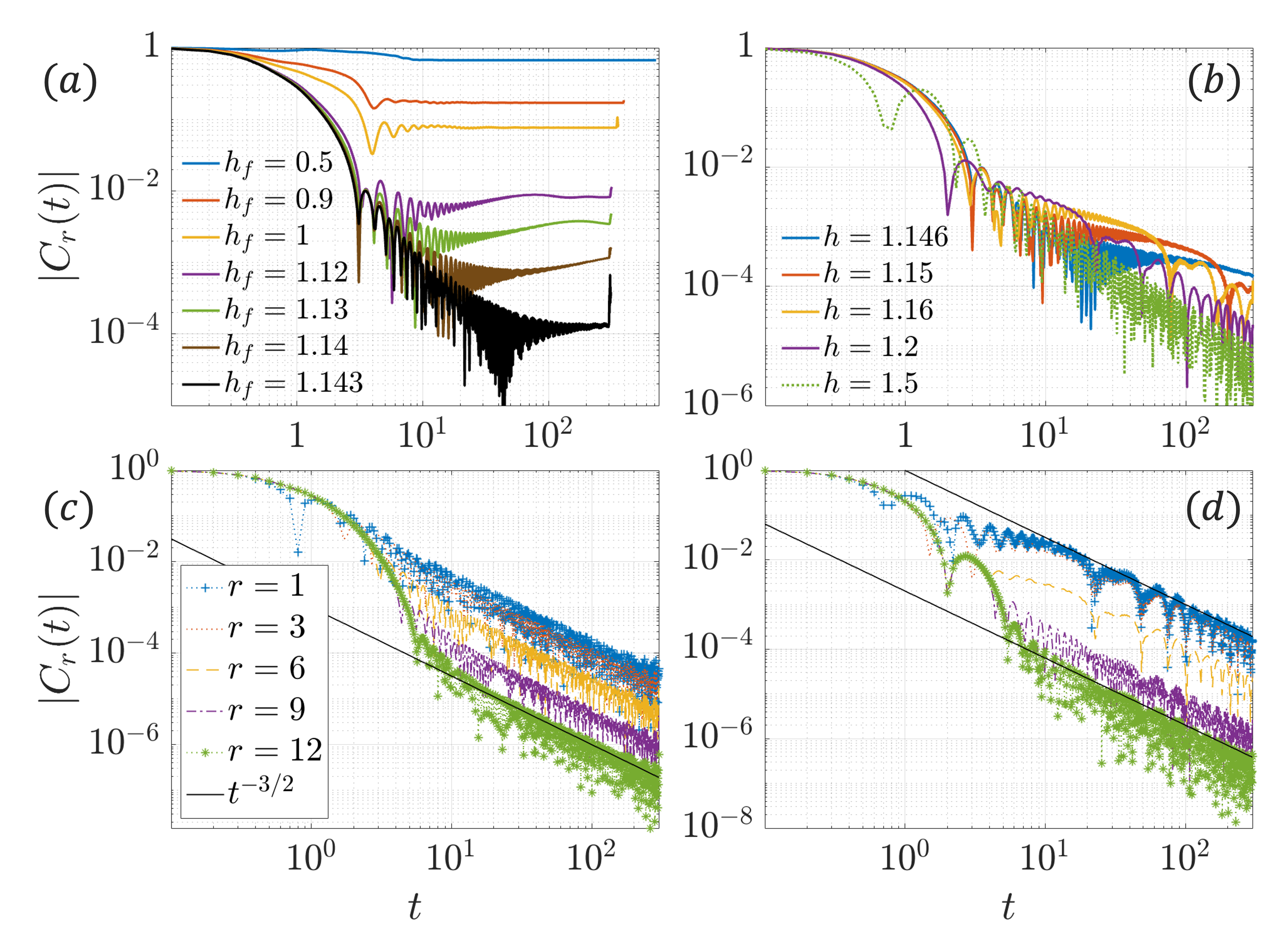}}\hfill
\caption{qMFT single-site magnetization of the near-integrable model with $\Delta=0.1$ (a) for $h \leq 1.143$, an interval of $h$ that is representative of a dynamically-ordered phase and (b) for $1.146 \leq h \leq 1.5 $, an interval of $h$ that is representative of a dynamically-disordered phase at a system size of $N=1440$ at $r=6$. The recurrence attempts observed here imply the breakdown of the cluster theorem. qMFT single-site magnetization of the near-integrable model for probe sites $r=1,3,6,9,12$ (c) at DCP $h_{dc}=1.1437$, (d) in the disordered phase sharing the same legend.}
\label{FigS21}
\end{figure}

In order to incorporate the qMFT formalism to our quench dynamics that is described in Sec.~\ref{clusterThmSec}, we need to express the n.n.n.~term in the TFIC Hamiltonian in the fermionic picture. This expression reads,
\begin{eqnarray}
%&=&-\Delta \sum_r \left(c_r - c_r^{\dagger}\right) \left(1 - 2 c_{r+1}^{\dagger} c_{r+1}\right) \left(c_{r+2} + c_{r+2}^{\dagger}\right), \notag \\
\text{n.n.n.}&=&\Delta \sum_r \left( c_r^{\dagger}-c_r \right) \left(1 - 2 c_{r+1}^{\dagger} c_{r+1}\right) \left(c_{r+2} + c_{r+2}^{\dagger}\right), \notag \\
&=& \Delta \sum_r \phi^-_r \phi^+_{r+1} \phi^-_{r+1}  \phi^+_{r+2}, \label{eq1}
\end{eqnarray}
where $\Delta > 0$ and $\phi_r^{\pm}$ stand for the auxiliary fermions.

In Hartree--Fock expansion, we assume $|\Delta| \ll |J|$, and write Eq.~\eqref{eq1} as
\begin{eqnarray}
&=& \Delta \sum_r \bigg [ \Braket{\phi^-_r \phi^+_{r+1}}_{t\rightarrow \infty} \phi^-_{r+1}  \phi^+_{r+2} \notag \\
&+& \phi^-_r \phi^+_{r+1} \Braket{\phi^-_{r+1}  \phi^+_{r+2}}_{t\rightarrow \infty} -  \phi^-_r \Braket{\phi^+_{r+1} \phi^-_{r+1}}_{t\rightarrow \infty}  \phi^+_{r+2} \notag \\
&-& \Braket{\phi^-_r \phi^+_{r+2}}_{t\rightarrow \infty} \phi^+_{r+1} \phi^-_{r+1} \bigg]. \label{MFT}
\end{eqnarray}
Here the $\Braket{\cdot}_{t\rightarrow \infty}$ means that we calculate the free fermion problem and obtain the correlators with respect to the state in the q.s.~regime when $t\rightarrow \infty$, instead of the ground state which would be used to calculate the equilibrium QCP. An analytical qMFT formalism was first introduced in Ref.~\cite{PhysRevLett.123.115701} to calculate a dynamical order parameter, based on two-point correlators in a periodic chain, by utilizing the momentum space representation where the infinite time limit can indeed be taken. In our numerics for an open-boundary chain, we treat the largest time point allowed by the cluster theorem, $t_l=\Delta x/(2v_q)$, as the asymptotically infinite time limit. Note that for an open-boundary chain one needs to carefully take the edges of the chain into account when calculating Eq.~\eqref{MFT}. Using the above expansion, we obtain an effective mean field Hamiltonian which has slightly stronger n.n.~coupling compared to the free problem, as well as new n.n.n~couplings. Further, the effective chemical potential slightly decreases, which is reasonable when we think about how the critical point shifts to favor order, e.g.,~for $\Delta=0.1$, $h_c = 1.1631\pm 0.0037$ (see Appendix~\ref{AppC}). 

When applying the cluster theorem to the qMFT of the weakly-interacting nonintegrable TFIC, one needs to estimate the lightcone (correlation) velocity $v_q$ of the model. While for the integrable TFIC this velocity is analytically known, this is not true when we introduce nonintegrability to the model. Here we approximate a quasiparticle velocity based on the analytical prediction of the integrable TFIC: $v_q=2h$ for $h\leq h_c$ and $v_q=2h_c$ for $h>h_c$ \cite{Calabrese_2012}. Since this is only an approximation, we sometimes exceed the time when the cluster theorem really breaks down. This means that the distant sites of the chain had already become correlated with one another. This time can be observed with a recurrence attempt in the results, e.g.,~Fig.~\ref{FigS21}a, which is also a sign of finite-size effects. 

Figures~\ref{FigS21}a~and~\ref{FigS21}b show single-site magnetization of the near-integrable model with $\Delta=0.1$ at probe site $r=6$ and for different $h$ across the QCP, $h_c$. One could easily notice the qualitative difference between single-site magnetizations in different phases: While there is a q.s.~regime present in Fig.~\ref{FigS21}a for $h$ sufficiently away from the QCP signaling a dynamically-ordered phase, the magnetization decays as a power-law in Fig.~\ref{FigS21}b for $h \geq h_c$ indicating a dynamically-disordered phase.

\subsection{Detection of QCP in the near integrable TFIC through qMFT analysis}

Now we study the single-site magnetization of the near integrable TFIC where $\Delta=0.1$ in Eq.~\eqref{Hamiltonian} treated with qMFT and when the system is quenched to the vicinity of QCP. This model has a QCP at $h_c = 1.1631\pm 0.0037$ which is calculated with DMRG (Appendix \ref{AppC}). 

\begin{figure}
\includegraphics[width=0.5\textwidth]{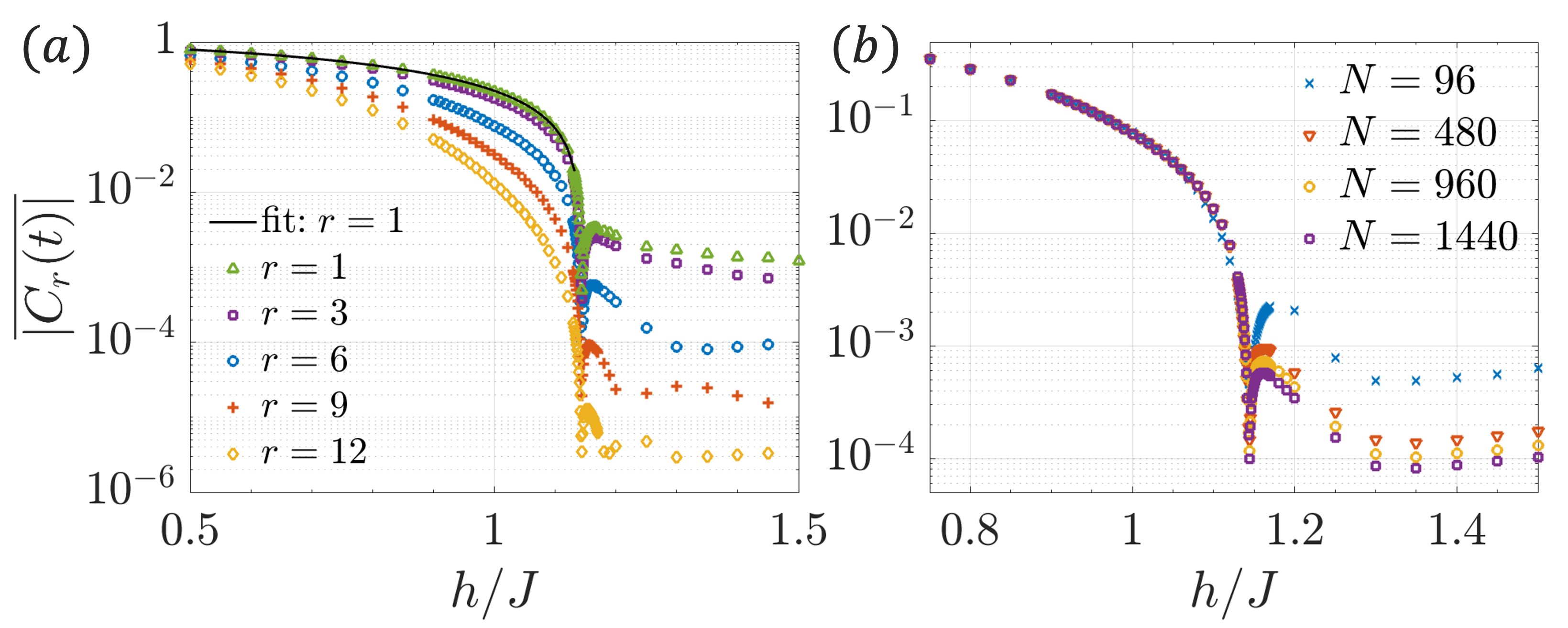}\hfill
\caption{qMFT nonequilibrium phase diagram of the near-integrable TFIC with $\Delta=0.1$ (a) for different sites $r=1,3,6,9,12$ in descending order at $N=1440$ exhibiting a kink at the DCP and solid-black line is a fit function for the edge magnetization; (b) for different system sizes $N=96$~--~$1440$ at site $r=6$.}
\label{Fig20}
\end{figure}

Local order profiles follow similarly to the case of integrable TFIC ($\Delta=0$) except for a shift in the QCP to favor order, $h_c > 1$, as expected (Fig.~\ref{Fig20}a). We focus on $\bar{C}_{6}(h)$ in Fig.~\ref{Fig20}b where we find a kink at $h_{dc}=1.1437\pm0.0001 < h_c$ whose magnetization decreases with increasing system size $\bar{C}_r(h_c) \propto N^{-1}$.  $N^{-1}$ scaling follows from the fact that the single-site magnetization at $h_{dc}$ decays as $\propto t^{-3/2}$ (see Fig.~\ref{FigS21}c and Appendix \ref{AppC}). The magnetization in the disordered phase decays in the same way in Fig.~\ref{FigS21}d. Consistently, this behavior is translated to the local order profiles as $\bar{C}_r(h \geq h_{dc}) \propto N^{-1}$ suggesting $\bar{C}_r(h \geq h_{dc}) = 0$ in the thermodynamic limit. On the contrary for $h < h_{dc}$, the dynamical order is persistent, i.e.,~$\bar{C}_r(h < h_{dc})$ value stays the same with increasing system size. 

The q.s. value of the edge magnetization, which our numerical results access for quenches far away from the vicinity of the transition, can be fitted well with a functional form that is reminiscent of that of the integrable TFIC, $C^{qs}_{1}(h)=\alpha (h_{dc}^{\nu}-h^{\nu})$ for $h \leq h_{dc}$ and zero otherwise. Furthermore, because a critically prethermal regime with self-similar dynamics is found also for the same model in \cite{letter}, all local order profiles in the ordered phase approach to QCP linearly in $h_n$ in the thermodynamic limit. Therefore, we observe a non-analytic behavior for the near-integrable TFIC when it is treated with qMFT. Subsequently, we call the location of this non-analyticity as a DCP.

We also apply energy gap analysis to the effective qMFT Hamiltonian with n.n.n. terms in Eq.~\eqref{eq1} and find the ground state gap of this effective Hamiltonian closes at $h_c^{\text{qmft}}=1.1438\pm 0.0011$ (see Appendix \ref{AppC}). Let us note that $h_c^{\text{qmft}} \sim h_{dc} < h_{c}$, suggesting that the DCP of the near-integrable model traces back to the QCP of the effective qMFT Hamiltonian. Therefore, we conclude that the small shift in $h_{dc}$ from $h_c$ likely originates from the qMFT method.

It is important to note that although the qMFT treatment gives rise to a q.s. regime for long intervals of time for quenches away from the vicinity of the DCP (Fig.~\ref{FigS21}a), this is not conclusive evidence for infinitely long-lived nonthermal behavior in a nonintegrable model. Indeed, MFT is not expected to adequately capture thermalization as it may neglect fluctuations that are essential for the latter. However as we will detail in the next subsection, $t-$DMRG confirms the presence of a transition in the thermodynamic limit. Since the relaxation time of the system observables critically diverges as we approach the QCP \cite{letter}, it is likely that an infrared cutoff at $t_l$ might not accurately capture q.s.~state in the infinite-limit near QCP. This could explain why our observed DCP at $h_{dc}=1.1437\pm0.0001$ for $\Delta=0.1$ is slightly smaller than the DCP found in Ref.~\cite{PhysRevLett.123.115701}, which is $1.15$. 

After determining the DCP, we obtain $\alpha=0.81$ and $\nu=1.81$ for the fit function of the edge magnetization. This fit function is plotted in Fig.~\ref{Fig20}a as a black-solid line. 

%Although it is possible that the DCP we obtain is actually the QCP and the small numerical difference between them is a mere artifact of the qMFT method, it is worth noting that MFT usually predicts a larger critical point than the physical one since it neglects fluctuations, whereas here $h_{dc}<h_c$. 

%Finally let us note the time evolution of the local magnetization around the DCP in Figs.~\ref{FigS11a} and~\ref{FigS11b} for $N=1440$ which highlights a qualitative change across the DCP. For example, up until $h=1.144$ which is denoted by the gray-dashed line in Fig.~\ref{FigS11a}, there is evidence of equilibration, whereas starting at $h=1.145$, which is denoted by the pink-dotted line, the response starts to develop a low-frequency oscillatory feature seen in the downward trend of its dynamics in Fig.~\ref{FigS11b}. This feature is captured as a nonanalyticity in the phase diagram as can be seen in Fig.~\ref{Fig20} for different sites but at the same location which uniquely defines a DCP. 

%Let us also note that long wavelength oscillations are characteristic of the decay at the QCP and in the disordered phase, Fig.~\ref{figS14a}. One can notice that the envelope of the nonequilibrium response in Fig.~\ref{FigS11b} decays in a power-law fashion, suggesting the absence of a q.s. regime. 

\subsection{\label{t-DMRG}$t$-DMRG calculations}

\begin{figure}
\centering
\subfloat[]{\label{fig3a}\includegraphics[width=0.24\textwidth]{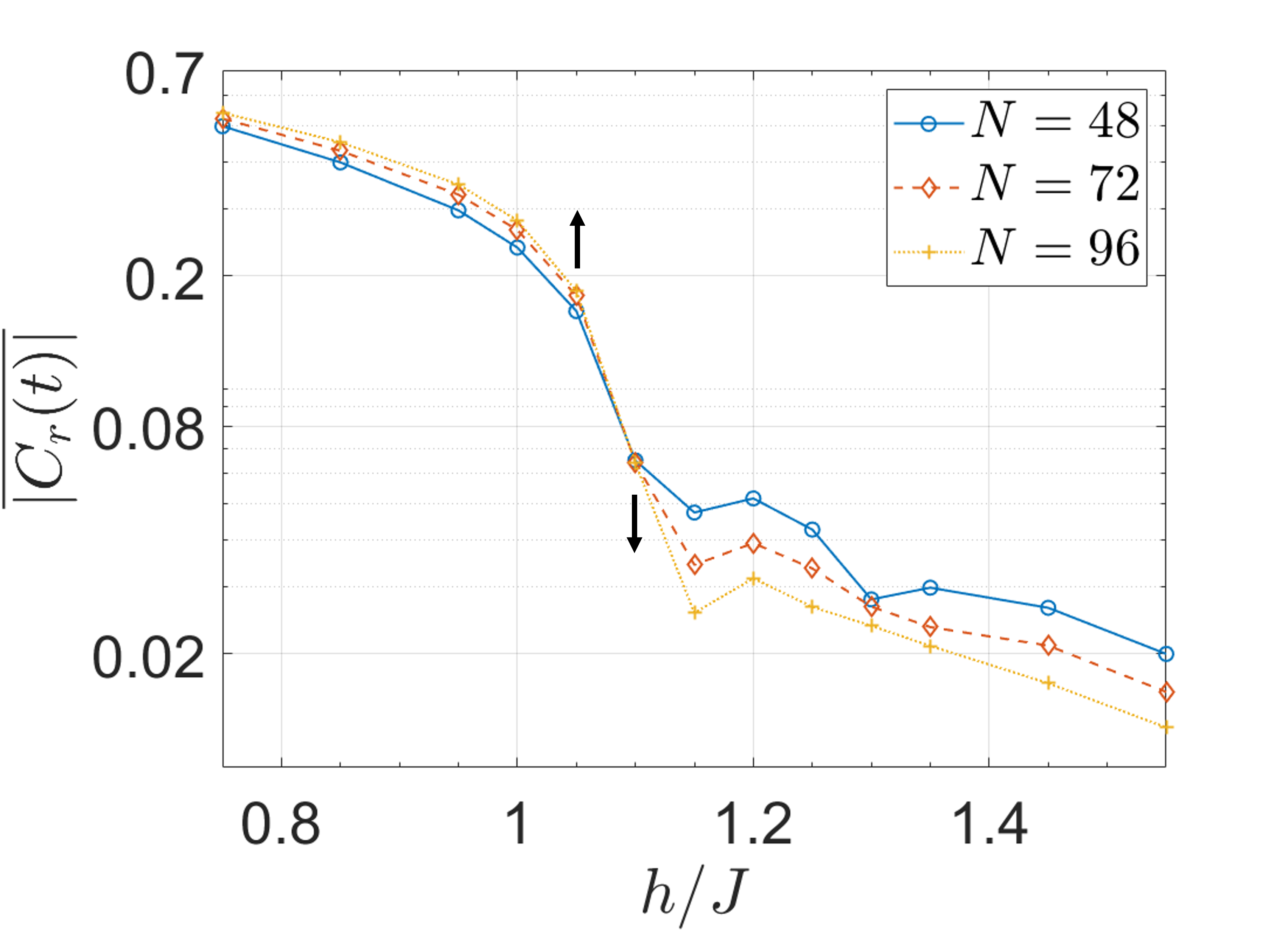}}\hfill 
\subfloat[]{\label{fig3b}\includegraphics[width=0.24\textwidth]{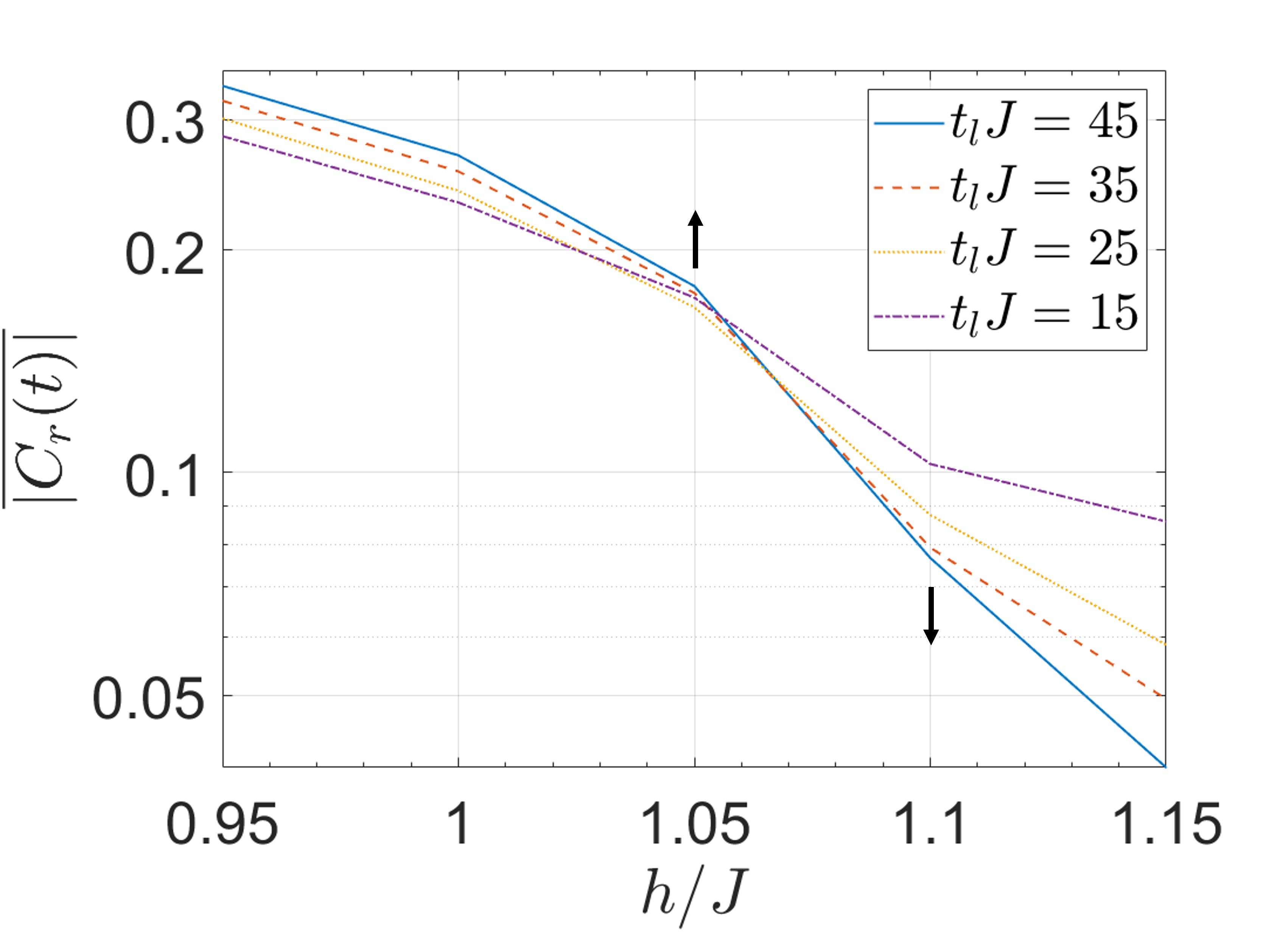}}\hfill 
\caption{Order profiles calculated with $t$-DMRG for weakly nonintegrable TFIC with $\Delta=0.1J$, (a) for different system sizes $N=48, 72,96$ at site $r=3$ when a (infrared) temporal cutoff of $t_l = N/2$ is applied. The order profile is tested against different infrared and ultraviolet temporal cutoffs and shown to be robust (see text). (b) Finite-time scaling analysis at a system size of $N=96$ for different infrared cutoffs, shown in the legend, with a fixed ultraviolet cutoff of $t^*=0$. The upward and downward arrows highlight $h$ points where the order grows or diminishes, respectively with increasing (a) system size and (b) simulation time.}
\label{Fig3}
\end{figure}

\begin{figure*}
\subfloat[]{\label{fig4a}\includegraphics[width=0.33\textwidth]{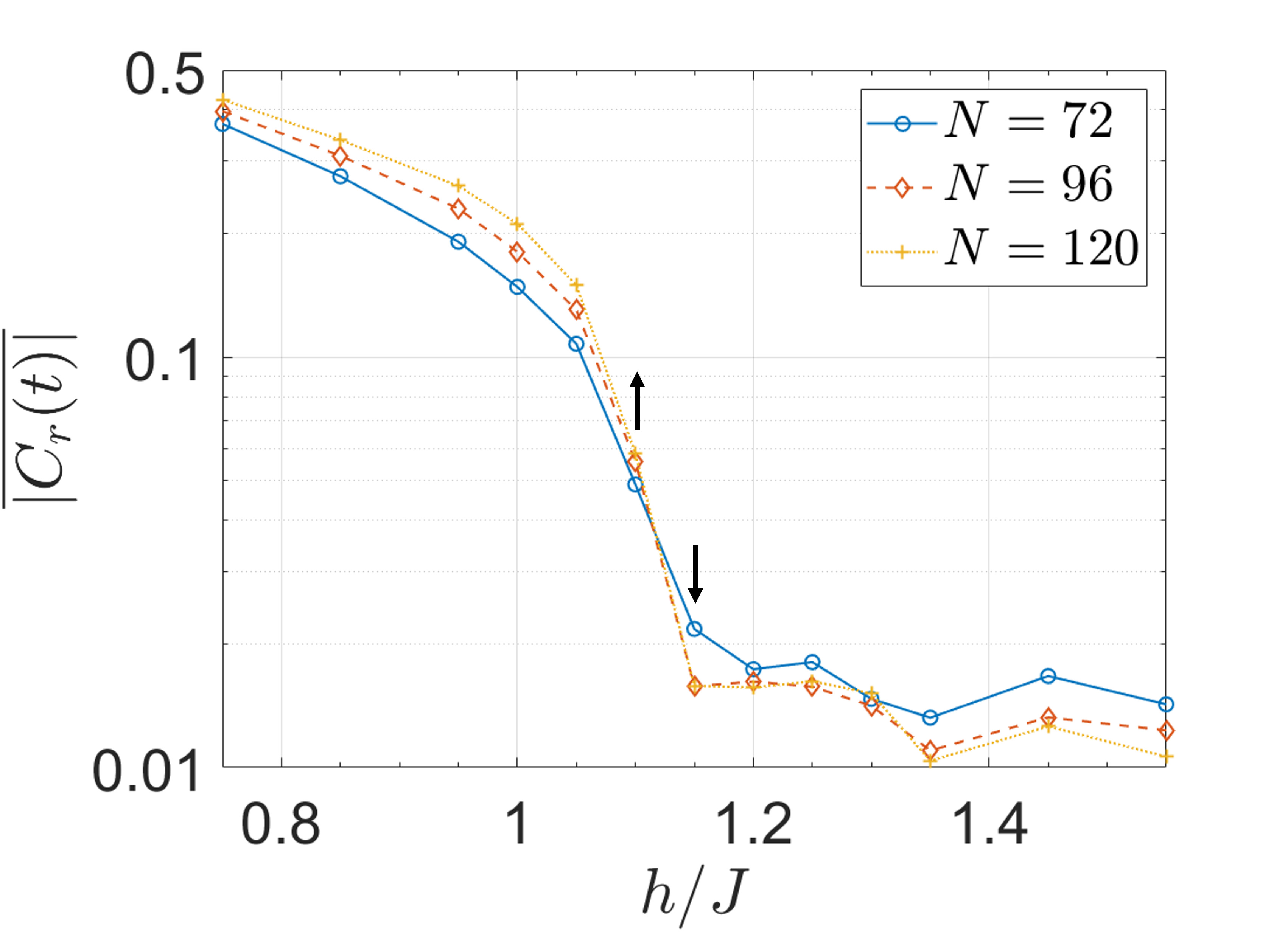}}\hfill
\subfloat[]{\label{fig4b}\includegraphics[width=0.33\textwidth]{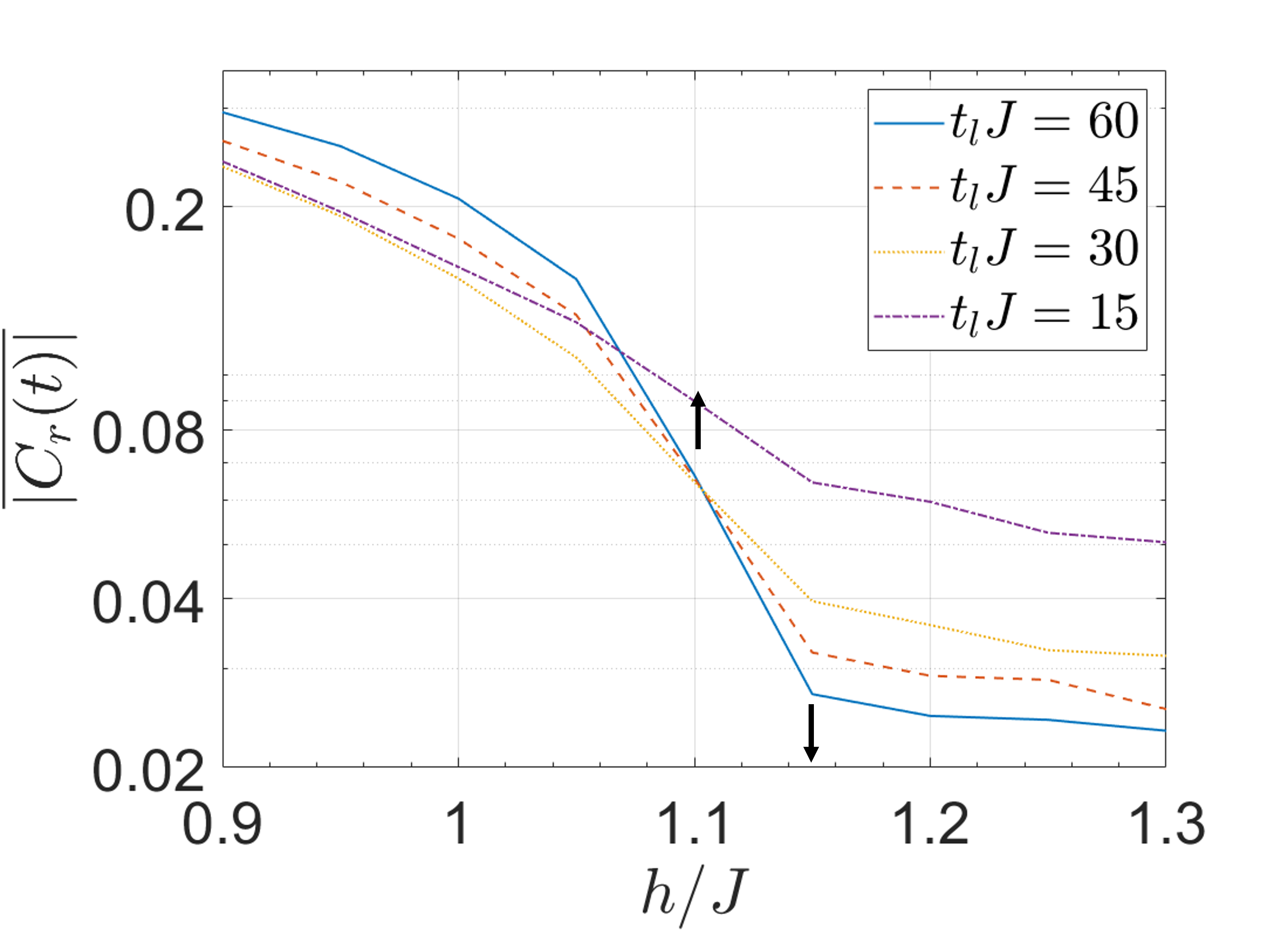}}\hfill
\subfloat[]{\label{fig4c}\includegraphics[width=0.33\textwidth]{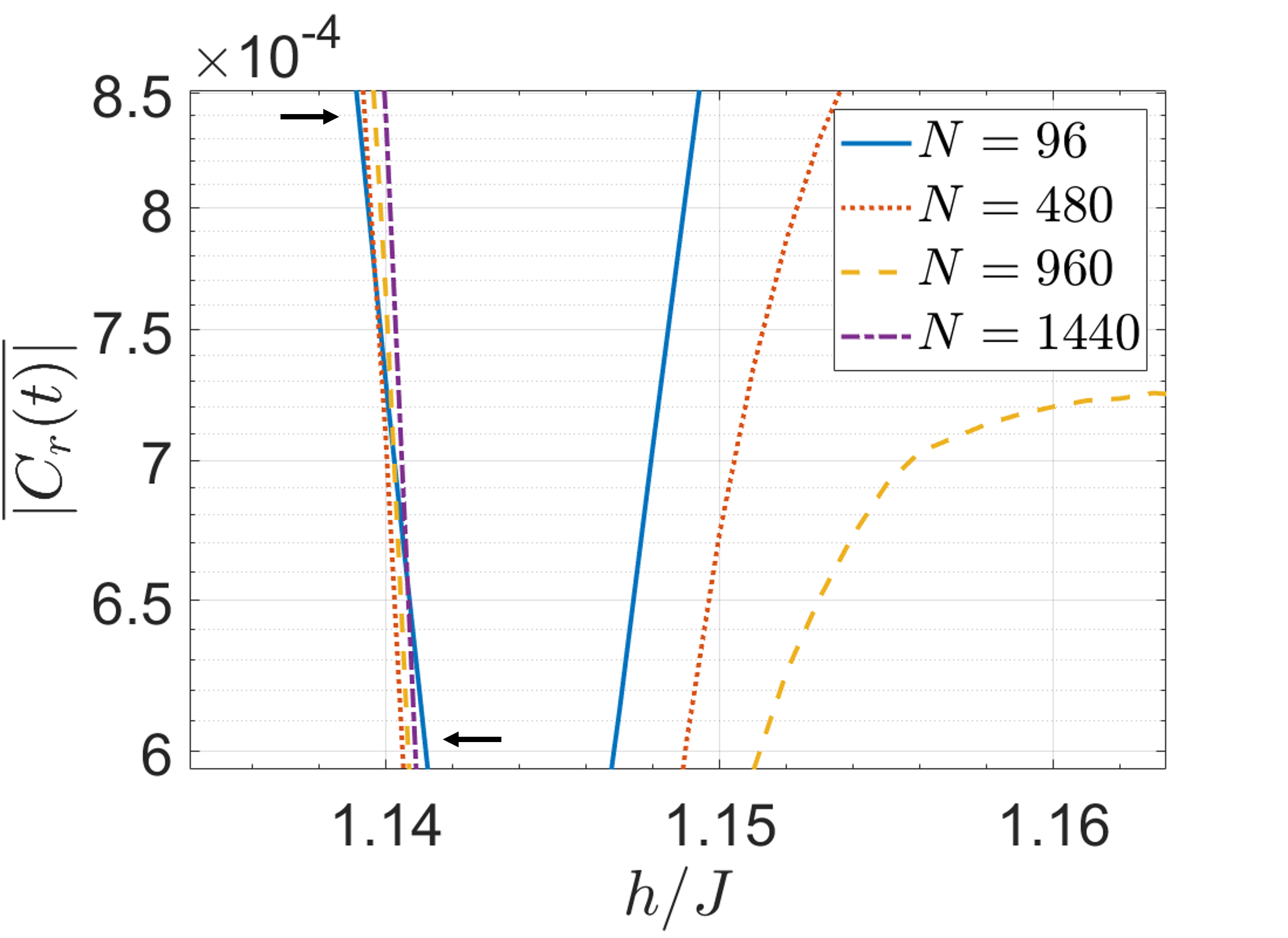}}\hfill
\caption{Order profiles calculated with $t$-DMRG for weakly nonintegrable TFIC with $\Delta=0.1J$, (a) for different system sizes $N=72,96,120$ at site $r=6$ when a (infrared) temporal cutoff of $t_l = N/2$ is applied. The order profile is tested against different infrared and ultraviolet temporal cutoffs and shown to be robust (see text). (b) Finite-time scaling analysis at a system size of $N=120$ for different infrared cutoffs, shown in the legend, with a fixed ultraviolet cutoff of $t^*=0$. The upward and downward arrows highlight the $h/J$ points where the order grows or diminishes, respectively, with increasing (a) system size and (b) simulation time. (c) Zoom over the phase diagram based on the qMFT result with arrows demonstrating how the phase diagram of the smallest system size $N=96$ crosses through those of the larger system sizes before the found DCP $h_{dc}\approx 1.1437$ with the qMFT analysis.}
\label{Fig4}
\end{figure*}

\begin{figure}
\subfloat[]{\label{fig5a}\includegraphics[width=0.24\textwidth]{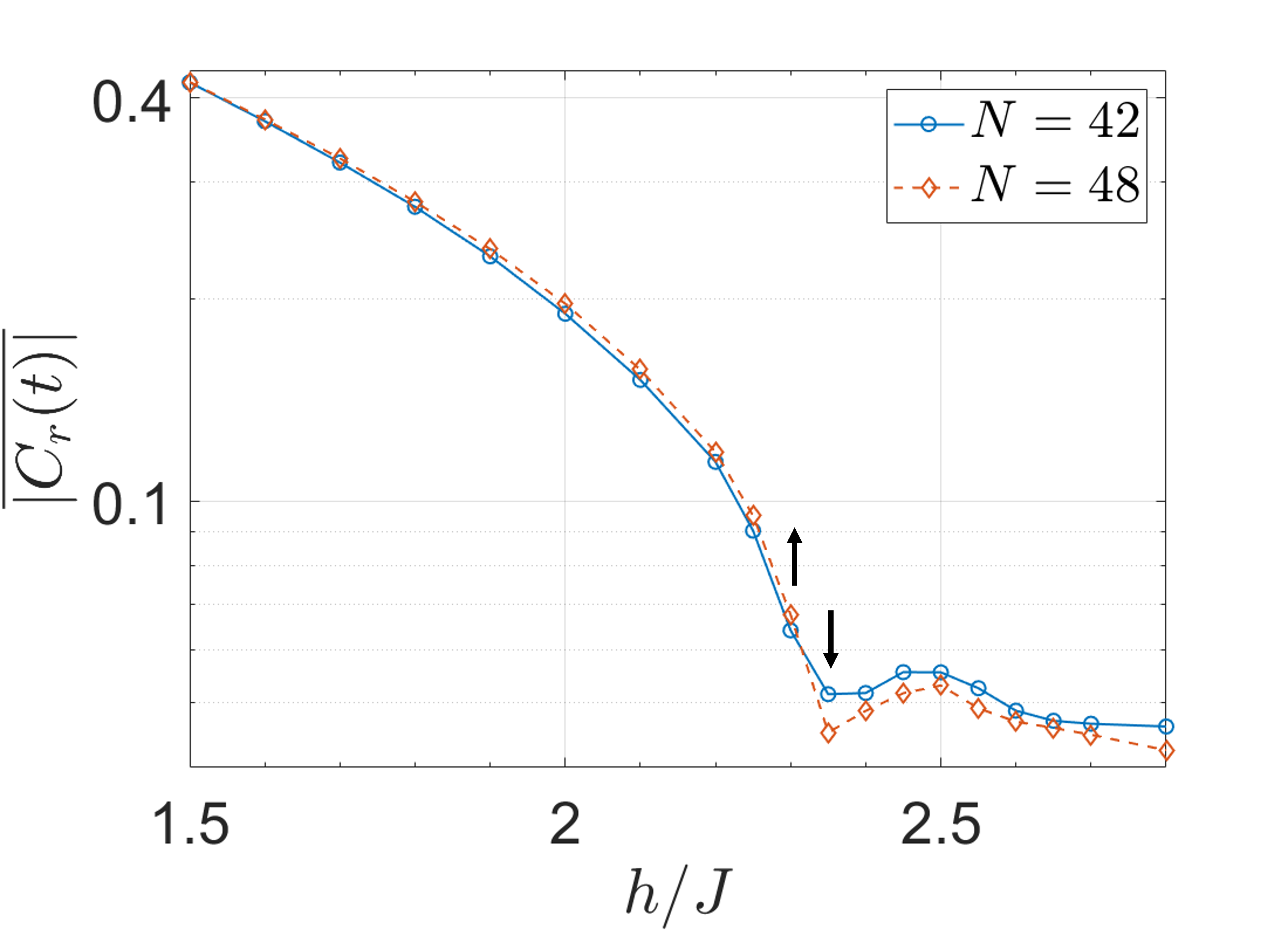}}\hfill
\subfloat[]{\label{fig5b}\includegraphics[width=0.24\textwidth]{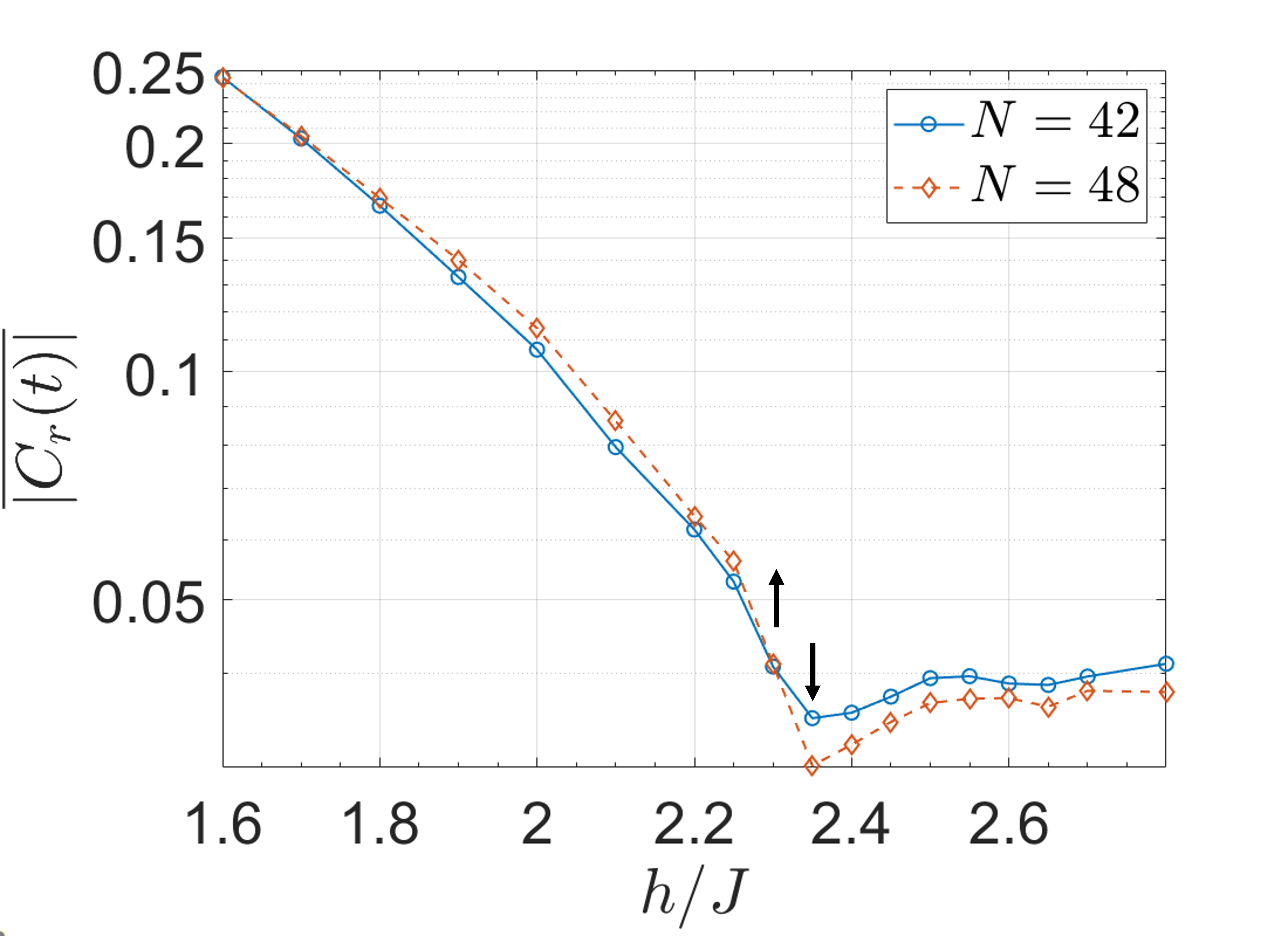}}\hfill
\caption{Order profiles calculated with $t$-DMRG for strongly nonintegrable TFIC with $\Delta=1$ for different system sizes $N=42,48$ at sites (a) $r=3$ and (b) $r=6$ when an infrared temporal cutoff of $t_l = N/3$ and ultraviolet temporal cutoff of $t^*=0$ are applied.}
\label{Fig5}
\end{figure}

\begin{figure*}
\subfloat[]{\label{fig6a}\includegraphics[width=0.33\textwidth]{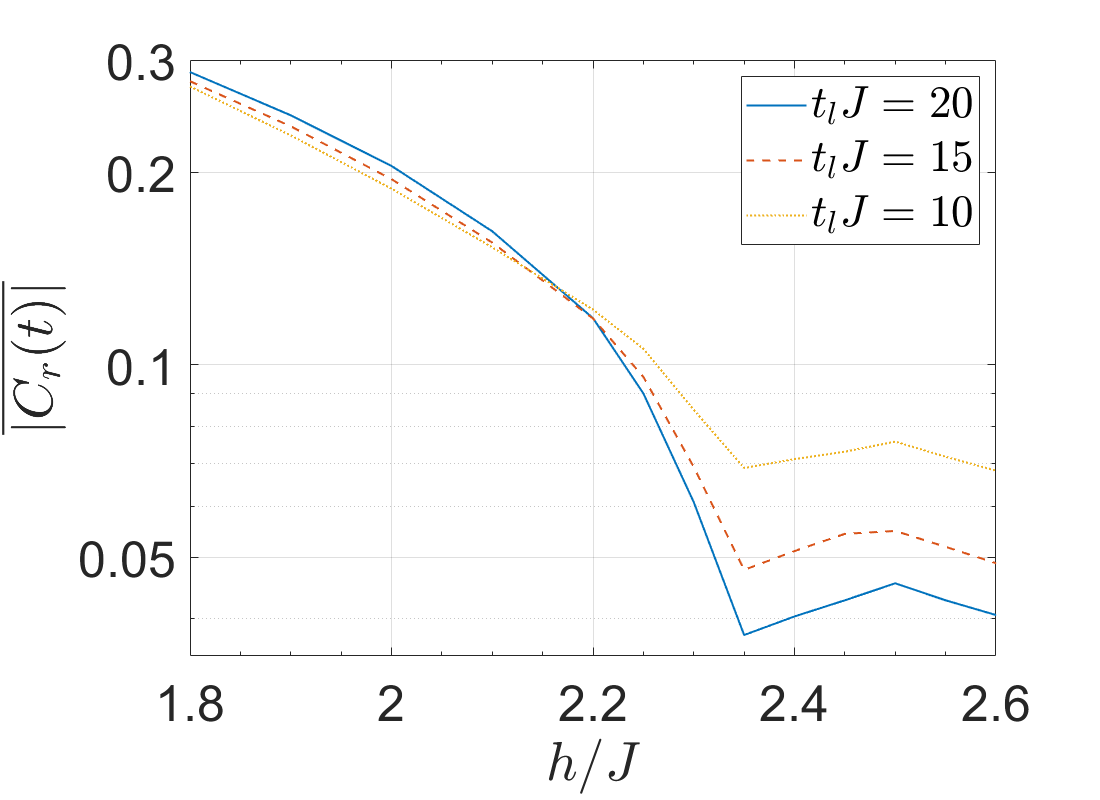}}\hfill
\subfloat[]{\label{fig6b}\includegraphics[width=0.33\textwidth]{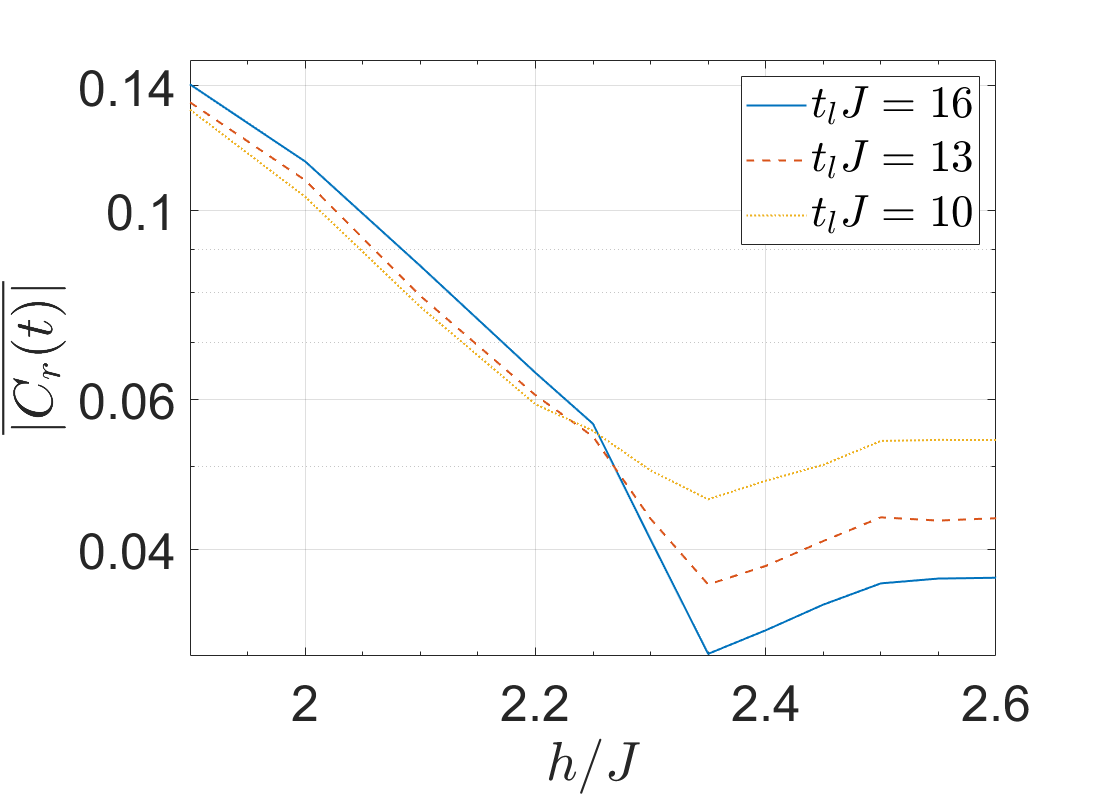}}\hfill
\subfloat[]{\label{fig6c}\includegraphics[width=0.33\textwidth]{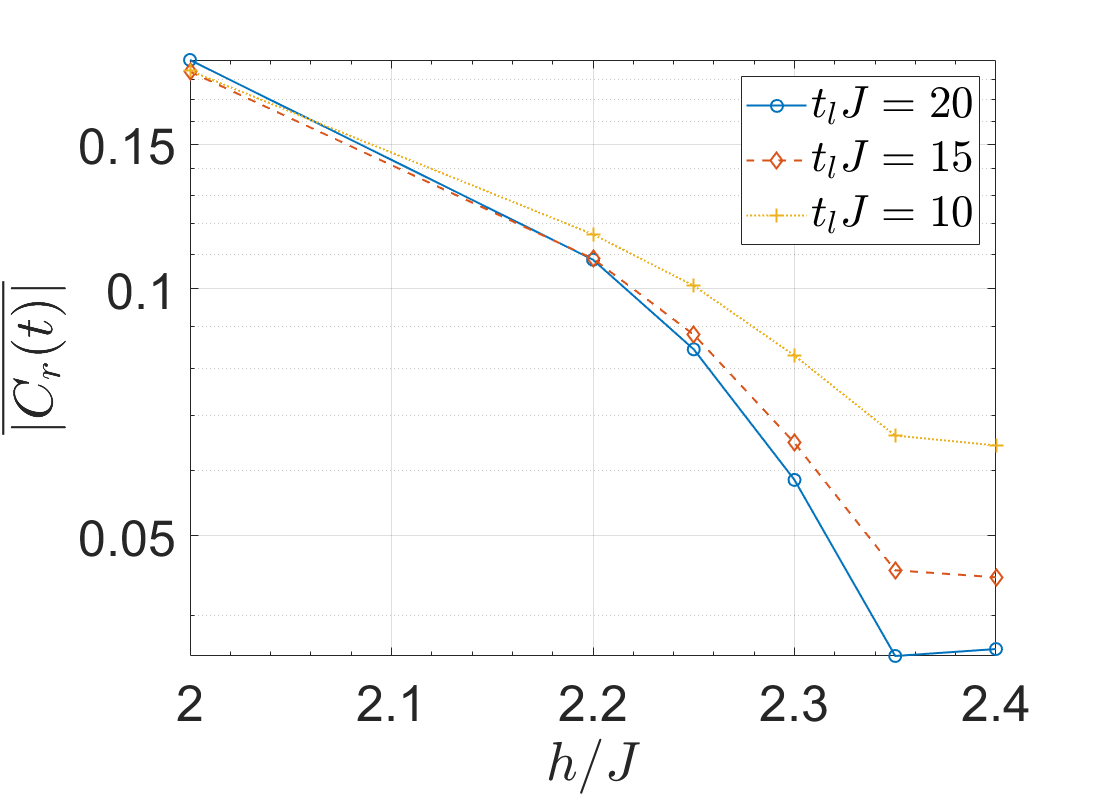}}\hfill
\caption{Order profiles calculated with $t$-DMRG for strongly nonintegrable TFIC with $\Delta=1$ based on finite-time analysis at system size $N=48$ for (a) site $r=3$ and (b) site $r=6$. The infrared cutoffs are shown in the legends and the ultraviolet cutoff is $t^*=0$. (c) Same parameters as (a), however the maximum bond dimension is set to $\chi_m=200$, instead of $\chi_m=100$.}
\label{Fig6}
\end{figure*}

We utilize the ITensor environment \cite{ITensor} to construct matrix product states (MPS) and Trotter decomposition for the time evolution of the MPS. We set a maximum bond dimension $\chi_m$ for the resulting compressed MPS, and set the initial truncation error cutoff for the compression of the MPS as $\epsilon \approx 10^{-8}$. The truncation error cutoff is adaptive: As the maximum bond dimension is reached for the resulting MPS, the error cutoff increases systematically up until a hard error threshold of $\epsilon \approx 10^{-5}$ to be able to access longer times. Setting a maximum allowed bond dimension thus introduces an error which grows with time. Consequently, we are confined to early times for which the above interval of the error thresholds is satisfied.

We study two nonintegrable models with $t$-DMRG: (i) Weakly interacting near-integrable model, $\Delta=0.1$, whose qMFT results are presented in the previous subsection and compared to the results obtained from the exact numerics in this subsection. (ii) A strongly nonintegrable model, $\Delta=1$ where qMFT is inadequate, and we therefore employ $t$-DMRG only. The latter is numerically exact, but within a given fidelity threshold the accessible evolution times are limited and naturally shorter than those achievable for the (near-)integrable model. 

Figs.~\ref{fig3a} and~\ref{fig3b} show the finite-size and finite-time scaling analysis, respectively for the weakly interacting TFIC at $r=3$. In the rest of this subsection, we use infrared cutoffs either parametric in the system size $t_l(N)$ (in finite-size analysis) or set to a fixed time $t_l$ (in finite-time analysis).

In Fig.~\ref{fig3a}, we observe a dynamical order profile and a crossing point for system sizes between $N=48$ and $N=96$. The crossing point is found to be in the interval of $h_{dc}/J\in (1.05,1.1)$ where the dynamical order builds up for $h/J \leq 1.05$ as the system size increases and it vanishes for $h/J \geq 1.1$. We use $t^*=10$ and $t_l=N/2$ as ultraviolet and infrared cutoffs in these figures, however the phase diagram is robust to changes in the temporal cutoffs, as we tested with values in the intervals $t^*\in [8,12]$ and $t_l\in [N/3,N/2]$. In Fig.~\ref{fig3b}, we employ a form of finite-time scaling analysis \cite{PhysRevLett.123.115701} where we see a crossing point between order profiles with different infrared cutoffs at a fixed ultraviolet cutoff $t^*=0$. The crossing point resides in the interval of $h_{dc}/J\in (1.05,1.1)$ agreeing with the finite-size analysis. For $h/J \leq 1.05$ and $h/J \geq 1.1$, the dynamical order grows or diminishes with longer simulation times, respectively. Given that the system sizes are constrained in $t$-DMRG compared to qMFT analysis, we observe a smaller DCP than what we have found with qMFT analysis. We repeat the same calculation for $r=6$, however with slightly larger system sizes. Figs.~\ref{fig4a} and~\ref{fig4b} demonstrate the finite-size and finite-time scaling analysis, respectively for this parameter set. In both cases, we find the crossing interval to be $h_{dc}/J\in (1.1,1.15)$. Finally, we zoom on the nonequilibrium phase diagram of the qMFT analysis in Fig.~\ref{fig4c} to demonstrate that the crossing is present in this method too, however for small system sizes, e.g.,~$N=96$ shown with black arrows around $h/J \approx 1.14$. On the contrary, all system sizes $N \geq 480$ collapse on the qMFT-predicted DCP suggesting that these finite but large system sizes effectively simulate the thermodynamic limit. Therefore, we conclude that the $t$-DMRG method of the simulated system sizes $N \leq 120$ supports the presence of a DCP, verifying the results of the qMFT method presented previously.

Next, we consider the strongly nonintegrable TFIC with $\Delta=1$, whose QCP lies at $h_c\approx2.46$ \cite{2020arXiv200412287D}. Figure~\ref{fig5a} and ~\ref{fig5b} show the local order profiles for $r=3$ and $r=6$, respectively. As the system becomes strongly interacting, the maximum system size and the evolution times become more limited. In this set of calculations, we study $N=42$ and $N=48$ with maximum evolution times of $t \approx 20$. Hence, we apply an infrared cutoff of $t_l=N/3$ and set an ultraviolet (UV) cutoff $t^*=0$ to increase the temporal range of data to average over. Both figures exhibit a crossing point similar to that of the discussion in the previous paragraph, in the interval of $h_{dc}/J \in (2.3,2.35) < h_c$ where for $h/J < 2.3$ the dynamical order increase with increasing system size, and for $h/J> 2.35$ decreases. Therefore, we conclude that (i) the crossing point seems to be independent of the measurement site, and (ii) the presence of a crossing point in $t$-DMRG data hints at the presence of a DCP. We note that more data with larger system sizes and longer simulations times are required to test the robustness of these results, determine the exact location of the DCP, and further look for a critically prethermal regime and dynamical scaling in this strongly nonintegrable model. Nevertheless, single-site observables close to an edge seem to be a probe of criticality generally in short-range models, not limited to noninteracting or weakly-interacting systems.

Finally, we apply the finite-time scaling analysis on the strongly nonintegrable TFIC with $\Delta=1$ shown in Figs.~\ref{fig6a} and~\ref{fig6b} for $r=3$ and $r=6$, respectively. Although there is a well-defined finite-time crossing point for this set of parameters, the crossing suggests $h_{dc}/J \approx 2.2$ for $r=3$ and $h_{dc}/J \in (2.25,2.3)$ for $r=6$ all of which is less than the result determined by the finite-size analysis. To check whether the results depend on the maximum bond dimension, we set $\chi_m=200$ for $r=3$, repeat the calculation on the relevant region in Fig.~\ref{fig6c}, and observe that the crossing point does not change. Although more data is necessary for conclusive results, we note that the mismatch between finite-size and finite-time analyses as well as the discrepancy between different sites in finite-time analysis might point to a change in the light cone structure, e.g.,~from linear to power-law, as the transverse field increases and approaches the equilibrium QPT in finite-size systems.

\section{\label{longrange}The quasi-stationary regime in long-range interacting nonintegrable TFIC}

The q.s. temporal regime also emerges in the long-range hard-boundary TFIC with power-law decaying interactions, as was previously noted in Ref.~\cite{Neyenhuise1700672} in the context of prethermalization. The Hamiltonian for the one-dimensional long-range TFIC reads,
\begin{eqnarray}
H = - \sum_{r,r'} J(r,r') \sigma_r^z \sigma_{r'}^z + h \sum_r \sigma_r^x, \label{HamiltonianLR}
\end{eqnarray}
where $J(r,r')=J/|r-r'|^{\alpha}$. In the limit where $\alpha = 0$, the model becomes integrable with all-to-all interactions, e.g. LMG model; whereas in the limit of $\alpha \rightarrow \infty$ the model reduces to short-range n.n.~TFIC. For $\alpha \geq 3$, the model belongs to the short-range Ising universality class \cite{PhysRevB.64.184106,PhysRevLett.111.147205}, and it has algebraic light cones for $\alpha > 2$ that approach linear cones as $\alpha \rightarrow \infty$ \cite{PhysRevLett.114.157201}. In this section, we provide numerical evidence on the presence of a q.s. regime in TFIC with various $\alpha$, and determine when the q.s. regime breaks down. Furthermore, we calculate the local order profiles at $\alpha=2.5$, which is a long-range model with algebraic light cones.

All data presented in this section is obtained from the time-dependent DMRG method \cite{RevModPhys.77.259,SCHOLLWOCK201196,Daley_2004,PhysRevE.71.036102,mptoolkit} with Krylov time evolution \cite{krylov}. We find convergence for a time-step of $\Delta t=0.01J$ and a fidelity threshold between $10^{-8}$~--~$10^{-6}$ for our most stringent calculations. Power-law profiles are approximated as a sum of five exponentials fitted over the length of the chain, as is optimal for a matrix product operator formulation \cite{Crosswhite2008}.

\begin{figure}
\centering
%\subfloat[]{\label{figS1a}\includegraphics[width=0.24\textwidth]{Fig6a.png}}\hfill 
\subfloat[]{\label{figS1b}\includegraphics[width=0.24\textwidth]{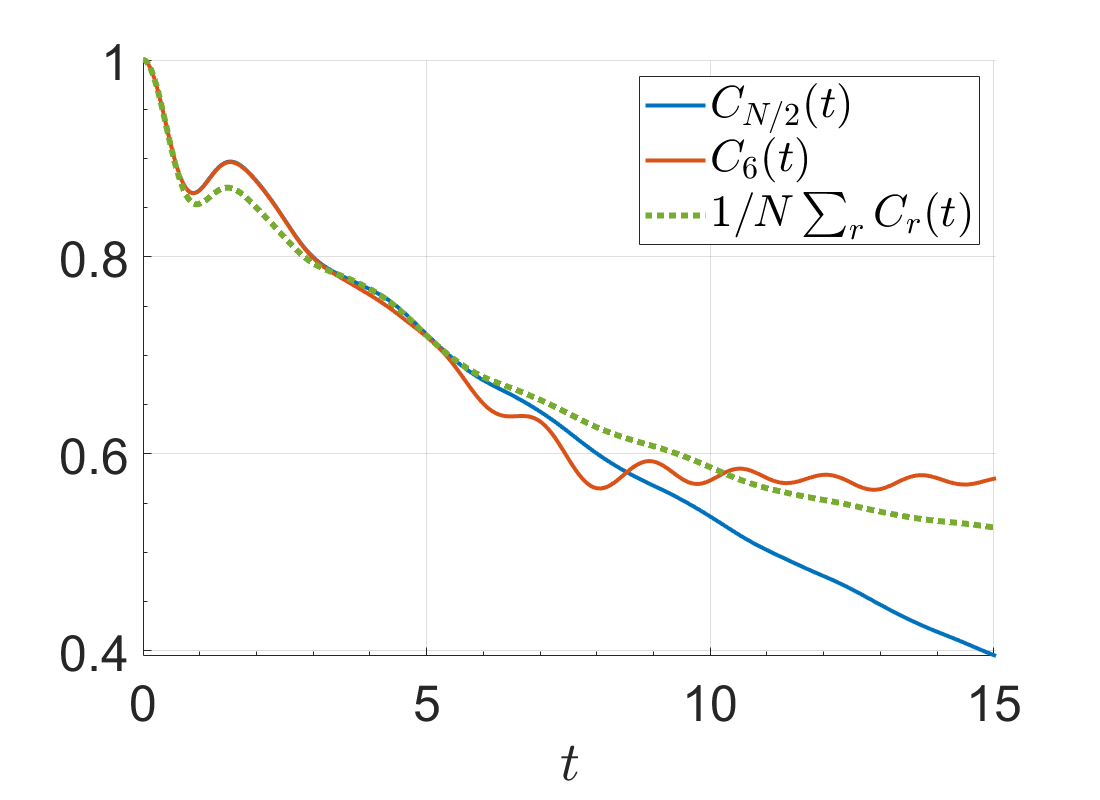}}\hfill 
\subfloat[]{\label{figS1c}\includegraphics[width=0.24\textwidth]{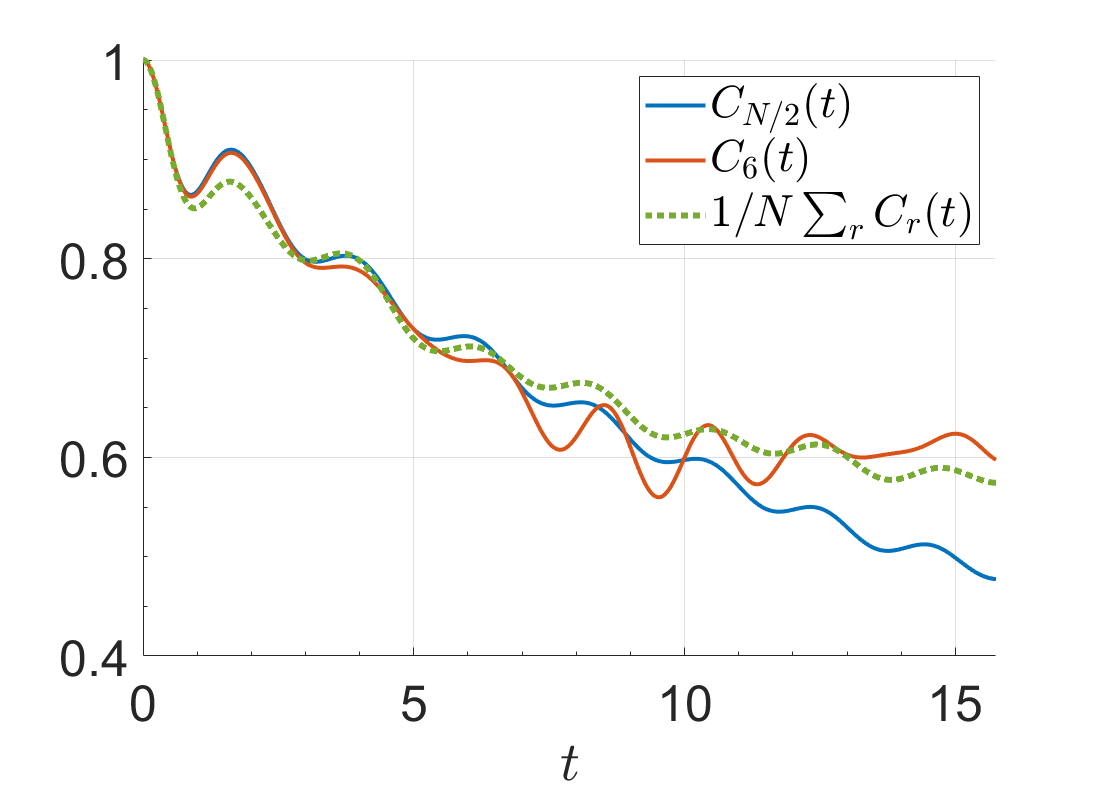}} \hfill 
\subfloat[]{\label{figS1e}\includegraphics[width=0.24\textwidth]{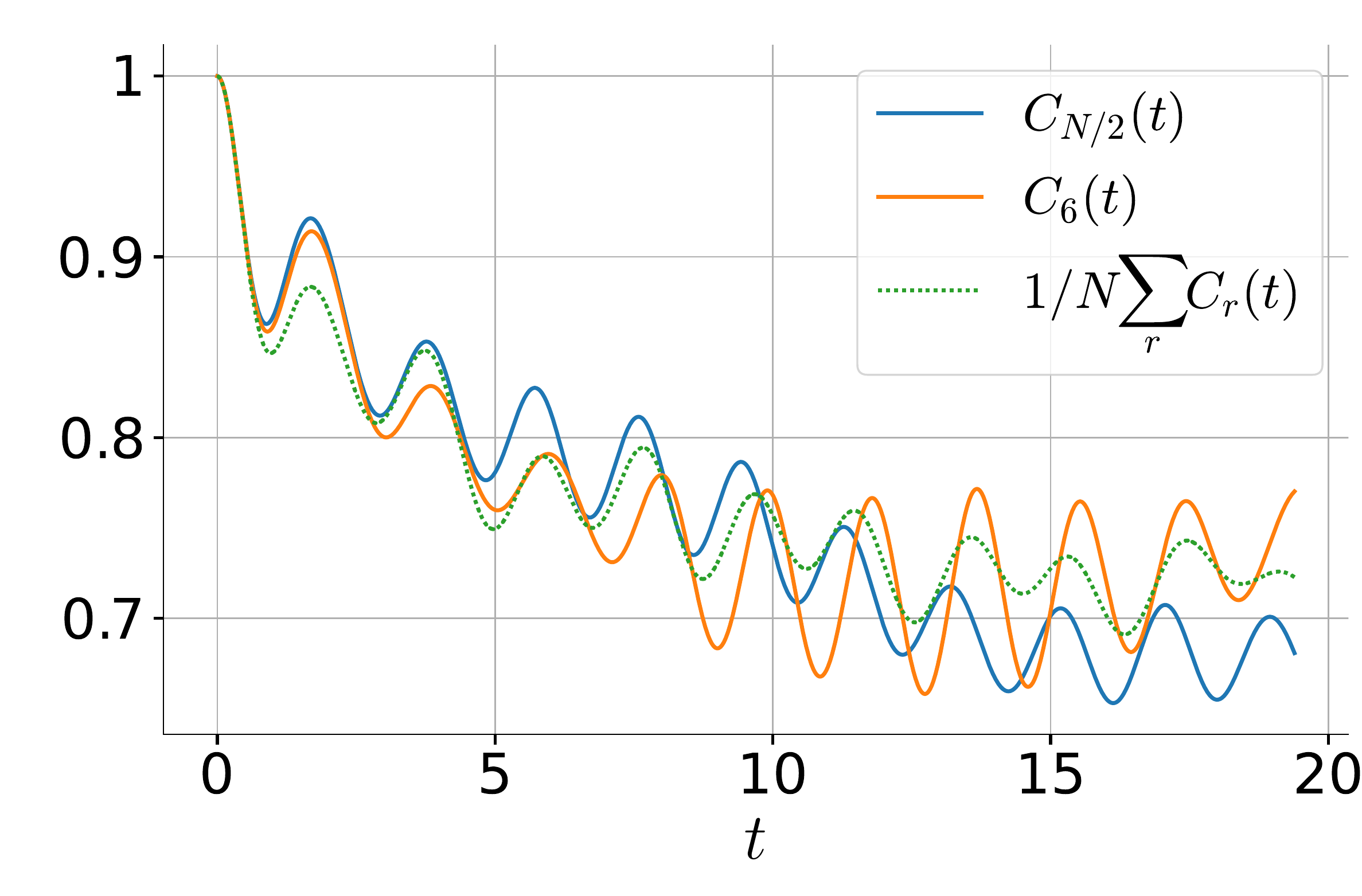}} \hfill 
\subfloat[]{\label{figS1d}\includegraphics[width=0.24\textwidth]{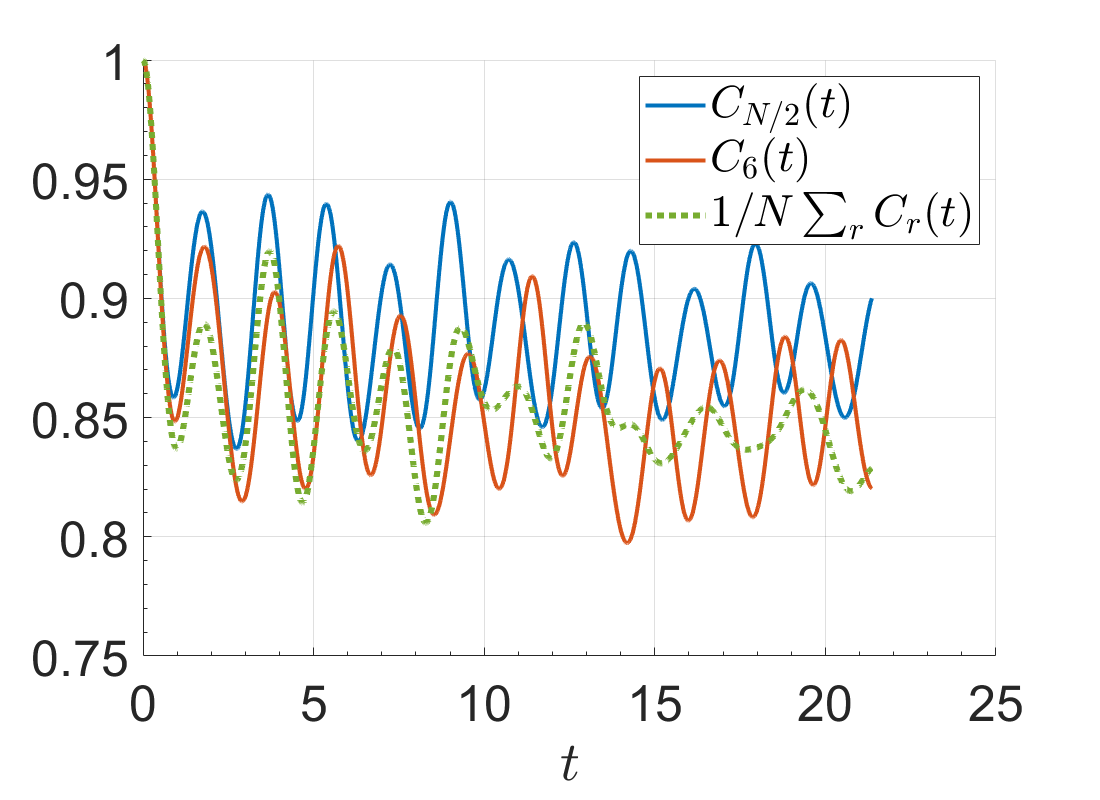}} \hfill
\caption{Single-site magnetization for $r=N/2$ (spin in the middle of the chain), $r=6$ (spin close to the boundary) and the total magnetization, when (a) $\alpha=4$, (b) $\alpha=3$, (c) $\alpha=2.5$ and (d) $\alpha=2$ at transverse field $h=0.5\tilde{h}_c$.}
\label{FigS1B}
\end{figure}

Fig.~\ref{figS1b} shows the edge magnetization of single-site observables near the edge, and in the middle of the chain, as well as the total magnetization evolving under Hamiltonian Eq.~\eqref{HamiltonianLR} with $\alpha = 4$ and $h=0.5\tilde{h}_c$ quenched from a polarized state. The unit notation $\tilde{h}_c$ will be explained shortly. We observe that the single-site observable close to the boundary (red) develops a q.s. regime whereas the observable in the middle of the chain (blue) exhibits decay. This behavior, consistently, is similar to what we have observed in the n.n.~TFIC model, c.f.~Sec.~\ref{NNNmodel}. Although the decay of total magnetization seems to be slowing down, the data is not conclusive to determine its long-time behavior. We notice the oscillatory q.s. regime for $\alpha \lesssim 4$ in Fig.~\ref{FigS1B} for the single-site observable near the edge, which was also observed for nonintegrable n.n.n.~TFIC \cite{2020arXiv200412287D}. Although the oscillations grow as we decrease $\alpha$, one could still observe the onset of a q.s. regime in the observable near the edge of the chain for $\alpha = 3$, compared to the observable in the middle of the chain (Fig.~\ref{figS1c}). The difference in the single-site magnetization of spins near the edge and in the middle of the chain decreases as we keep decreasing $\alpha$. Fig.~\ref{figS1e}, for $\alpha=2.5$, demonstrates for the first time an onset of a q.s. regime for a spin in the middle of the chain. However, the onset of the q.s. regime is delayed in the middle of the chain compared to near the edge, and this observation points to the locality of the underlying Hamiltonian that is still preserved to an extent. In contrast, there is no boundary effect observed at $\alpha=2$ (Fig.~\ref{figS1d}), where we do not see any difference in the general trend of the nonequilibrium responses. This is physically intuitive, because the model possesses algebraic light cones for $\alpha > 2$ instead of a logarithmic cone, pointing to the importance of the locality in the formation of a q.s. regime. As discussed previously, the origin of the q.s. regime could be traced back to the asymmetric motion of quasi-particles in opposite directions, and the reflection from the nearest boundary. When the model is no longer integrable, even though quasi-particles do not exist one could still refer to correlation speeds measured through linear or sublinear light cones. In conclusion, the presence of a q.s. regime does not depend on the integrability, but rather depends on the locality of the underlying Hamiltonian. In this sense, this boundary induced temporal regime acts as a signature of linear, nearly-linear or sublinear light cones.

\begin{figure}
\subfloat[]{\label{FigS2}\includegraphics[width=0.24\textwidth]{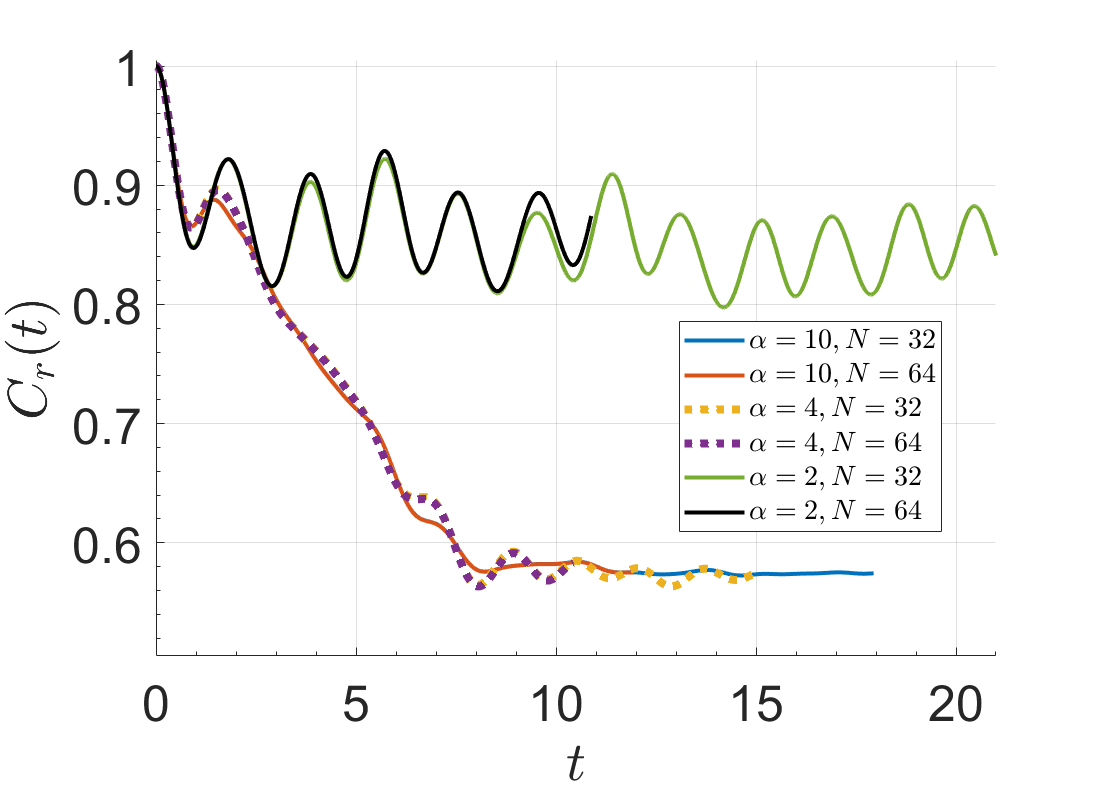}}\hfill 
\subfloat[]{\label{Fig10}\includegraphics[width=0.24\textwidth]{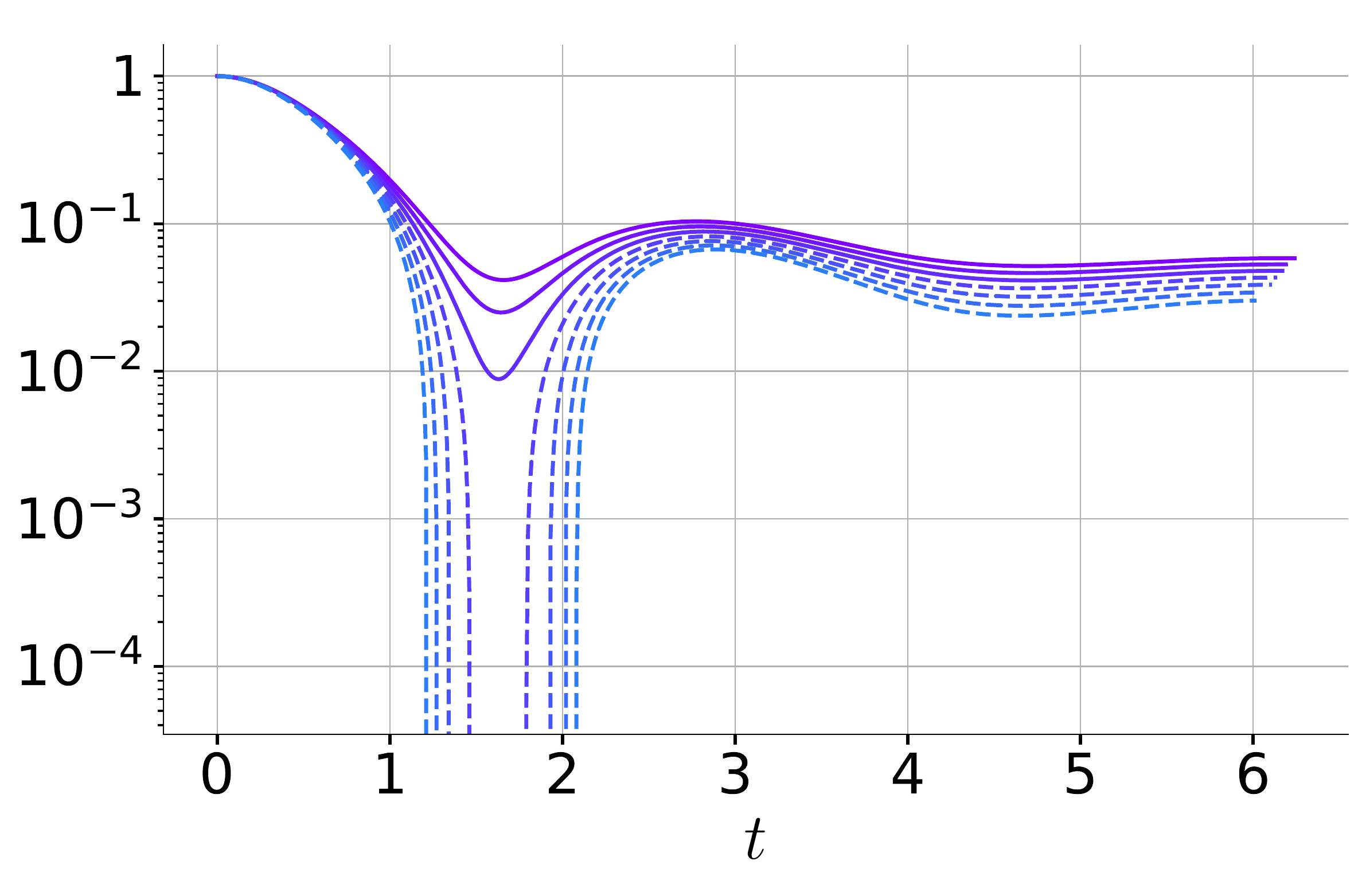}}\hfill 
\caption{(a) Single-site magnetization at $r=6$ for different $\alpha=2,4,10$ and different system sizes $N=32,64$ at transverse field $h=0.5\tilde{h}_c$. (b) Total magnetization $\sum_{1}^N \sigma_r^z / N$, quenched from a polarized state to $h/\tilde{h}_c=1.04, 1.05, 1.06, 1.07, 1.08, 1.09, 1.1$, from top to bottom. The dashed line-style indicates curves which cross the time axis. The last magnetization that does not cross the time axis is $h/\tilde{h}_c=1.06$, meaning that the DCP should lie in the interval of $(1.06,1.07)\tilde{h}_c$.}
\end{figure}

To demonstrate that the q.s. regime is not a finite-size effect, we show in Fig.~\ref{FigS2} the single-site magnetizations of a spin close to the boundary at $r=6$ for different system sizes and different $\alpha$. For a given $\alpha$, one can determine the time at which finite-size effects kick in by observing when the data for different system sizes $N=32,64$ no longer overlap. Note that for $\alpha=10$ and $\alpha=4$, the q.s. regime develops before the finite-size effects appear.

Finally, we set $\alpha=2.5$, which yields a long-range model \cite{PhysRevB.64.184106} with algebraic light cones \cite{PhysRevLett.114.157201}, and study the resulting local order profiles due to the q.s. regime. In all of our $t$-DMRG calculations on long-range TFIC, we Kac-normalize the interaction term by dividing it by $\sum_{1}^L 1/r^\alpha$, and hence approximate the QCP as $\tilde{h}_c=J$, which in the thermodynamic limit is a good approximation when $\alpha>1$ \cite{Jaschke2017}. This is performed because QCP is actually not well-defined for a finite-size system, and finite-size fluctuations are expected to shift the ``critical point'' to smaller values. Therefore, we first determine the DCP based on the total magnetization \cite{halimeh2020local}. Fig.~\ref{Fig10} shows the total magnetization for various transverse field strengths. One sees that the total magnetization crosses the time axis within the interval $h_{dc}/\tilde{h}_c\in (1.06,1.07)$, which pins down the DCP of the model quenched from a polarized state \cite{halimeh2020local}. The observation that $h_{dc} > \tilde{h}_c$ points to the fact that the QCP is initially underestimated.
\begin{figure}
\centering
%\subfloat[]{\label{fig7a}\includegraphics[width=0.33\textwidth]{Fig7a.pdf}} \hfill 
\subfloat[]{\label{fig7b}\includegraphics[width=0.24\textwidth]{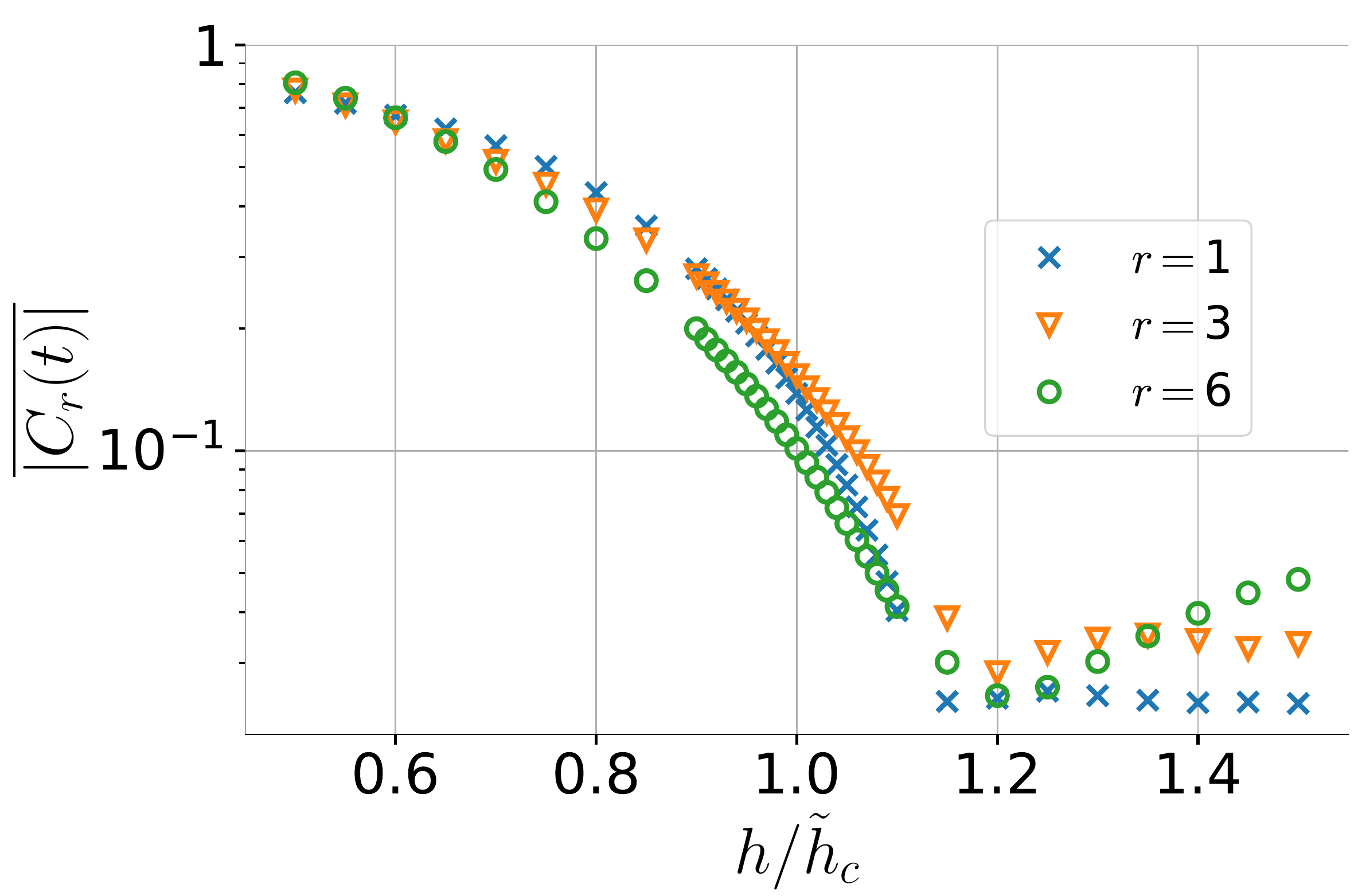}} \hfill 
\subfloat[]{\label{fig7c}\includegraphics[width=0.24\textwidth]{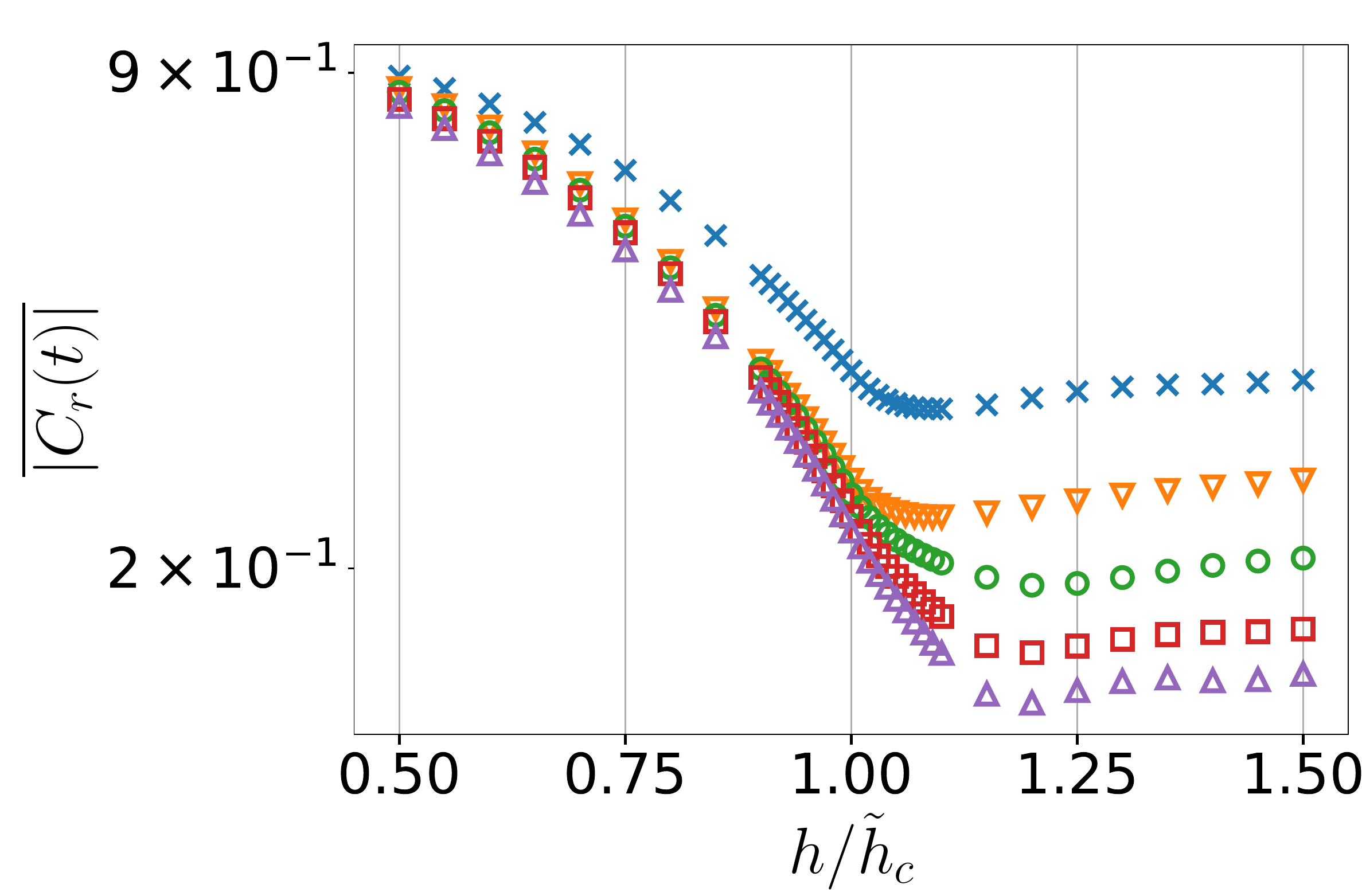}} \hfill
\caption{(a) Single-site order profiles calculated with $t$-DMRG for the long-range TFIC, with $\alpha=2.5$ and system size $N=32$, at different sites $r=1,3,6$ when a UV temporal cutoff of $t^*=2$ and an infrared temporal cutoff of $t_l=6$ are applied. (b) Finite-time analysis curves for $r=3$ are calculated with $t^*=0$ and $t_l=2,3,4,5,6$, from top to bottom. Since there is no crossing point, finite-time analysis fails.}
\label{FigS1}
\end{figure}

Since the accessible simulation times are significantly limited, we set the infrared and ultraviolet cutoffs as $t_l =6$ and $t^*=2$ (Fig.~\ref{fig7b}). Remarkably, we find that the single-site order profiles at different sites dip at the same $h/\tilde{h}_c$ value, suggesting a crossover in the interval of $h_{cr}/\tilde{h}_c\in (1.15,1.2)$. We note that unlike in previous sections and Ref.~\cite{letter} where we observe $\overline{|C_{1}(t)|}>\overline{|C_{3}(t)|}>\overline{|C_{6}(t)|}> ...$ in the ordered phase, here for the long-range model we do not observe such an ordering. This is because of the presence of algebraic, instead of linear, light cones for $\alpha=2.5$, which wash away the differences in the single-site magnetization of nearby sites, and hence make the system less locally connected. Strengthening this argument, we also find that the finite-time analysis fails for this model. Fig.~\ref{fig7c} shows nonequilibrium responses at $r=3$ between a UV cutoff $t^*=0$ and various infrared cutoffs $t_l=2,3,4,5,6$, from top to bottom. The curves do not intersect at any point, unlike what we observed for n.n.n.~TFIC models. 

\begin{figure*}
\centering
\subfloat[]{\label{fig12a}\includegraphics[width=0.33\textwidth]{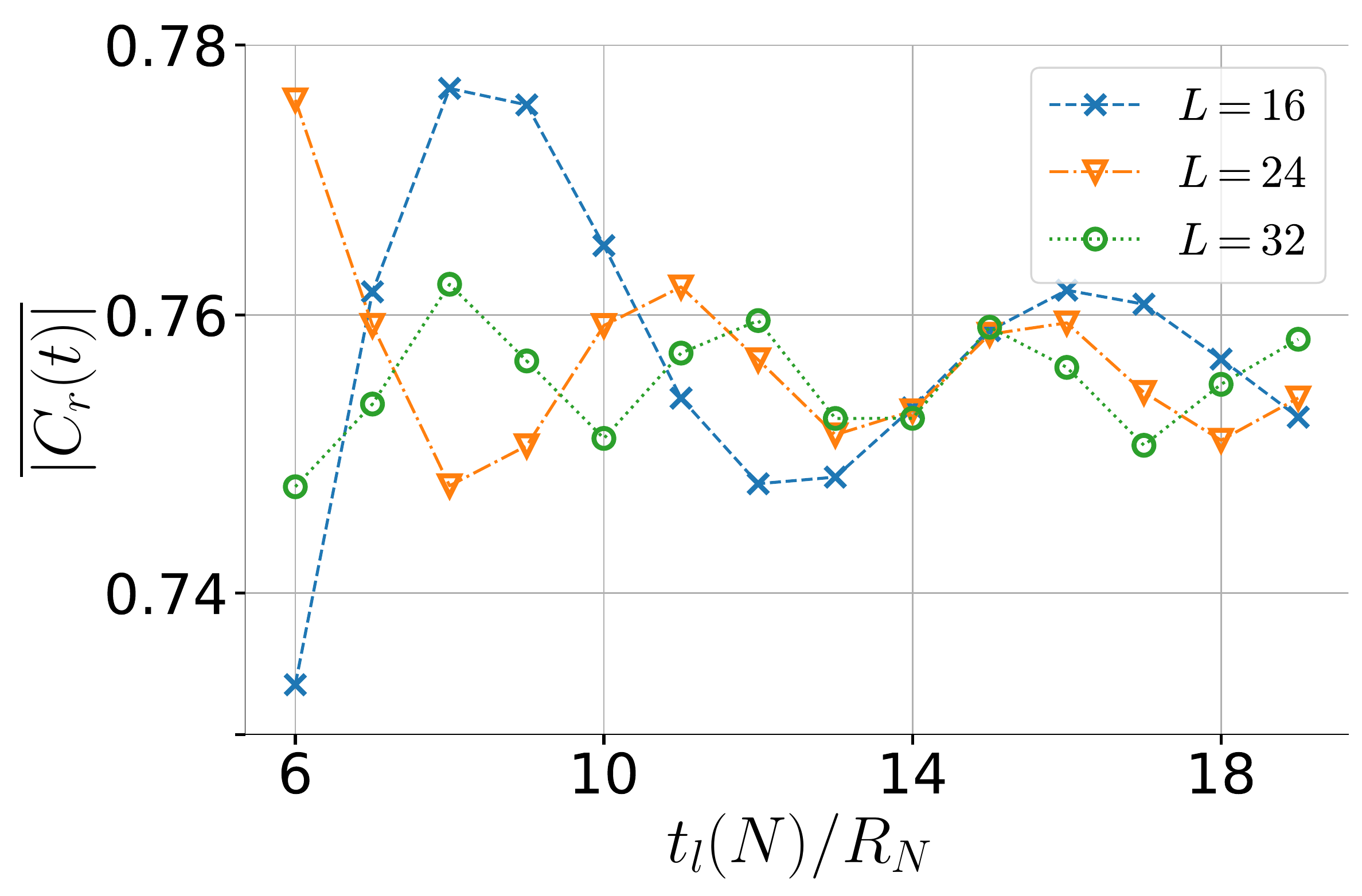}} \hfill
\subfloat[]{\label{fig12b}\includegraphics[width=0.33\textwidth]{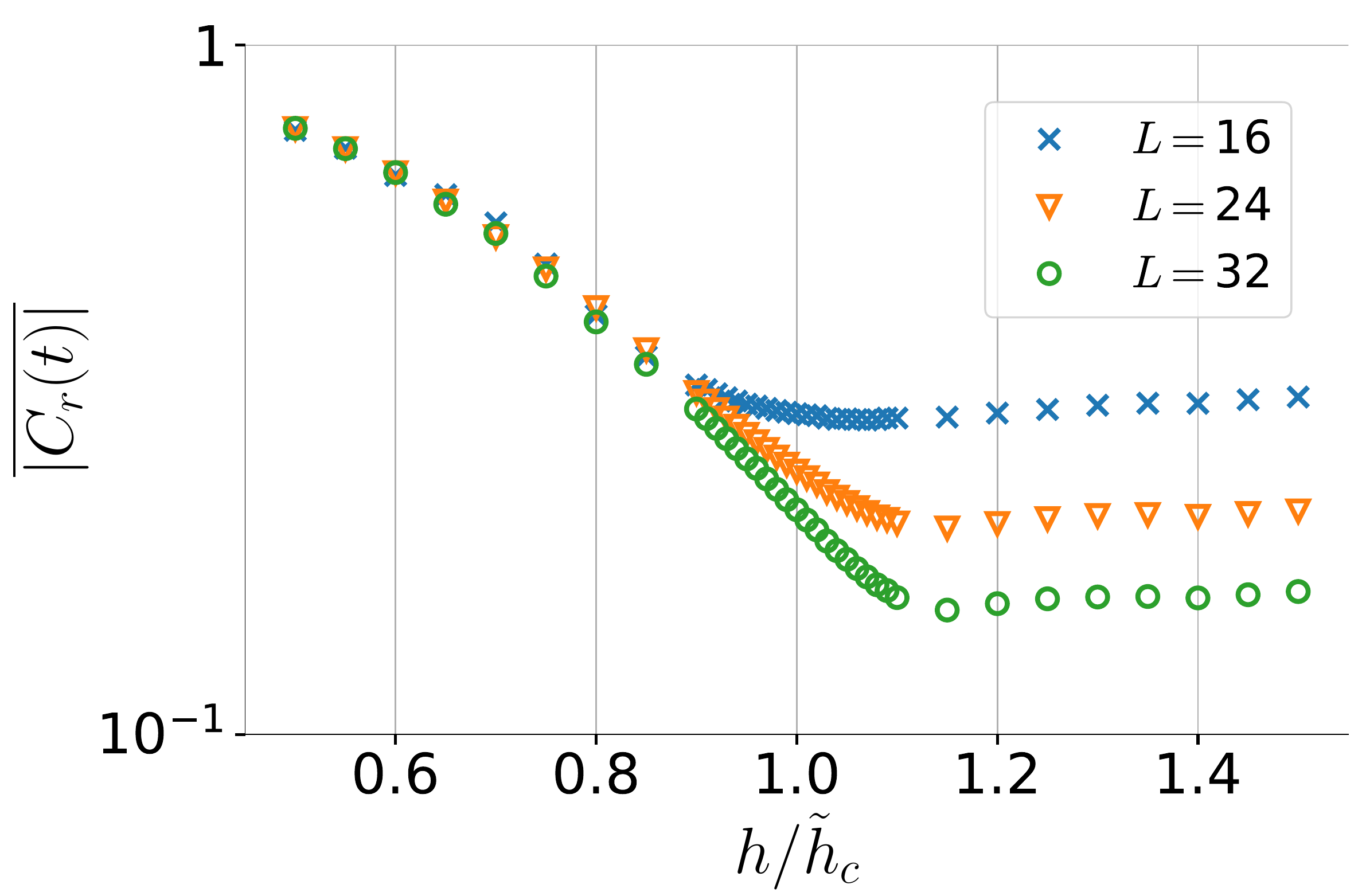}} \hfill
\subfloat[]{\label{fig12c}\includegraphics[width=0.33\textwidth]{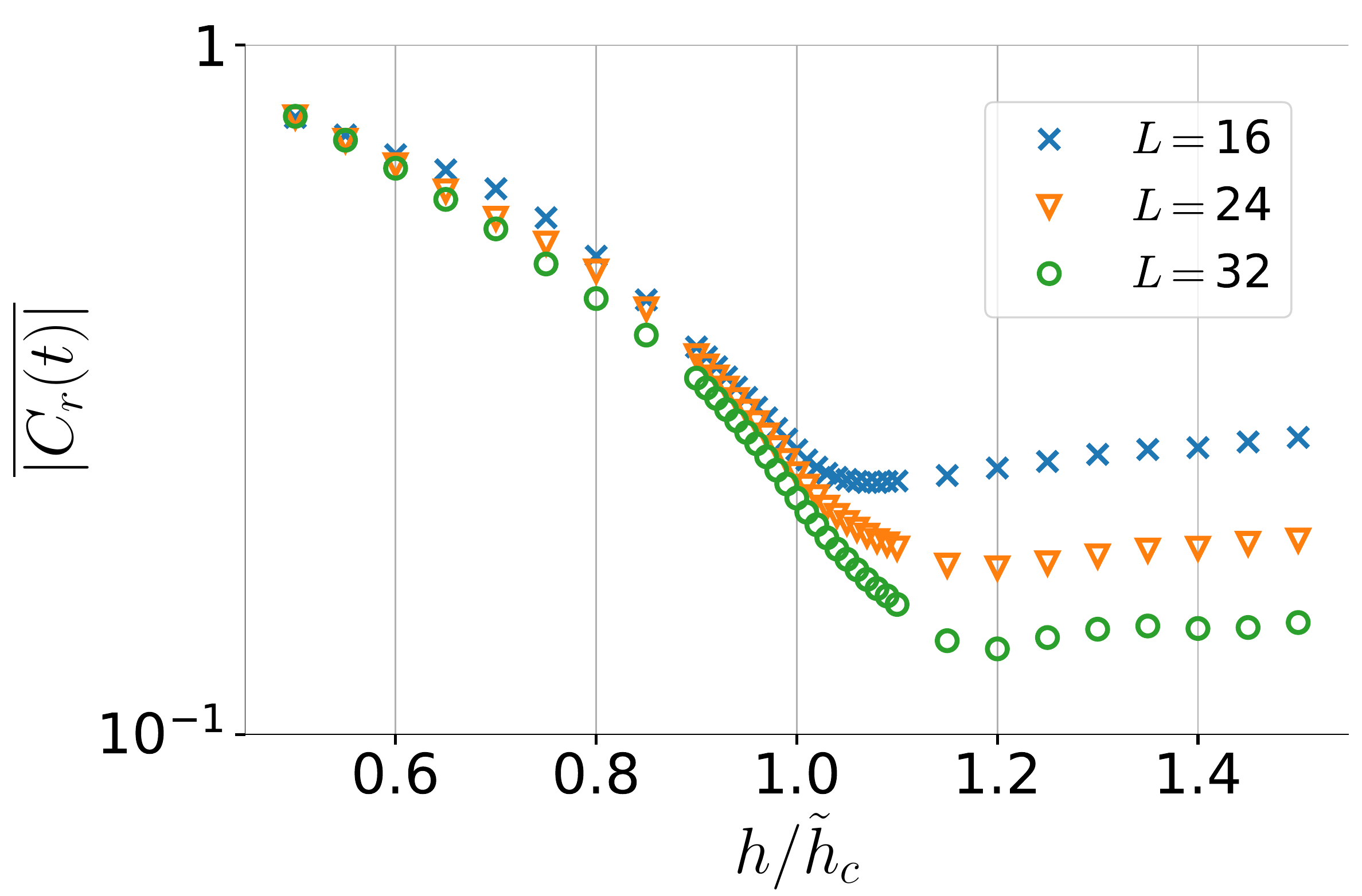}} \hfill
\caption{Order profiles calculated with $t$-DMRG for the long-range TFIC at different system sizes $N=16,24,32$ with $\alpha=2.5$ (a) at site $r=1$ and $h/\tilde{h}_c=0.5$ with respect to different infrared cutoffs parametrized as $t_l(N)/R_N$; (b) at $r=1$ with respect to $h/\tilde{h}_c$ with an infrared cutoff of $t_l(N,h)$; (c) at $r=3$ with respect to $h/\tilde{h}_c$ with an infrared cutoff of $t_l(N,h)$. All subfigures use a UV cutoff of $t^*=0$.}
\label{Fig12}
\end{figure*}

The crossover region might be probing the actual QCP $h_c$, as it appears above the DCP determined with total magnetization at $h_{dc}/\tilde{h}_c\in (1.06,1.07)$, and this is expected in long-range models for quenches from the ferromagnetic phase \cite{PhysRevB.96.134427,PhysRevB.96.104436,PhysRevB.97.174401}. To test this idea, we apply finite-size analysis with $N=16$ and $N=24$. This analysis accentuates the shortness of accessible times as the system size increases: A system size-dependent infrared cutoff $t_l(N)=6 R_N$, where $R_N=N/32$, corresponds to $t_l=4.5$ and $t_l=3$ for $N=24$ and $N=16$, respectively, throughout the entire set of transverse-field values. An additional challenge in the long-range models is the oscillations with large amplitudes \cite{PhysRevB.95.024302} as seen in Fig.~\ref{figS1e}. This feature particularly becomes significant as we approach the boundary. Fig.~\ref{fig12a} demonstrates how the order profile at $h/\tilde{h}_c=0.5$ would change with different infrared cutoffs $t_l(N)/R_N$ and system sizes. The order profiles are also affected by the oscillations, in particular for early infrared cutoffs, e.g.,~$t_l(N)=6R_N$. As the averaging interval increases with increasing $t_l$, we notice that the oscillations are damped and the dependence on the cutoff becomes less pronounced. Again due to oscillations, comparison between the order profiles of different system sizes becomes less reliable, as can also be seen in Fig.~\ref{fig12a} where the ordering between different system sizes keeps changing as we change $t_l(N)/R_N$. While the trend of the data suggests that the oscillations continue to be damped and the ordering possibly converges as we increase the infrared cutoff, we cannot confirm this conclusion due to limited simulation times. Therefore, both to alleviate the effects of oscillations and to utilize the accessible times as much as possible, we adopt an infrared cutoff that is a function of both system size and the transverse field $t_l(N,h)=R_N t_{max}(N=32, h)$ where $t_{max}(N=32, h)$ is the longest time accessible in our $t$-DMRG calculations for the largest considered system size $N=32$ at each $h$. While this adaptive approach helps to decrease the effect of oscillations on the order profiles, in particular deep in the ferromagnetic phase, it does not completely eliminate the effect, which is possible only by increasing the simulation times. Figs.~\ref{fig12b} and~\ref{fig12c} show the order profiles of different sizes at $r=1$ and $r=3$, respectively. We observe that the order profile of $N=32$ tends to decrease at much smaller $h$ than where the crossover resides. This is exactly because of the limited simulation times and hence the application of very early infrared cutoffs as we increase the transverse field. It is also worth noting that finite-size scaling for models with power-law interaction profiles faces the additional challenge that the profile tails will look different for different system sizes, and this becomes particularly nontrivial near criticality. As a result of the above, whether the observed crossover profile is actually a transition cannot be answered conclusively.

\section{\label{Conc}Conclusions and Outlook}

We have studied the q.s. regime at sites close to the edge in both integrable and nonintegrable TFICs, and shown the necessity of the zero modes to stabilize this temporal regime. Based on analytical arguments and numerical evidence, we demonstrated the presence of non-analyticity in the dynamical order profiles of integrable TFIC and near-integrable TFIC treated with qMFT. This suggested a dynamical critical point that coincides with the QCP in the integrable TFIC and slightly shifts from the corresponding QCP in the near-integrable model. Therefore, we conclude that the single-site observables are able to extract QCP in short-range interacting models independently of integrability, measurement location and initial state so long as $r\ll N/2$ and $h_i < h_c$.

We performed $t$-DMRG calculations to obtain the order profiles of single-site observables at various sites in open-boundary nonintegrable TFICs. The models that we studied range from locally connected with next nearest neighbor interactions to power-law decaying interactions. Our qMFT and $t-$DMRG results for the location of the DCP in near-integrable model agree, as we found $h_{dc}^{\Delta=0.1}\in (1.1,1.15)$ with $t-$DMRG for system sizes $N \leq 120$. Finite-size analysis for the strongly nonintegrable model determined the DCP to be in the interval of $h_{dc}^{\Delta=1}\in (2.3,2.35)$, independent of the measurement sites, and to be shifted from the QCP at $h_c\approx 2.46$. Whether this shift is related to finite-size effects, or due to interactions as argued in \cite{PhysRevLett.123.115701}, is a question that needs further elaboration in the future. We find that the finite-time analysis with single-site observables matches well with the finite-size analysis in the near-integrable model, whereas there is a mismatch between the two analyses in the strongly nonintegrable model. We argue that this is likely because of a change in the light cone structure as the transverse field increases, which is yet to be explored. We demonstrated the presence of a q.s. regime even in long-range power-law interacting TFIC for $\alpha=3$ and $\alpha=2.5$, where in the former one could still observe a significant difference in the single-site magnetization of the spin in the middle and close to an edge, pointing to nearly-linear light cones hosted in the system. Increasing the range of interactions to $\alpha=2.5$ decreases the differences between sites, emphasizing the long-range nature of the model and expectantly the finite-time analysis fails. The order profiles at $\alpha=2.5$ for different sites dip at the same transverse field $h_{cr}/\tilde{h}_c<1.2$ suggesting a crossover, which is found to be larger than the calculated DCP based on total magnetization at $h_{dc}/\tilde{h}_c=(1.06,1.07)$. Given that the crossover dip appears after $h_{dc}/\tilde{h}_c$, and since the calculations were based on an estimate of QCP $\tilde{h}_c$, it is an interesting direction to check whether the crossover point $h_{cr}$ probes the actual QPT. Our data is significantly limited to early times, especially as we approach the transition boundary, and hence is currently inconclusive to answer this question.

Finally, our setup is experimentally convenient, because (i) open-boundary chains are a more natural setup than their periodic counterparts in most quantum simulators, and (ii) spatially minimal probes are  readily accessible in modern quantum simulators \cite{2009Natur.462...74B}. Most theoretical works have naturally focused on infinite or periodic chains to utilize the translational symmetry \cite{PhysRevB.95.024302,PhysRevLett.123.115701,PhysRevResearch.2.033111}, which removes site-dependency of the local order profiles within the ordered phase. In this sense, our work complements the literature via explicitly demonstrating the potential of spatially minimal measurements in open-boundary chains and hence exploiting the boundary effects in probing criticality.

\begin{acknowledgments}
The authors thank Philipp Hauke for suggestions on the manuscript. C.B.D.~thanks Kai Sun for extensive and fruitful discussions; Norman Yao and L.-M. Duan for helpful suggestions. C.B.D.~was supported by the National Science Foundation under Grant EFRI-1741618 and ITAMP grant at Harvard University. This work is part of and supported by Provincia Autonoma di Trento, the ERC Starting Grant StrEnQTh (project ID 804305), the Google Research Scholar Award ProGauge, and Q@TN — Quantum Science and Technology in Trento.
I.P.M.~acknowledges support from the Australian Research Council (ARC) Discovery Project Grants No.~DP190101515 and DP200103760.
\end{acknowledgments}
\bibliographystyle{apsrev4-1}
%\bibliography{Bibliography} % The references (bibliography) information are stored in the file named "Bibliography.bib"


\pagebreak
\appendix

\begin{widetext}
\section{\label{AppA}Independency of the results from temporal cutoffs}

\begin{figure}
\centering
\subfloat{\label{figS6a}\includegraphics[width=0.33\textwidth]{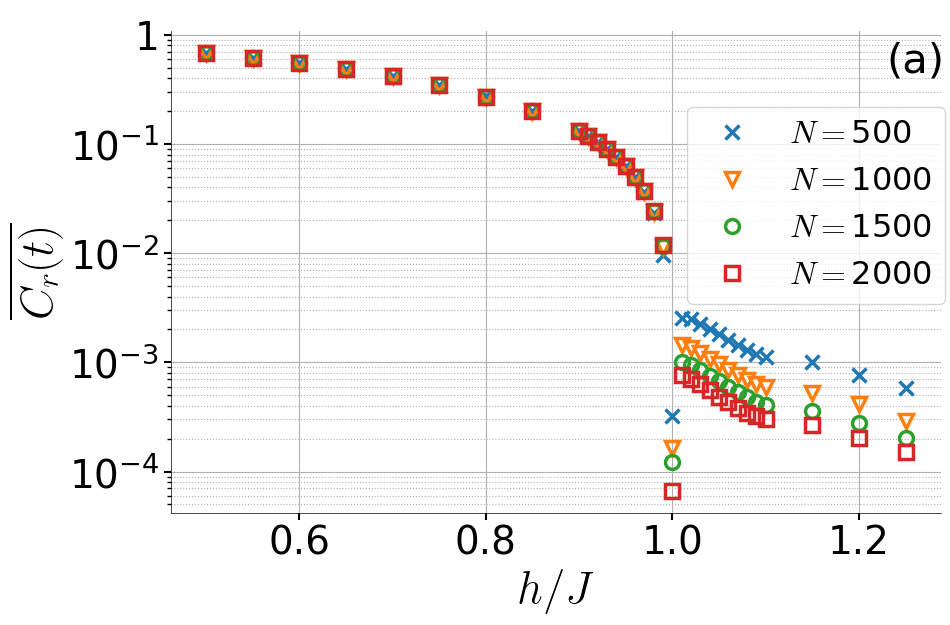}}\hfill 
\subfloat{\label{figS6b}\includegraphics[width=0.33\textwidth]{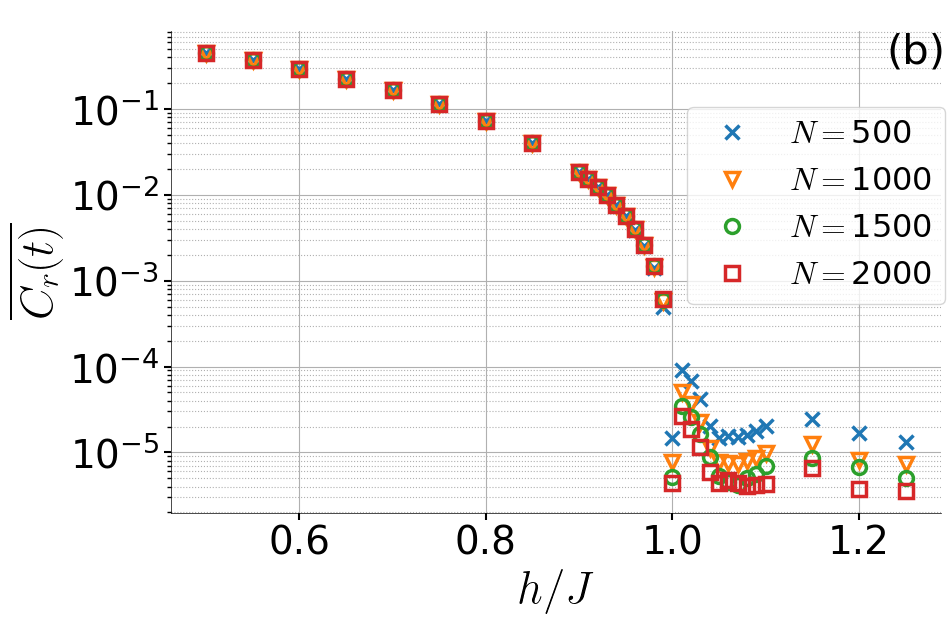}}\hfill 
\subfloat{\label{figS6c}\includegraphics[width=0.33\textwidth]{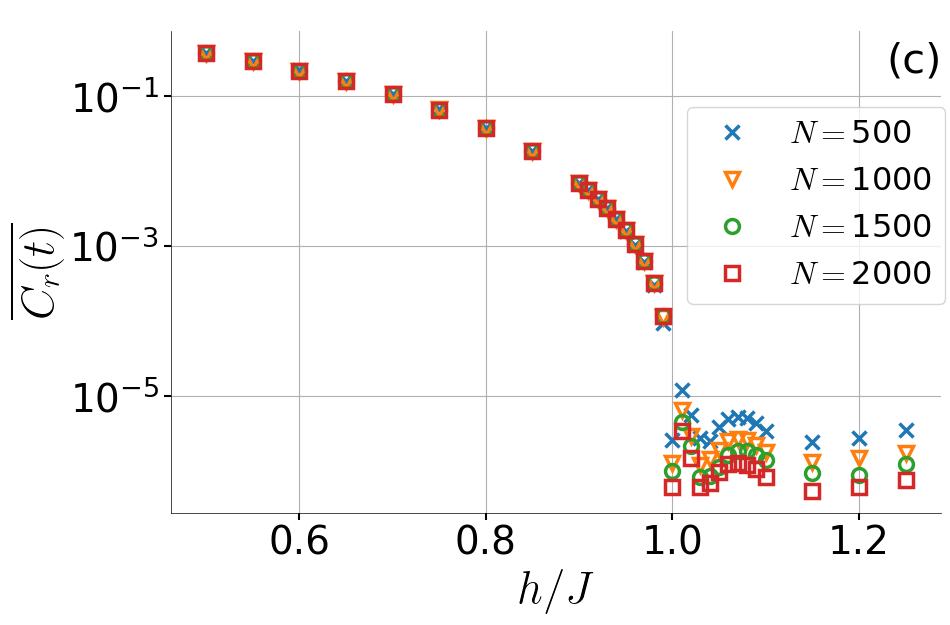}}\hfill
\subfloat{\label{figS4a}\includegraphics[width=0.33\textwidth]{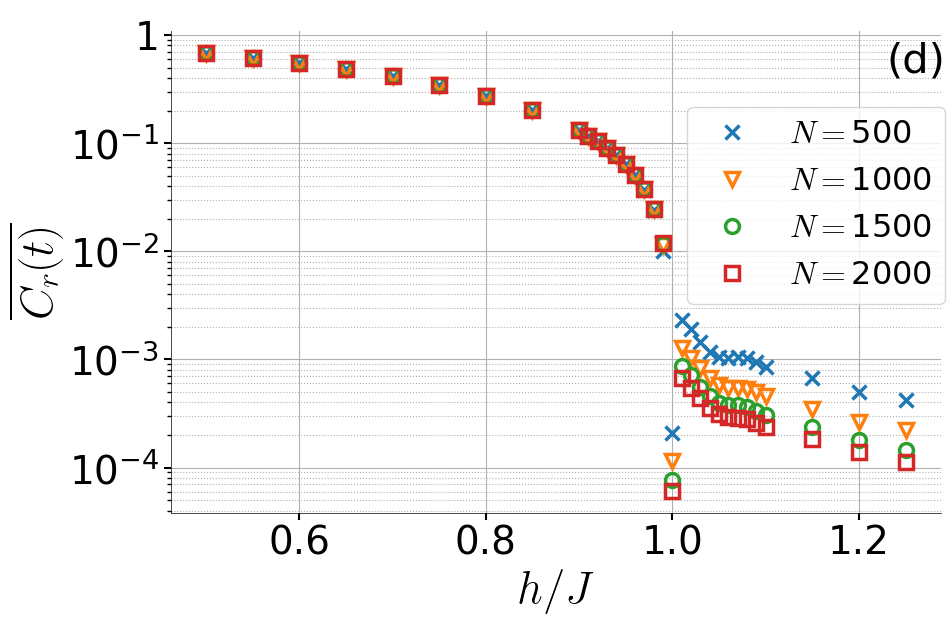}}\hfill 
\subfloat{\label{figS4b}\includegraphics[width=0.33\textwidth]{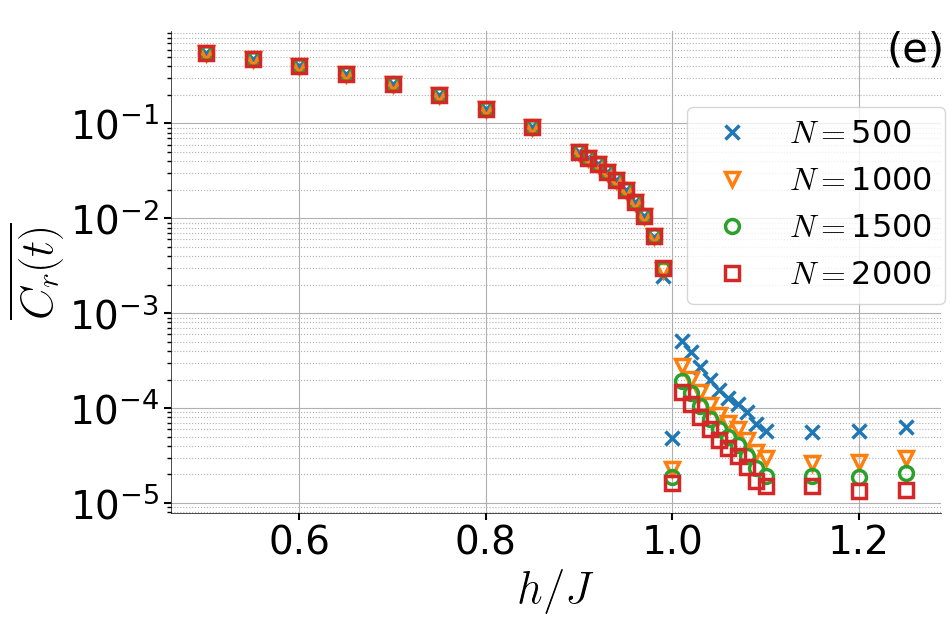}}\hfill 
\subfloat{\label{figS4c}\includegraphics[width=0.33\textwidth]{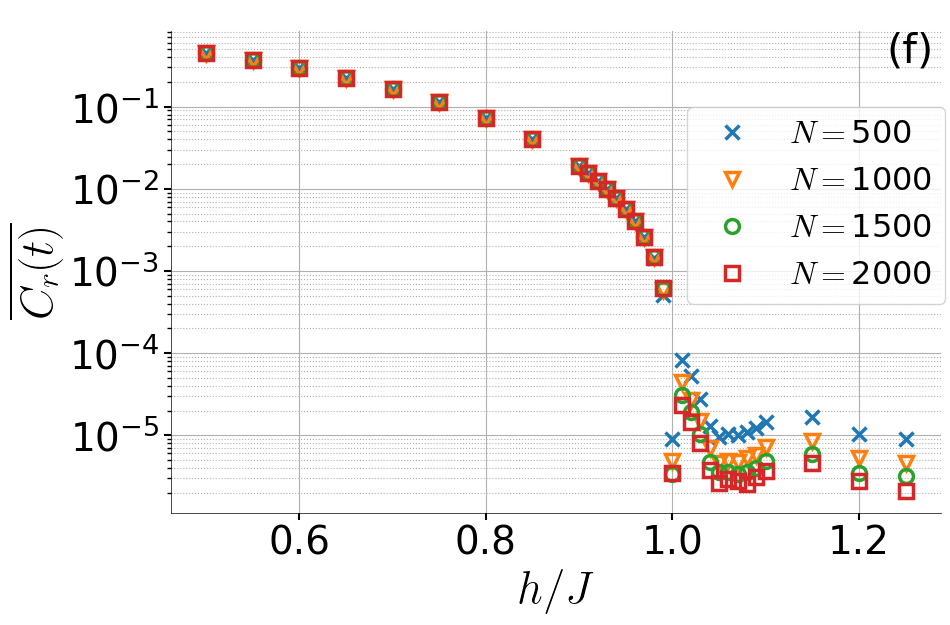}}\hfill 
\subfloat{\label{figS4d}\includegraphics[width=0.33\textwidth]{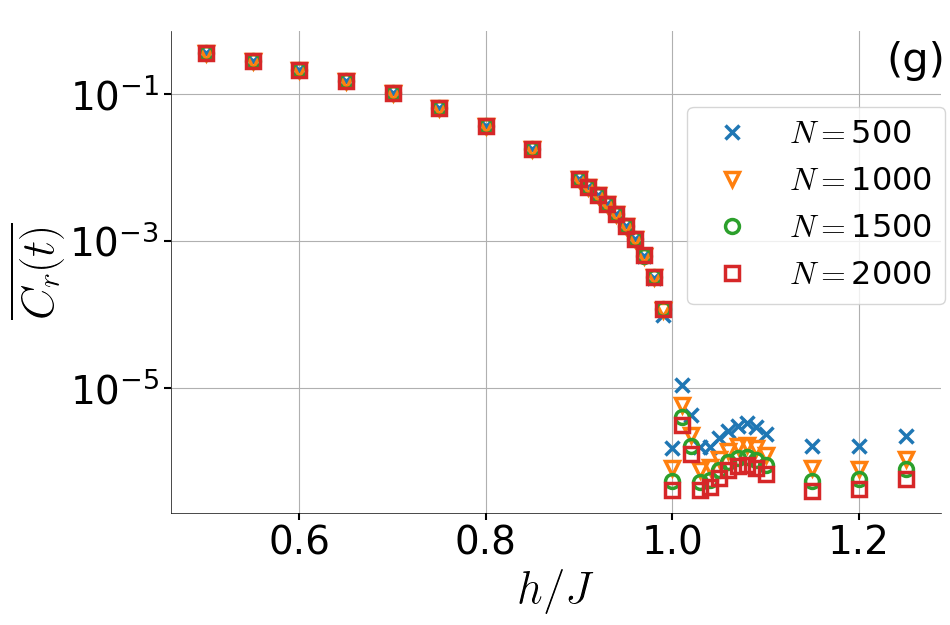}}\hfill
\subfloat{\label{figS5a}\includegraphics[width=0.33\textwidth]{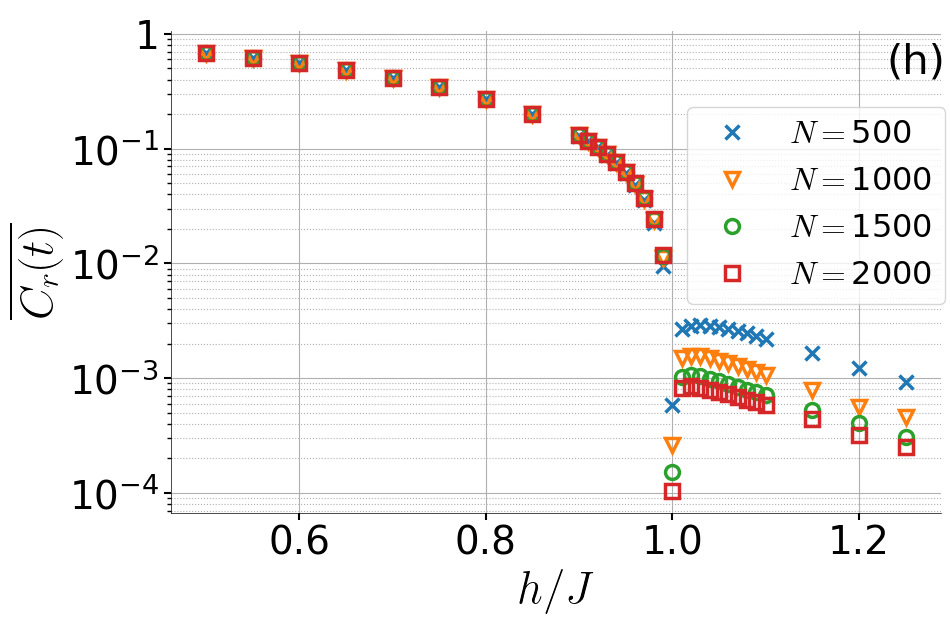}}\hfill 
\subfloat{\label{figS5b}\includegraphics[width=0.33\textwidth]{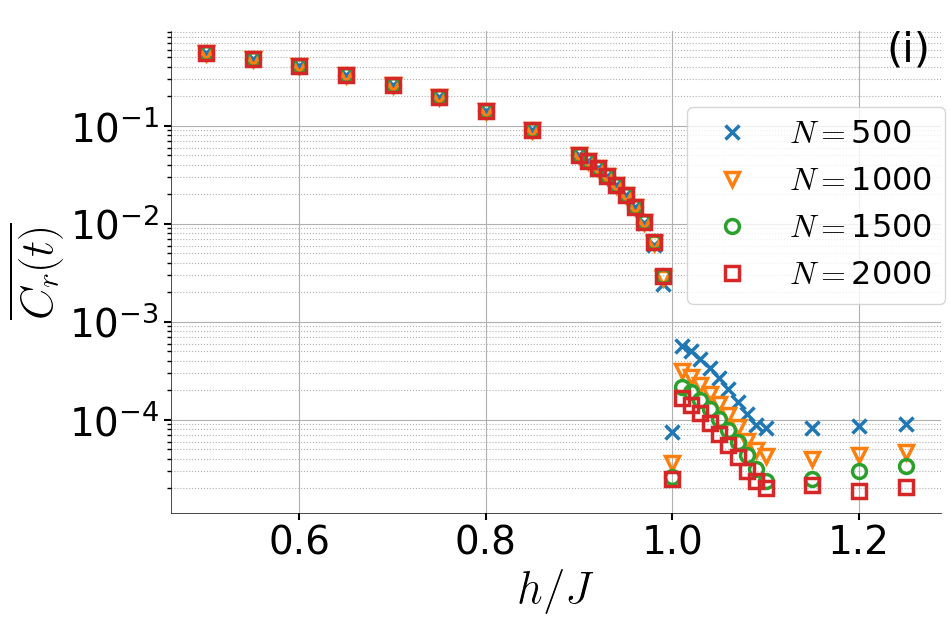}}\hfill 
\subfloat{\label{figS5c}\includegraphics[width=0.33\textwidth]{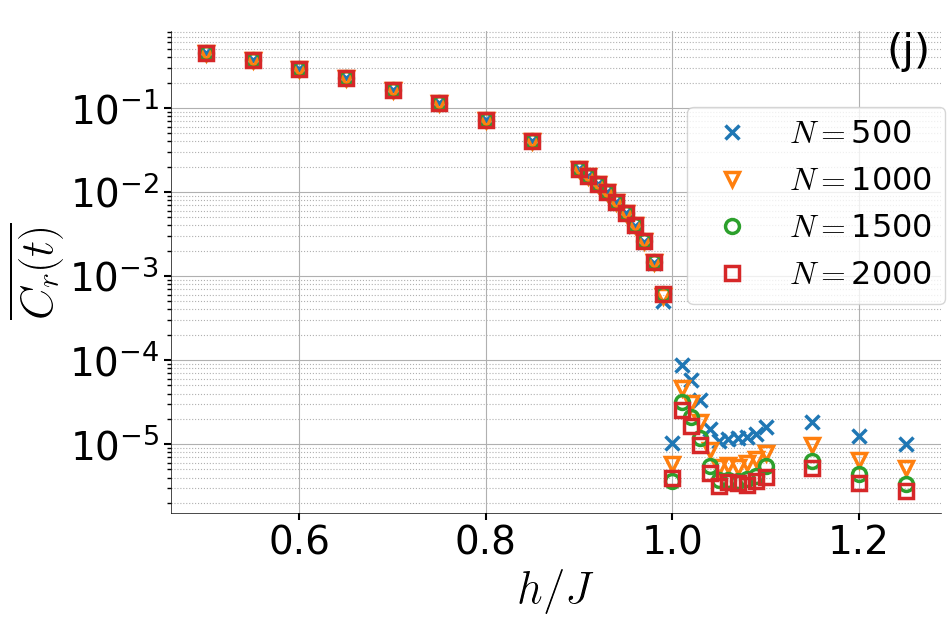}}\hfill 
\subfloat{\label{figS5d}\includegraphics[width=0.33\textwidth]{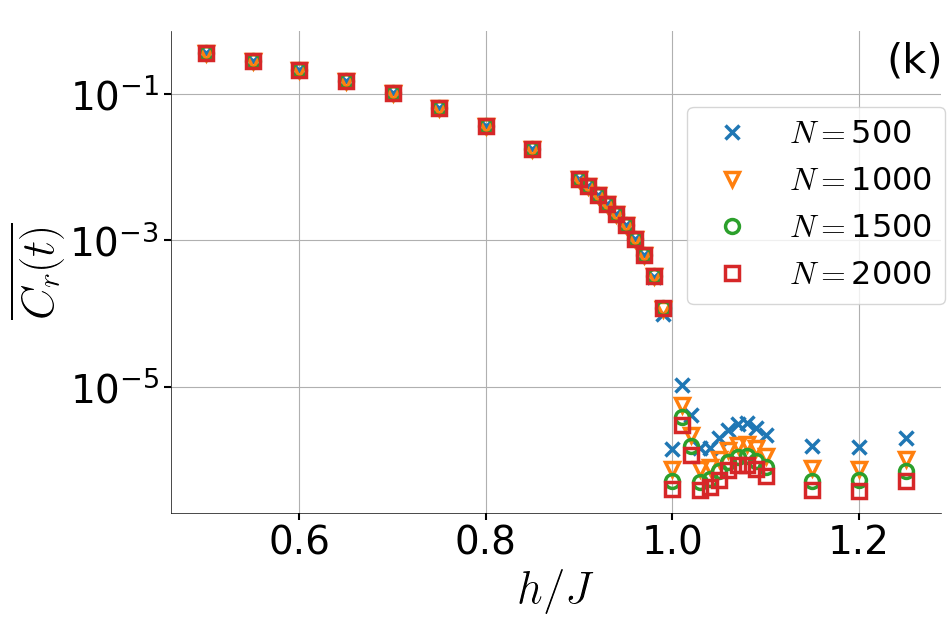}}\hfill
\subfloat{\label{figS7a}\includegraphics[width=0.33\textwidth]{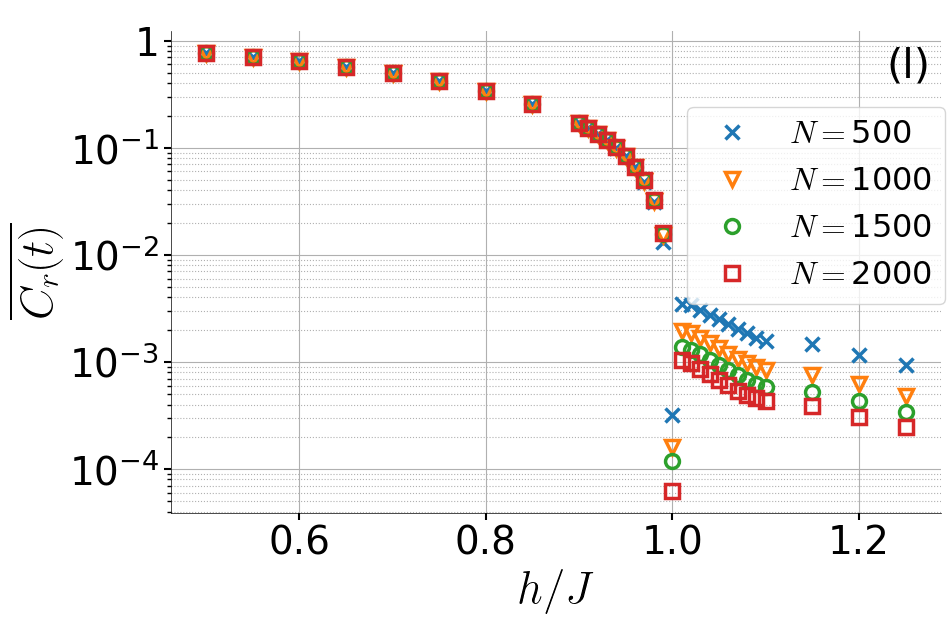}}\hfill 
\subfloat{\label{figS7b}\includegraphics[width=0.33\textwidth]{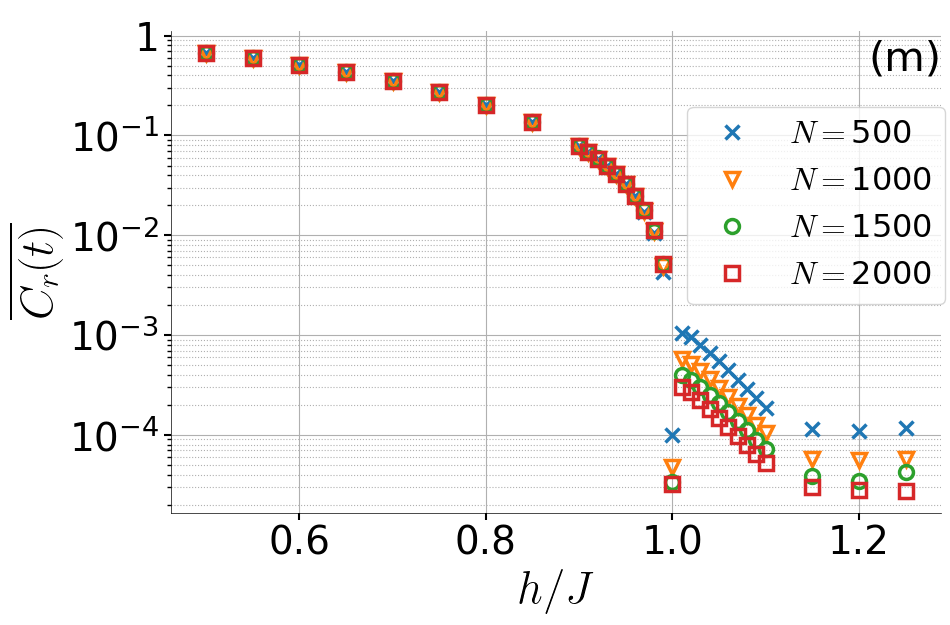}}\hfill 
\subfloat{\label{figS7c}\includegraphics[width=0.33\textwidth]{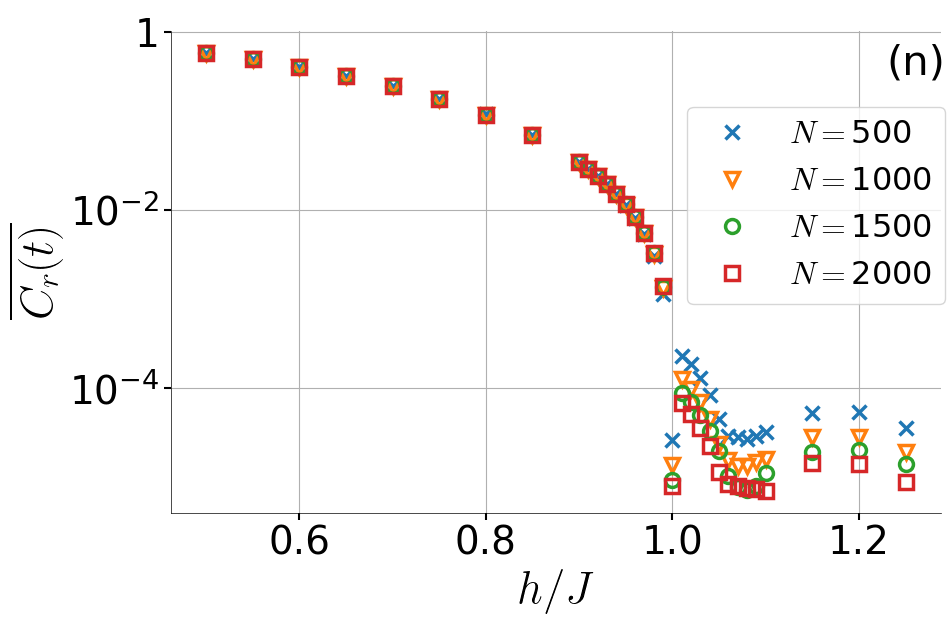}}\hfill 
\subfloat{\label{figS7d}\includegraphics[width=0.33\textwidth]{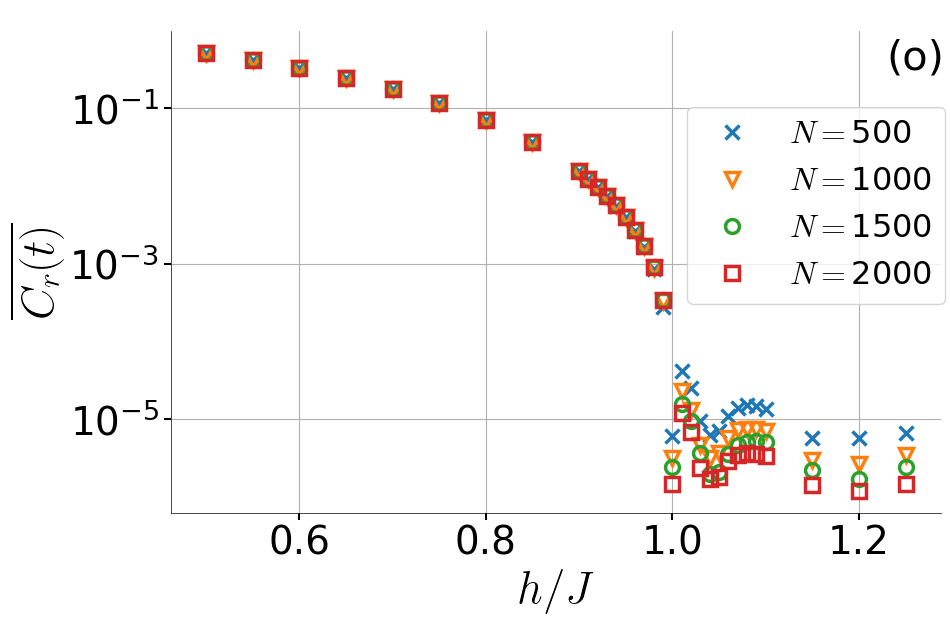}}
\caption{The single-site nonequilibrium phase diagrams of the integrable TFIC with a UV temporal cutoff of $t^*=10$ for (a) $C_{3}(t)$, (b) $C_{9}(t)$ and (c) $C_{12}(t)$; with a UV temporal cutoff of $t^*=20$ for (d) $C_{3}(t)$, (e) $C_{6}(t)$, (f) $C_{9}(t)$ and (g) $C_{12}(t)$; with a UV temporal cutoff of $t^*=2 \alpha \Delta x/v_q$ where $\Delta x= r-1$ with $r$ being the single-site observable location and $\alpha=2$ is a tuning parameter for (h) $C_{3}(t)$, (i) $C_{6}(t)$, (j) $C_{9}(t)$ and (k) $C_{12}(t)$. (l-o) The single-site nonequilibrium phase diagrams with an initial state as the ground state of an initial Hamiltonian with $h_i=0.1$ and a UV temporal cutoff of $t^*=10$ for (l) $C_{3}(t)$, (m) $C_{6}(t)$, (n) $C_{9}(t)$ and (o) $C_{12}(t)$. The behavior is qualitatively the same as the results of $h_i=0$.}
\label{f:phaseDiags_vs_tUV_and_hi}
\end{figure}

We first test whether our results depend on the choice of ultraviolet (UV) cutoff $t^*$. The results in the main text are produced with a fixed UV cutoff of $t^*=10$, for all $h$. However, none of our results depend on the choice of UV cutoff: Subfigures (a-k) in Fig.~\ref{f:phaseDiags_vs_tUV_and_hi} all show the same qualitative behavior for single-site nonequilibrium phase diagrams for various choices of UV cutoff. Figs.~\ref{figS6a}~--~\ref{figS6c} complement the $t^*=10$ data in the main text by showing the nonequilibrium phase diagrams of the single-site magnetization at $r=3,9,12$. Figs.~\ref{figS4a}~--~\ref{figS4d} exhibit another fixed UV cutoff of $t^*=20$, whereas Figs.~\ref{figS5a}~--~\ref{figS5d} demonstrate the results of a parametric UV cutoff for all studied sites. The parametric UV cutoff is determined as follows: We roughly estimate the onset of the q.s. regime as the time required for the quasiparticles to reflect back from the edge closest to the observation site. Therefore, the estimate can be mathematically stated as $t^*=2\alpha \Delta x/v_q$, where the distance $\Delta x=r-1$ is the distance between the observation site, $r=3,6,9,12$ and the closest edge site, $r'=1$, in our case. The parameter $\alpha$ is a tuning parameter, as our analytical formula is only an estimate. In fact we find that $\alpha=2$ presents phase diagrams qualitatively the same as those of other $t^*$, for all $r$ that we studied. 

The long-time asymptotic value of $J_1(4t)/2t$ is $t^{-3/2}$, which is what we observe in Fig.~\ref{figS14a} not only for the edge but for all probe sites $r$. Let us note that the same holds for the near-integrable model treated with qMFT (Fig.~\ref{FigS21}c) as well as the disordered phase in both models (Figs.~\ref{figS14b} and~\ref{FigS21}d). Therefore, it is straightforward to find how the time-averaged magnetization, e.g.,~single-site dynamical order parameters, scale with the system size at the QCP and in the disordered phase. 

\begin{eqnarray}
\overline{C(t,h\geq h_c)} &=& \frac{1}{(N/v_c-t^*)} \int_{t^*}^{N/v_c} t^{-3/2}dt = -2 \frac{1}{(N/v_c-t^*)} t^{-1/2}\bigg \vert_{t^*}^{N/v_c} = -2 \frac{1}{(N/v_c-t^*)} \left(\frac{1}{\sqrt{N/v_c}} - \frac{1}{\sqrt{t^*}} \right) \notag \\
&=& -2 \frac{(4h_c)^{3/2}}{N^{3/2}-4t^* h_c \sqrt{N}} + 2 \frac{4h_c}{(N-4t^*h_c)\sqrt{t^*}}. \label{EqA1}
\end{eqnarray}
Note that $t^*$, the ultraviolet cutoff, has to be introduced in the first line to prevent divergence which is nonphysical. Series expansion for Eq.~\eqref{EqA1} when $N\rightarrow \infty$ gives,
\begin{eqnarray}
\overline{C(t,h \geq h_c)} = \frac{8}{\sqrt{t^*}}\left(\frac{N}{h_c}\right)^{-1} - 16 \left(\frac{N}{h_c}\right)^{-3/2} + 32 \sqrt{t^*} \left(\frac{N}{h_c}\right)^{-2} + \mathcal{O}(N^{-5/2}). \label{EqA2}
\end{eqnarray}

According to Eq.~\eqref{EqA2}, the dynamical order at the QCP decreases as $\propto 1/N$ for all probe sites $r$ and becomes $0$ in the thermodynamic limit. We numerically confirm this prediction in Figs.~\ref{figS6d},~\ref{figS4e} and~\ref{figS5e} with fits of the form $N^{-b}$ where $b \approx 1$. Let us note that the analytical expression is only exact asymptotically in time and the numerics take into account the early-time behavior of $|C_r(t)|$ too. Hence, it is not surprising to see $b \approx 1$ instead of $b=1$ in the numerics. 

\begin{figure}
\centering
\subfloat{\label{figS6d}\includegraphics[width=0.24\textwidth]{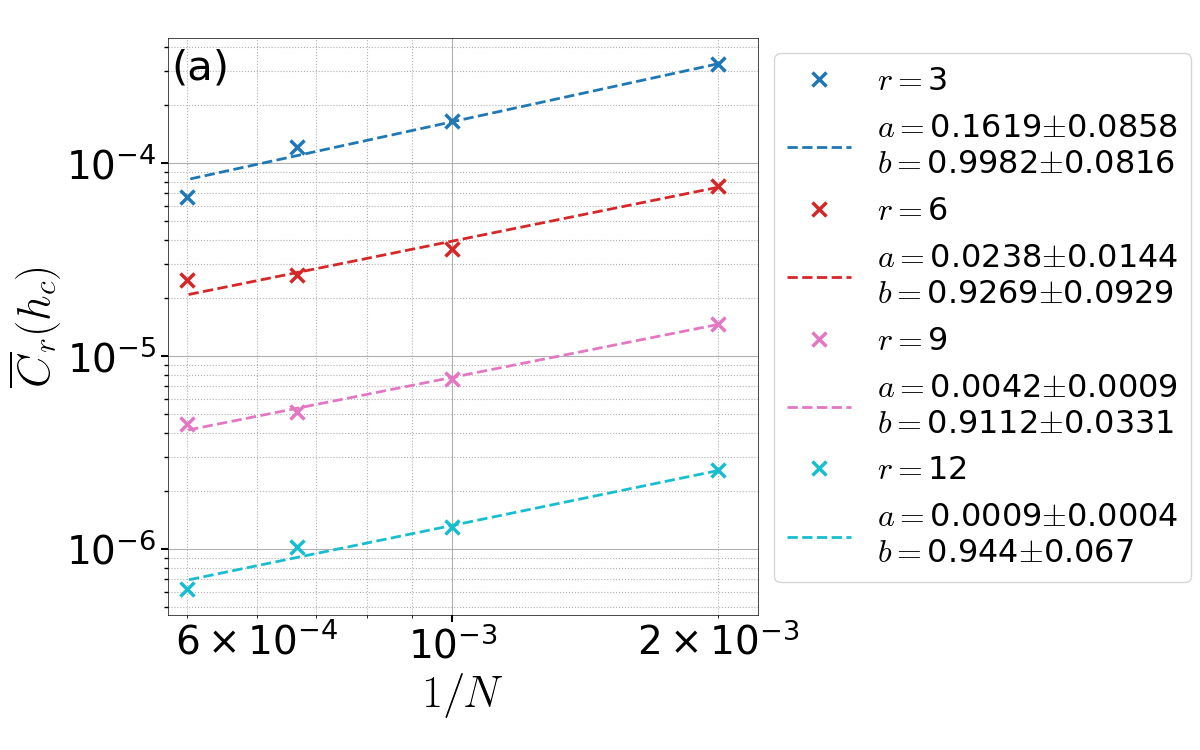}}\hfill
\subfloat{\label{figS4e}\includegraphics[width=0.24\textwidth]{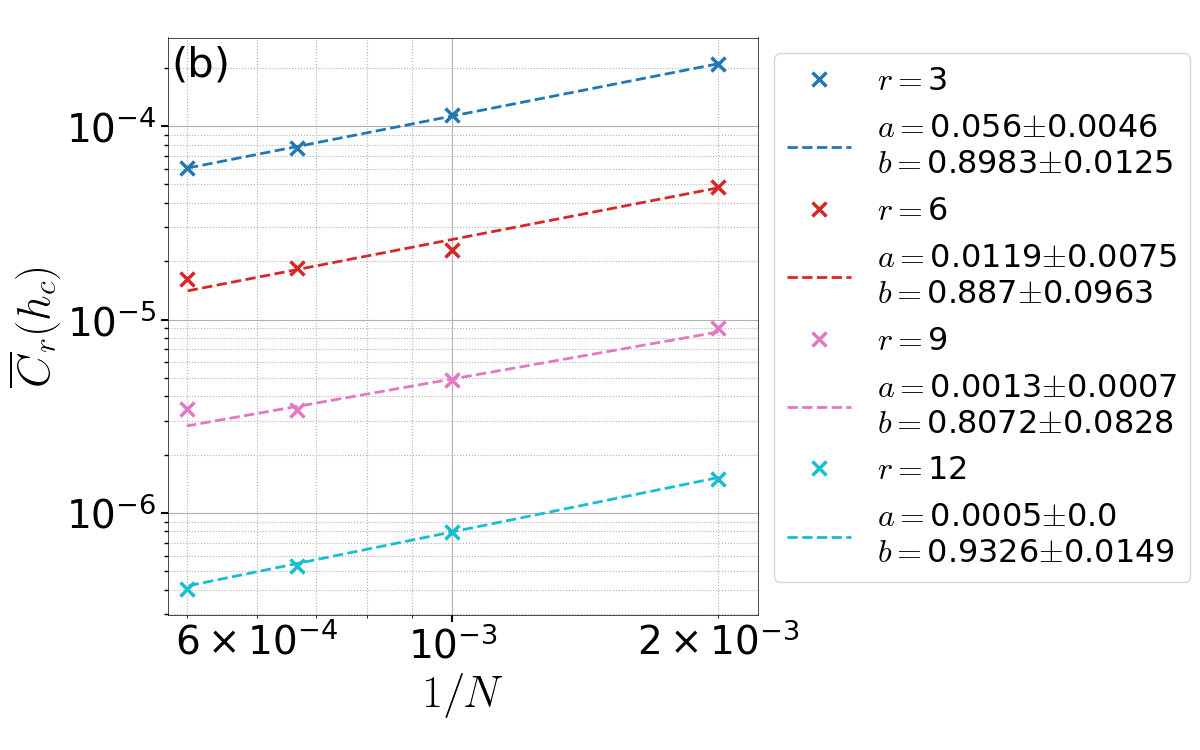}}\hfill 
\subfloat{\label{figS5e}\includegraphics[width=0.24\textwidth]{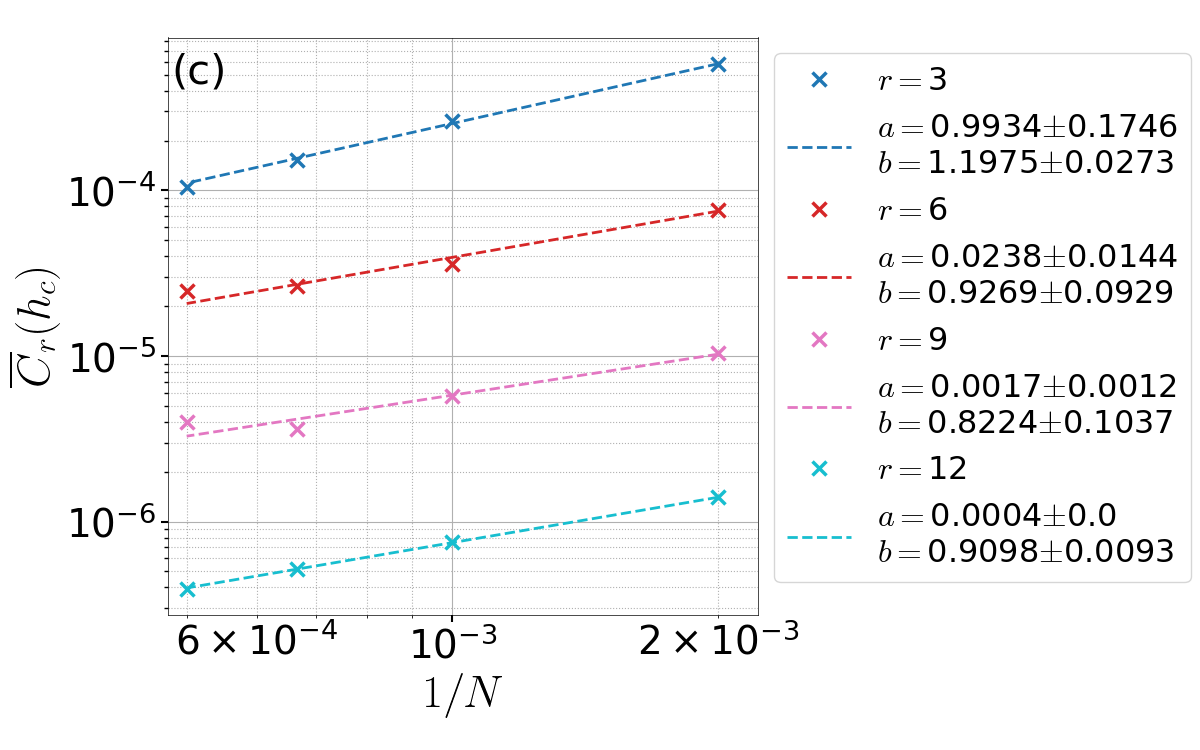}}\hfill 
\subfloat{\label{figS7e}\includegraphics[width=0.24\textwidth]{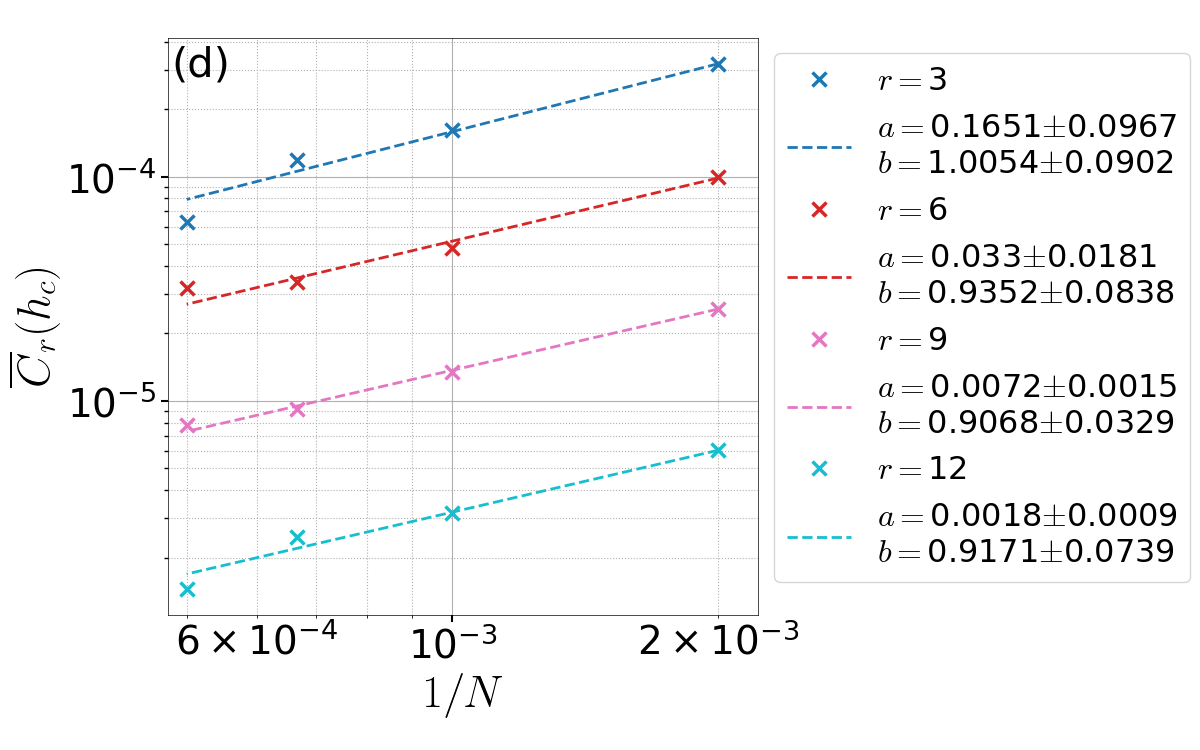}}\hfill 
\caption{The system size scaling of different single-site magnetization at the critical point $h_c$ with an initial state at $h_i=0$ and cutoff  (a) $t^*=10$, (b) $t^*=20$, (c) $t^*=2 \alpha \Delta x/v_q$ and $\alpha=2$, and (d) for an initial state at $h_i=0.1$ and a UV temporal cutoff of $t^*=10$. A downward trend can be seen with $a N^{-b}$ where $b \approx 1$ in all curves. In each subfigure, the curves from up to bottom represent the single sites $r=3, 6, 9,12$, respectively, as seen in the legend.}
\label{FigS6}
\end{figure}

\begin{figure}
\centering
\subfloat[]{\label{figS3a}\includegraphics[width=0.24\textwidth]{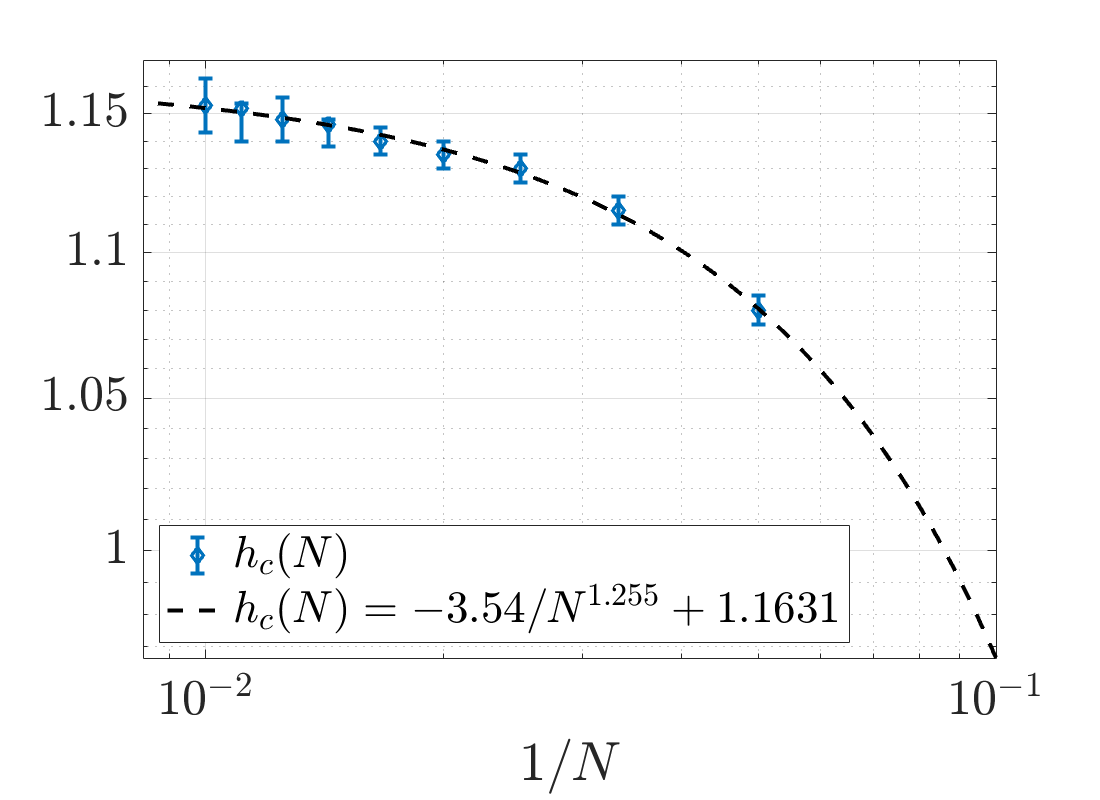}}\hfill
\subfloat[]{\label{figS3b}\includegraphics[width=0.24\textwidth]{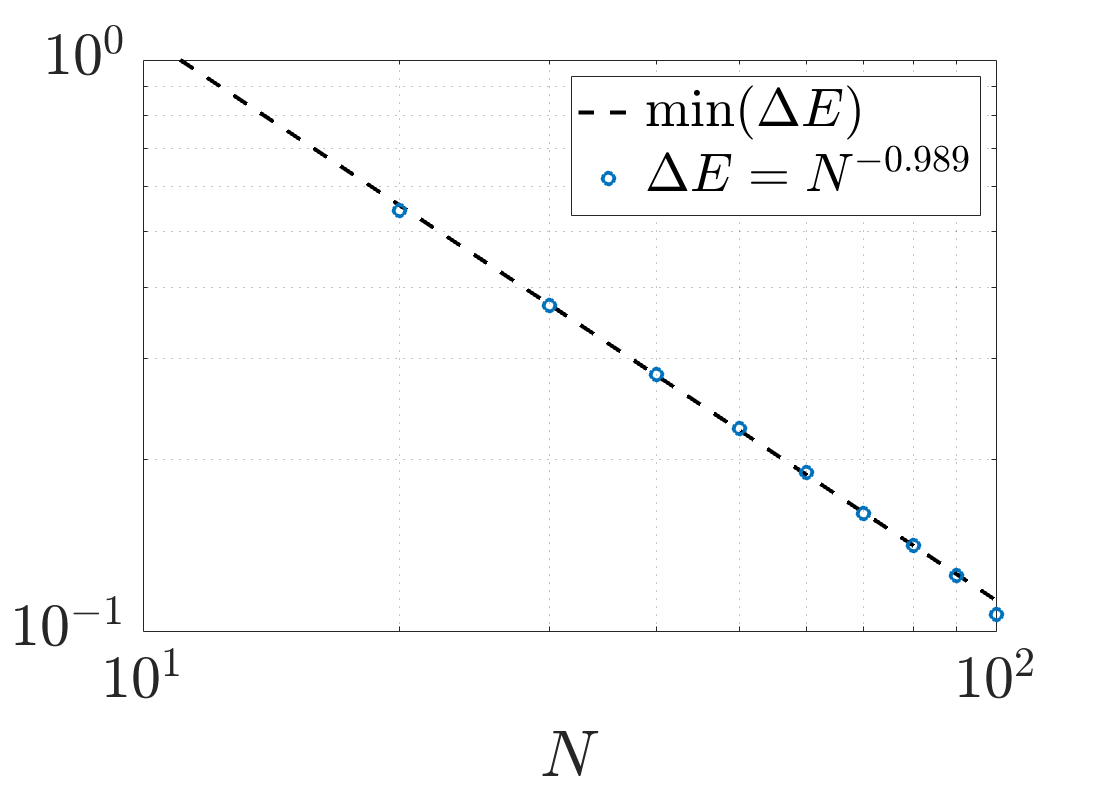}}\hfill 
\subfloat[]{\label{figS3c}\includegraphics[width=0.24\textwidth]{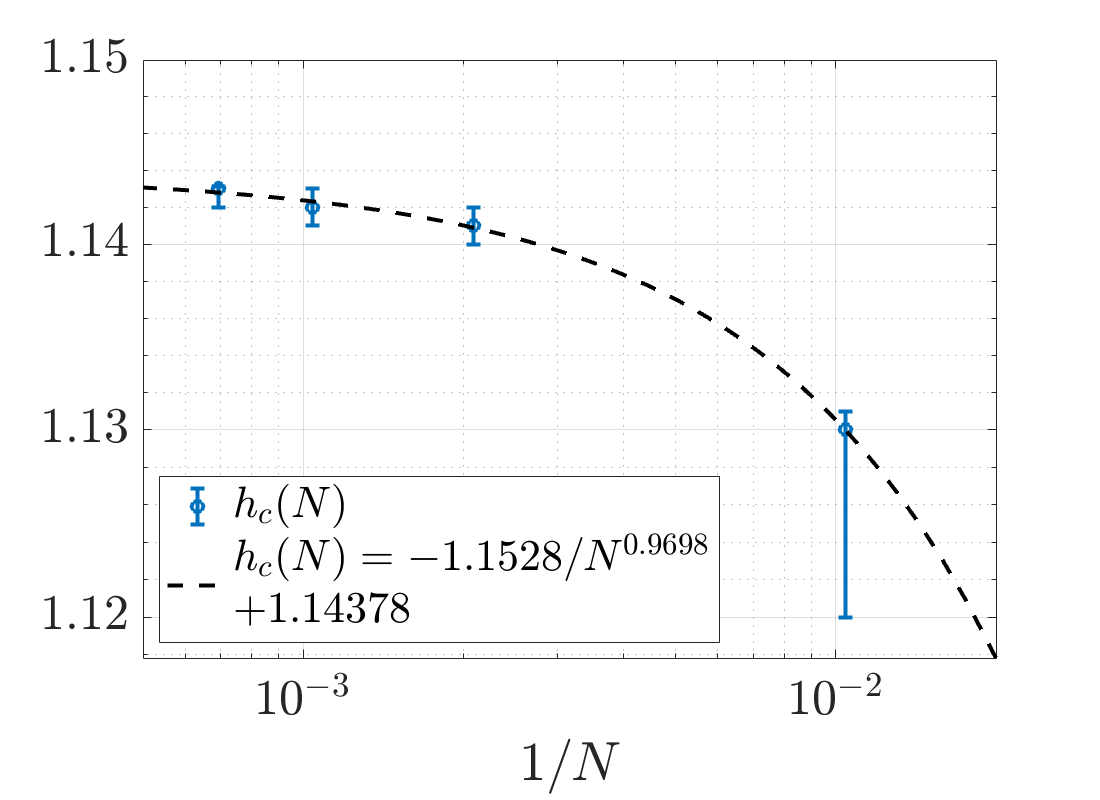}}\hfill 
\subfloat[]{\label{figS3d}\includegraphics[width=0.24\textwidth]{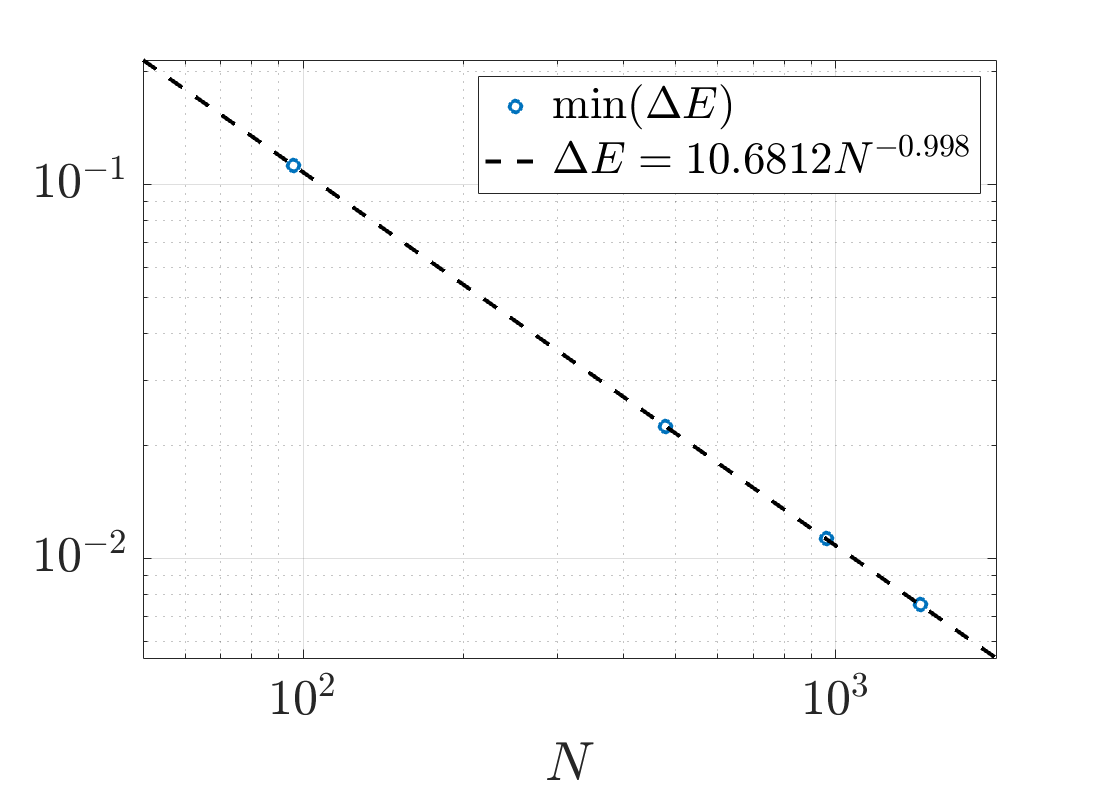}}\hfill 
\caption{The energy gap analysis (a-b) for the near integrable model with DMRG, (c-d) for the effective noninteracting Hamiltonian derived through qMFT method. (a,c) The critical points are marked $h_c=1.1631\pm 0.0037$ and $h_c^{\text{qmft}}=1.1438 \pm 0.0011$, respectively. (b,d) The energy gap scales with system size as $1/N$ in both.}
\label{FigS3}
\end{figure}

\section{\label{AppB}Initial state independence of the results}

In this subsection, we change the initial state to the ground state of an initial Hamiltonian with $h_i=0.1$, and test whether any of our results depend on the initial state. Figs.~\ref{figS7a}~--~\ref{figS7d} show the single-site nonequilibrium phase diagrams computed with this initial state. We do not observe a change in the qualitative behavior. The single-site magnetization at the critical point still decreases with increasing system size in Fig.~\ref{figS7e} by exhibiting $b\approx 1$. 

The analytical expression for the q.s. value of the edge magnetization in the ordered phase is,
\begin{eqnarray}
C^{qs}_{1}(h,h_i)=\frac{(1-h^2)(1-2h_i)^{1/2}}{1-2 h h_i}, \label{edgeM}
\end{eqnarray}
for $h,h_i<1$. Let us rewrite it in terms of the reduced control parameter $h_n=(h_c-h)/h_c$ as
\begin{eqnarray}
C^{qs}_{1}(h_n,h_i)=\frac{(2-h_n)h_n(1-2h_i)^{1/2}}{1+2(h_n-1) h_i}.
\end{eqnarray}
In the vicinity of the transition, $h_n \rightarrow 0$, we can expand this expression and find
\begin{eqnarray}
C^{qs}_{1}(h_n \rightarrow 0,h_i) &=& a_1(h_i) h_n + \mathcal{O}(h_n^2) %a_2(h_i) h_n^2 + a_3(h_i) h_n^3 + \cdots,\notag 
\\
a_1(h_i) &=& 2 + h_i + \frac{3}{4} h_i^2 + \frac{5}{8} h_i^3 + \cdots, \notag
%a_2(h_i) &=& -1 - \frac{5}{2} h_i - \frac{27}{8} h_i^2 - \frac{65}{16} h_i^3 + \cdots, \notag\\
%a_3(h_i) &=& h_i + \frac{7}{2} h_i^2 + \frac{55}{8} h_i^3 + \cdots.\notag
\end{eqnarray}
Therefore, one can see that $C^{qs}_{1}(h_n) \propto h_n$ in the vicinity of the transition, $h_n \rightarrow 0$, regardless of the choice of initial state. The initial state only changes the coefficient in front of $h_n$, which is nonuniversal.

\section{\label{AppC}Energy gap analysis}

Here we present the ground state energy gap analysis of the near integrable model $\Delta=0.1$ calculated with DMRG. Then we discuss the single-particle energy gap analysis for the effective Hamiltonian produced by the qMFT method.

Figs.~\ref{figS3a} shows how the location of the ground state energy gap minimum scales with system size. This finite-size scaling analysis computed with DMRG gives $h_{c} \approx 1.1631$. Meanwhile the energy gap at the QCP scales as $N^{-1}$ which is shown in Fig.~\ref{figS3b}. 

qMFT method provides us an effective Hamiltonian for the near-integrable model. In Fig.~\ref{figS3c} we focus on the minimum of the energy gap of this Hamiltonian. We observe that location of the minimum scales with the system size giving $h_{c}^{\text{qmft}}\approx 1.14378$. The energy gap at $h_{c}^{\text{qmft}}$ scales with the system size as $N^{-1}$. This result is consistent with the observation that a critically prethermal regime appears as we suddenly quench to the vicinity of the DCP of the near-integrable model in Ref.~\cite{letter}. Because the DCP of the near-integrable model, $h_{dc}\approx 1.1437$ is almost equal to the QCP of the effective qMFT Hamiltonian. Then the duration of the prethermal regime is actually governed by the QCP dynamics of the effective qMFT Hamiltonian in~\cite{letter}.

\section{\label{AppD}Changing the boundary conditions}

\begin{figure}
\centering{\includegraphics[width=0.35\textwidth]{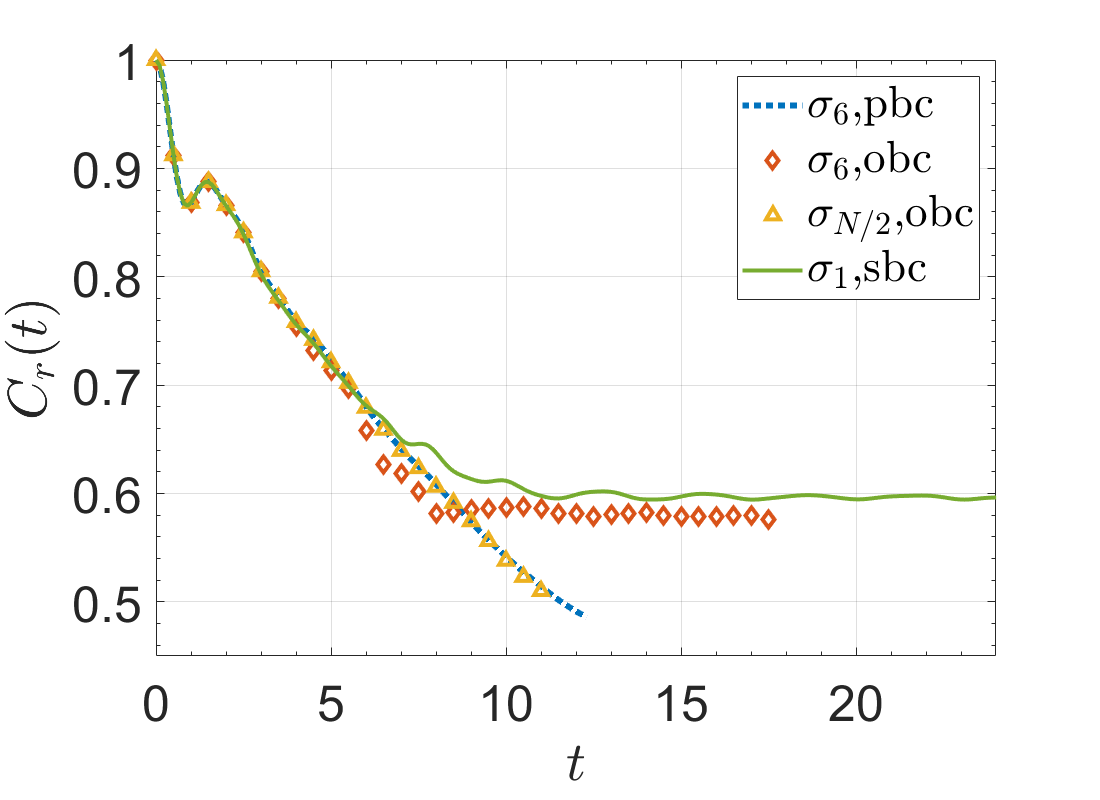}}\hfill 
\caption{Different boundary conditions are compared for the integrable TFIC at $h=0.5$ and system size $N=24$. One can obtain a q.s. regime with smooth boundary conditions too. OBC, PBC and SBC stand for open, periodic and smooth boundary conditions, respectively.}
\label{FigS12}
\end{figure}

We demonstrate that the q.s. temporal regime emerges not only when we introduce hard boundaries \cite{PhysRevLett.106.035701}, but also for smooth boundaries. A smooth boundary condition can be applied by smoothly turning off the Hamiltonian parameters towards the edges of the chain \cite{PhysRevLett.71.4283}. Fig.~\ref{FigS12} shows the single-site nonequilibrium responses of the integrable TFIC with hard boundarieas (red-diamonds), smooth boundaries (green-solid) and periodic boundary condition (blue-dotted). As shown before \cite{2020arXiv200412287D}, the middle of a hard-boundary chain (yellow triangles) acts like an arbitrary site in a periodic chain. We thus conclude that the q.s. regime is robust against altering the boundary conditions, so long as they remain open. 

\section{\label{AppE}Benchmarking the qMFT method}

\begin{figure}
\centering
\subfloat[]{\label{figS0a}\includegraphics[width=0.33\textwidth]{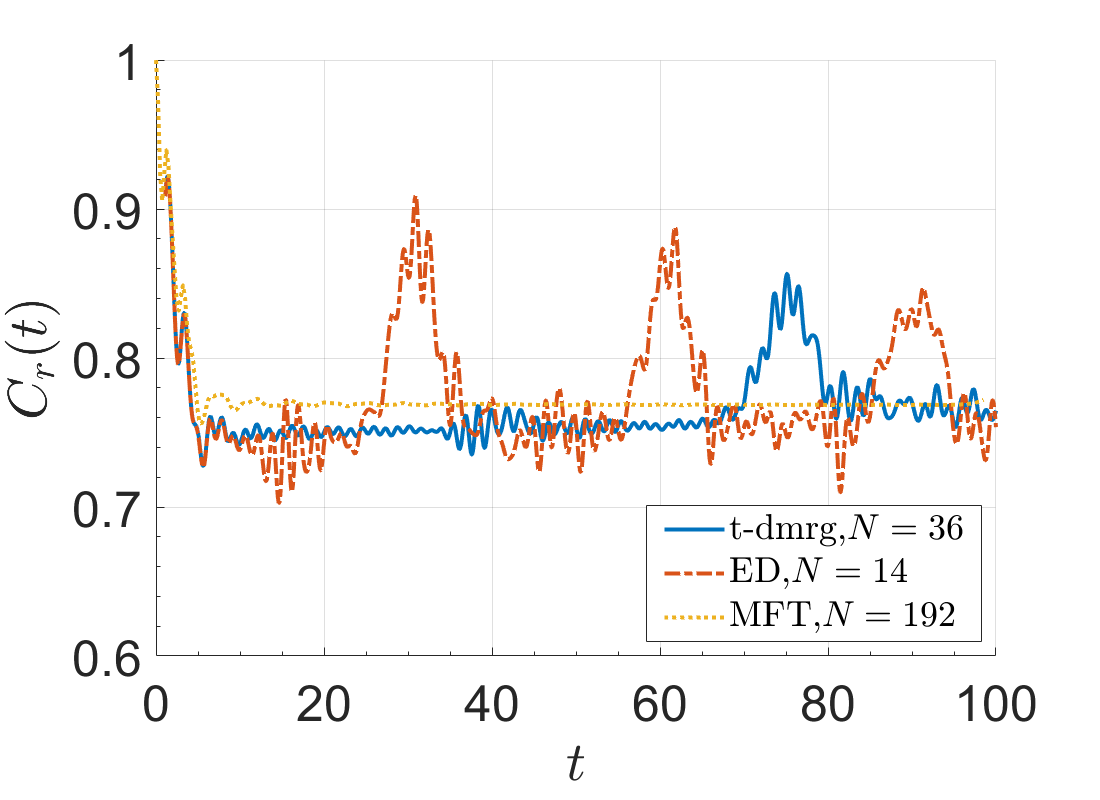}}\hfill 
\subfloat[]{\label{figS0b}\includegraphics[width=0.33\textwidth]{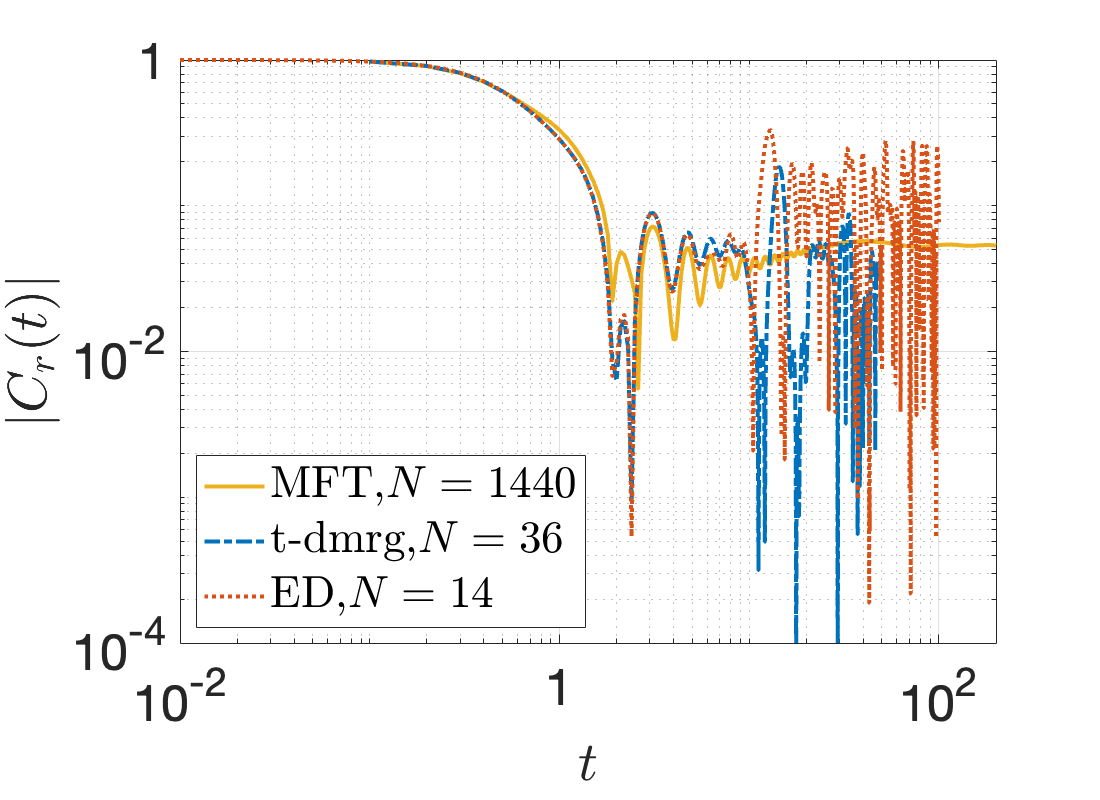}}\hfill 
\subfloat[]{\label{figS0c}\includegraphics[width=0.33\textwidth]{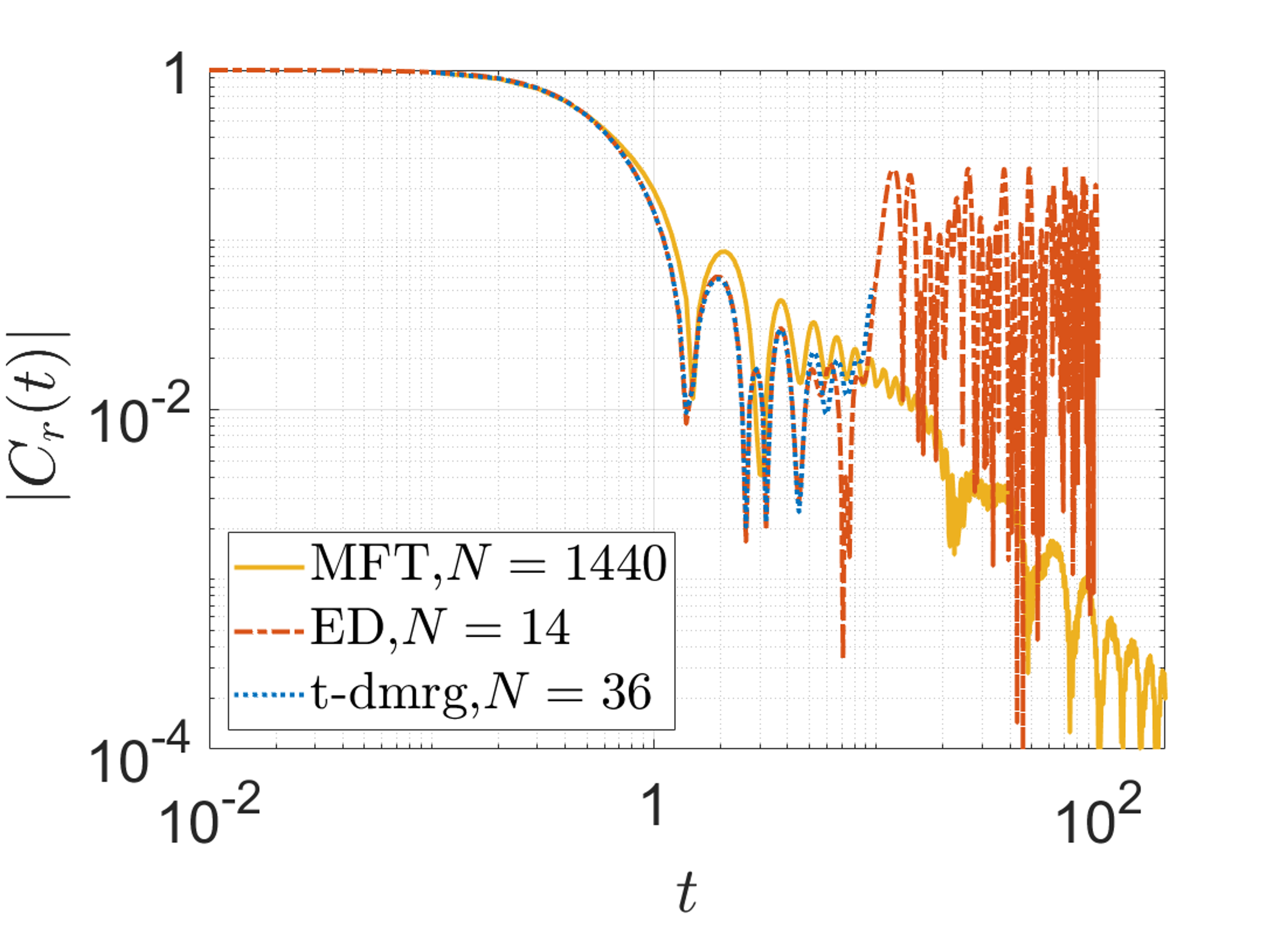}}\hfill 
\caption{Benchmarking qMFT analysis. All subfigures compare the results of qMFT, $t$-DMRG and exact diagonalization (ED) results (see individual legends for system sizes) for $C_{3}(t)$. The external fields are (a) $h=0.5$, (b) $h=1.1$ and (c) $h=1.2$. In all subfigures, the qMFT nonequilibrium response matches sufficiently well with the nonequilibrium responses of the exact methods, in particular in early times close to QPT and in the disordered phase where exact methods are constrained either by finite-size (ED algorithm) or finite bond dimension and short simulation times (t-DMRG algorithm).}
\label{FigS0}
\end{figure}

\begin{figure}
\centering
\subfloat[]{\label{fig2e}\includegraphics[width=0.45\textwidth]{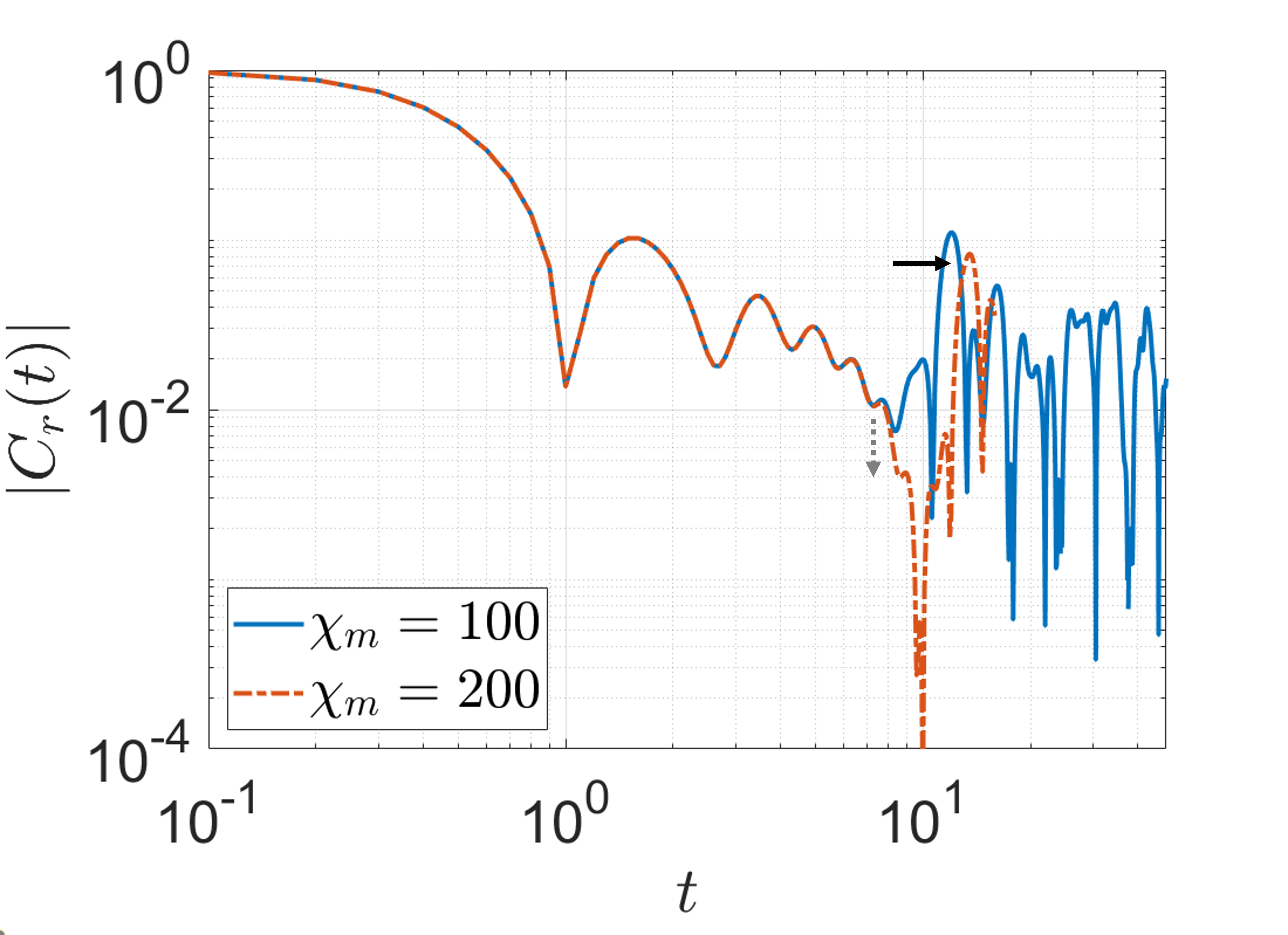}}
\subfloat[]{\label{fig2f}\includegraphics[width=0.45\textwidth]{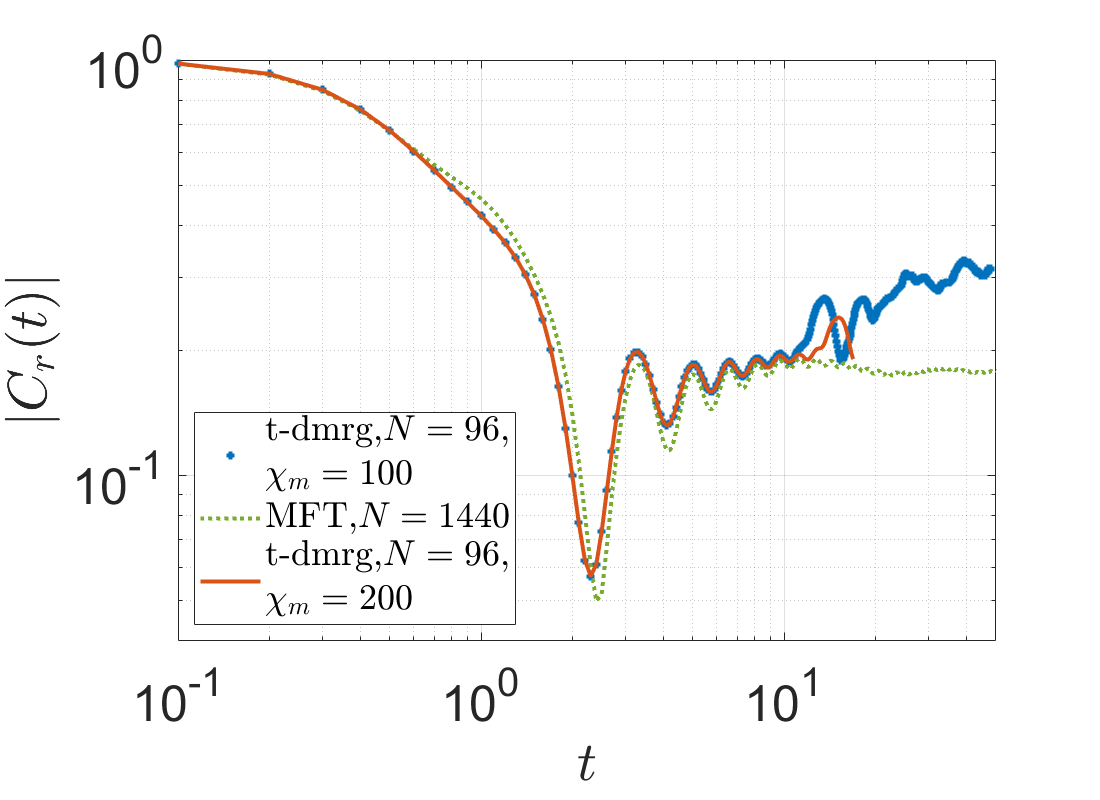}}\hfill 
\caption{Non-equilibrium responses calculated with t-DMRG for weakly interacting TFIM with $\Delta/J=0.1$ at $N=96$ are compared with respect to different maximum bond dimensions, $\chi_m=100$ and $\chi_m=200$ which is introduced as an approximation in the t-DRMG algorithm to simulate for longer times, see Sec.~\ref{t-DMRG}. The comparison is shown in subfigures (a) at $h/J=1.3$ and (b) at $h/J=1$. (a) With increasing bond dimension, a downward moving envelope appears in the t-DMRG result, emphasized by the dotted-gray arrow, agreeing with the qMFT results in the disordered phase. The solid-black arrow points to the effects of finite bond dimension. (b) The qMFT result at $N=1440$ is compared with the t-DMRG results. We notice the convergence of the t-DMRG result with greater $\chi_m$ (red-solid) to the qMFT result (green-dotted) for a small time interval that was accessible, before showing the effects of finite bond dimension.}
\label{Fig2}
\end{figure}

Fig.~\ref{FigS0} compares the results of qMFT, $t$-DMRG and exact diagonalization (ED) at $h=0.5$ and $h=1.1$ in the ordered phase and $h=1.2$ in the disordered phase. We observe that the qMFT analysis can even capture the correct frequency of the oscillations in early times and the general trend of the nonequilibrium response successfully, although it does not totally match with the exact methods, which is expected due to the fact that it is an approximate method that averages out the interactions. In Fig.~\ref{figS0a}, both the ED and t-DMRG responses show finite-size effects in the form of quantum revivals (solid-blue and red-dashed-dotted). The finite-size effects of these methods can also be clearly seen in Figs.~\ref{figS0b} and~\ref{figS0c}. As we approach the QCP and in the disordered phase, the match between the exact methods and the qMFT does not survive past $t=10$. This is either because of the finite-size effects, as apparent in ED response, or the finite maximum bond dimensions set in the t-DMRG algorithm, see Sec.~\ref{t-DMRG} for more details. In order to demonstrate the effects of finite bond dimension, we compare the nonequilibrium responses calculated with t-DMRG for different maximum bond dimensions. Fig.~\ref{Fig2} shows these responses at $h/J=1.3$ in subfigure (a) and at $h=1$ in subfigure (b). We observe that the response with bigger maximum bond dimension, $\chi_m=200$ departs from the response with $\chi_m=100$ at around $t\approx 10$, exhibiting a downward envelope and hence agreeing with the qMFT nonequilibrium responses in the disordered phase, see Fig.~\ref{fig2e}. This feature is shown with a grey-dotted arrow. As time increases, the nonequilibrium response with $\chi_m=200$ starts to exhibit features similar to those of the response with $\chi_m=100$, e.g.,~recurrences, as pointed out with a black-solid arrow. We note that these are effects of finite maximum bond dimension set in our t-DMRG algorithm. One could predict that as the maximum bond dimension $\chi_m$ increases, such effects will occur later in time, and instead the response will follow closely to that of qMFT response with an oscillatory downward trend. However, there is a trade-off between the maximum bond dimension and the maximum accessible simulation time, and simulating longer times while achieving a satisfactory precision for the response is simply out of our computational reach. Fig.~\ref{fig2f} shows the same comparison between different $\chi_m$ alongside with the qMFT result at $h=1$, which is a point in the ordered phase. The $t$-DMRG result with $\chi_m=100$ shows an upward trend as time increases, in contrast to the qMFT result which reaches a plateau, as expected from our theory based on noninteracting fermions. We also plot the response with $\chi_m=200$ and observe that the nonequilibrium response of greater maximum bond dimension approaches that of the qMFT results, departing from the t-DMRG result with $\chi_m=100$ for a small interval of time. Additionally we notice that the response with $\chi_m=200$ (red-solid) also demonstrates effects of finite maximum bond dimension in later times with a recurrence similar to what is observed in the response with smaller bond dimension $\chi_m=100$ (blue-star). Therefore, we argue that as $\chi_m$ increases, the t-DMRG result should converge to the qMFT result. This observation reveals an interesting feature where the approximate qMFT result can be obtained in full quantum dynamics only with sufficiently large maximum bond dimensions. 
\end{widetext}

\end{document}